# Estimation of treatment policy estimands for continuous outcomes using off treatment sequential multiple imputation

Thomas Drury [1], Juan J Abellan [2], Nicky Best [1], Ian R. White [3]

1. GSK, London UK
2. GSK, London – now working for EMA, Amsterdam, NL
3. MRC Clinical Trials Unit at UCL, University College London

## Abstract

The estimands framework outlined in ICH E9 (R1) describes the components needed to precisely define the effects to be estimated in clinical trials, which includes how post-baseline "intercurrent" events (IEs) are to be handled. In late-stage clinical trials, it is common to handle intercurrent events like "treatment discontinuation" using the treatment policy strategy and target the treatment effect on all outcomes regardless of treatment discontinuation. For continuous repeated measures, this type of effect is often estimated using all observed data before and after discontinuation using either a mixed model for repeated measures (MMRM) or multiple imputation (MI) to handle any missing data. In basic form, both of these estimation methods ignore treatment discontinuation in the analysis and therefore may be biased if there are differences in patient outcomes after treatment discontinuation compared to patients still assigned to treatment, and missing data being more common for patients who have discontinued treatment. We therefore propose and evaluate a set of MI models that can accommodate differences between outcomes before and after treatment discontinuation. The models are evaluated in the context of planning a phase 3 trial for a respiratory disease. We show that analyses ignoring treatment discontinuation can introduce substantial bias and can sometimes underestimate variability. We also show that some of the MI models proposed can successfully correct the bias but inevitably lead to increases in variance. We conclude that some of the proposed MI models are preferable to the traditional analysis ignoring treatment discontinuation, but the precise choice of MI model will likely depend on the trial design, disease of interest and amount of observed and missing data following treatment discontinuation.

## 1 Introduction

The ICH E9 (R1) addendum, "Estimands and sensitivity analysis for clinical trials", introduces the estimands framework and argues it should be used to precisely describe the effects to be estimated in clinical trials [1]. The framework defines the five components required for an estimand: a target population, an outcome variable, specified treatment conditions, a summary measure, and a list of intercurrent events (IEs) with the strategies used to handle them. IEs are events that occur after initiation of treatment and impact the interpretation or existence of the primary outcome variable of interest. The guideline discusses several strategies to deal with IEs, including the treatment policy approach, which aims to assess the effect of treatment on outcomes for all patients regardless of the IE occurring.

A fundamental IE in most clinical trials is discontinuation from the protocol-assigned treatment. Adopting a treatment policy approach for this IE implies that outcomes after treatment discontinuation are relevant for the effect of interest. In this setting, patient discontinuation from treatment and patient withdrawal from the study (also referred to as withdrawal from follow up) should be regarded as separate events, with the former being the IE and the latter creating missing data.

When a treatment policy strategy is chosen to deal with treatment discontinuation, patients should be encouraged to provide data after discontinuing from treatment until they complete follow up. However, many patients are likely to withdraw from the study at the same time as they discontinue from treatment or at some intermediate time before the primary assessment, creating a missing data problem.

In this setting it is typical for estimation methods to use all the observed outcome data pre and post discontinuation and rely on a basic missing at random (MAR) assumption. This assumption – which we will refer to as a '*common MAR*' assumption – states that, given the data collected up to the time of study withdrawal, any missing outcomes would be comparable with observed outcomes from similar patients remaining in the trial. Importantly, under this common MAR assumption, post-baseline factors other than time of study withdrawal and outcome data until that timepoint are not used to inform about the similarity between patients.

In reality, this type of MAR assumption is likely to be implausible if either the probability of missingness and/or the expected outcomes differ between the pre and post discontinuation data. When patients withdrawing from a study are no longer on protocol-assigned treatment, it may be more sensible to assume their missing data are more similar to observations collected after treatment discontinuation.

This paper has two aims. First, we propose a set of multiple imputation (MI) models for continuous repeated outcomes which use observed pre and post discontinuation outcomes to impute missing post-discontinuation data, for use in trials with a continuous endpoint where treatment discontinuation is handled with a treatment policy strategy. The MI models are easy to implement in standard software and flexible, allowing for different behaviour before and after discontinuation. Second, we use a simulation study to assess the performance of the models for estimating the treatment effects in different scenarios. We compare results with the analysis of the full simulated data (i.e. without missing values), standard repeated measures analyses, and other MI models. Based on the simulation study we also make recommendations about which models may be more suitable in different situations.

MI methods that use past data and treatment discontinuation status have been considered for recurrent-event data [2] and event time data [3]. Other MI models have also been proposed for continuous data [4-7]. This work adds to the toolbox of potential estimation methods and is intended for study teams that are targeting treatment policy estimands for continuous endpoints to help them specify suitable models for imputing missing data when post discontinuation outcomes are only partially collected. Example SAS code for all models is provided via GitHub [11].

For clarity, this paper refers to patient outcomes before treatment discontinuation as “on-treatment” and outcomes after treatment discontinuation as “off-treatment”. The label off-treatment is used because patients no longer receive the initially randomized treatment, but it is acknowledged that off-treatment patients could go on to receive rescue or alternative treatments as part of a clinical trial protocol.

The paper is organised as follows. Section two discusses the planning for a clinical trial which motivated this work, section three specifies the methods including the imputation models, section four outlines the characteristics of the simulation study with information on the data generation models, section five details the results of the simulation study, section six includes a discussion and finally section seven includes the conclusions we draw from this work.

## 2 Motivation

This work derives from the planning of a Phase III trial in patients with a respiratory disease to assess the potential for a new triple combination medicine to improve lung function over time. The trial planned to assess forced expiry volume in one second ($FEV_1$) at repeated timepoints and the patient outcome of interest was the change from baseline in $FEV_1$ at the final timepoint. The design was a randomized parallel group trial with one group randomized to receive the triple combination and another group randomized to receive control. In the estimands framework, the trial planned to estimate:

*Difference between Active and Control for mean change in $FEV_1$ from baseline to the final timepoint for patients with the respiratory disease of interest regardless of treatment discontinuation*.

Although the trial protocol planned to collect outcome data from patients who discontinued randomized treatment, it was expected that some patients would still withdraw from the study, and therefore the estimation strategy planned to use MI with on- and off-treatment outcomes to deal with the inevitable missing data problem. This created the question of how elaborate any MI model should be in order to accurately capture the on and off treatment behaviour when building a robust estimate of the intervention effects. This became the starting point for the assessment of the models in this work. The paper is written from the perspective of providing justification for the selection of analysis methods in a statistical analysis plan before the data are collected.

# 3 Methods

## 3.1 Notation

We consider the general setting of a two-group parallel trial where $Z$ denotes the randomized groups, with $Z = C$ and $Z = A$ denoting the groups of patients randomized to receive control and active respectively. All notation applies to each individual in the trial and therefore no patient level index is included. We define $Y_j$ as the actual outcome value at the $j$-th timepoint, $j = 0, \ldots, J$, with 0 representing baseline and $J$ the timepoint of the key assessment (for the motivating trial $J = 3$). We assume $Y_0$ is always observed. Let $D_j$ be a variable indicating whether the patient is still on randomized treatment ($D_j = 0$) or off treatment ($D_j = 1$) at the $j$-th timepoint. We assume a monotone discontinuation pattern and so any further timepoints are also off treatment. If $D_J = 0$ the patient completes the trial on treatment. Also let $P_j$ be a categorical variable representing the pattern of treatment discontinuation history up to the $j$-th timepoint $j = 0, \ldots, J$. As we assume that treatment discontinuation is monotone each $P_j$ has $j + 1$ levels with $P_J$ representing the $J + 1$ discontinuation patterns at the final timepoint.

## 3.2 Estimand

Applying the treatment policy strategy to treatment discontinuation in our motivating trial targets the estimand:

$$\Delta_j = E\left[Y_j - Y_0 \middle| Z = A\right] - E\left[Y_j - Y_0 \middle| Z = C\right],$$

which can be interpreted as the difference in expected change from baseline at timepoint $j$ for patients randomized to receive active compared with patients randomized to receive control regardless of treatment discontinuations. The group-specific expected changes from baseline values on the right-hand side are also of interest where they provide further context to the treatment effects.

## 3.1 Estimation

With complete data, a standard estimation method for the quantity above would be a regression model for change from baseline ($Y_j - Y_0$) with model terms for treatment group $Z$ and baseline $Y_0$.

However, with incomplete data it is usual to consider either Mixed Models for Repeated Measures (MMRM) or to use MI in a three-stage process: creating a set of complete datasets by imputing any missing outcomes, analysing the change from baseline $Y_j - Y_0$ in each dataset with the complete data model specified above, and combining the estimates using Rubin's rules [8]. Both of these estimation approaches are considered in this work and the models we use are specified below.

In the proposed MI models, outcomes for patients off treatment may differ from those on treatment in mean and/or in variance. The core assumption underlying all the imputation models is that missing outcomes would behave like outcomes of different groupings of 'similar' patients who are still in the trial. For simplicity the models are defined using pseudo regression notation reflecting how they are specified programmatically. Full algebraic specifications for each model are available in appendix 1. The interaction "star" operator (*) between model terms indicates that separate terms are included for each level (or combination) of the categorical factors. In this notation, the model used for analysing complete data and imputed data, termed ANCOVA, is:

$$Y_J - Y_0 = Intercept + Z + Y_0.$$

### 3.1.1 MMRM

This analysis uses all the available data and makes no distinction between the on and off treatment outcomes and relies on the common MAR assumption. The model includes terms for Treatment by Timepoint and Baseline by Timepoint interactions and can be denoted as:

$$Y = Intercept + Z + Timepoint + Y_0 + Z * Timepoint + Y_0 * Timepoint$$

### 3.1.2 Sequential Multiple Imputation models using previous outcomes

These proposed MI models are designed to be used sequentially for each timepoint $j = 1, \dots, J$, using previous outcomes, with separate parameters at each timepoint $j$ and separate imputation for each treatment group. All models have a common intercept together with slope parameters on $Y_0, \dots, Y_{j-1}$ at each timepoint. Some models use indicators $D_1, \dots, D_j$ as covariates to distinguish between on and off treatment status and others use variables containing the pattern $P_j$ of treatment discontinuation. Terms specified as $D_j * Y_j$ and $P_j * Y_j$ indicate separate $Y_j$ covariate slope terms for each level of the categorical covariates.

**Common Intercepts Common Slopes (CICS).**

$$Y_j = Intercept + Y_0 + \cdots + Y_{j-1}$$

The CICS model is the simplest. It imputes missing outcomes for timepoint *j* based on a single intercept and single set of slopes for previous outcomes whether on- or off-treatment.

**On/Off Intercepts Common Slopes (OICS).**

$$Y_j = Intercept + D_j + Y_0 + Y_1 + \cdots + Y_{j-1}$$

The OICS model extends CICS by allowing separate intercept terms for on- and off-treatment outcomes when imputing missing values for timepoint $j$.

**Pattern Intercepts Common Slopes (PICS).**

$$Y_j = Intercept + P_j + Y_0 + Y_1 + \cdots + Y_{j-1}$$

PICS extends OICS by including separate intercept terms for each treatment discontinuation pattern up to timepoint $j$.

**On/Off Intercepts On/Off Slopes (OIOS).**

$$Y_j = Intercept + D_j + Y_0 + \cdots + Y_{j-1} + D_j * Y_0 + \cdots + D_j * Y_{j-1}$$

The OIOS model allows separate on- and off-treatment intercept and slope parameters. The slopes are separated according to the discontinuation status of the subject at the *current imputation timepoint j*. The discontinuation status of the current timepoint is used as it maintains the principle of "no interactions without main effects".

**Pattern Intercepts On/Off Slopes (PIOS).**

$$Y_j = Intercept + P_j + Y_0 + Y_1 + \cdots + Y_{j-1} + D_0 * Y_0 + \cdots + D_{j-1} * Y_{j-1}$$

PIOS extends PICS. In addition to a separate intercept for each discontinuation pattern, it also allows the slopes to differ according to the discontinuation status of the patient at the time of the *previous outcomes*. As the pattern intercept term automatically includes each prior discontinuation status their inclusion for the slopes does not violate the "no interactions without main effects" principle.

**Pattern Intercepts Pattern Slopes (PIPS).**

$$Y_j = Intercept + P_J + Y_0 + Y_1 + \cdots + Y_{j-1} + P_J * Y_1 + \cdots + P_J * Y_{j-1}$$

The PIPS model is the most flexible proposed and is structured so that all intercept and slope parameters are estimated with outcomes from patients with the same final treatment discontinuation pattern.

There are other potential imputation models or variations to the six proposed above. However, for this work it was thought these six would cover a suitable range from very basic to very flexible imputation models. The idea was this range would allow conclusions to be drawn about the level of complexity required to achieve robust estimates given the likely data characteristics of the proposed trial. Table 1 summarises each model highlighting the similarities and differences.

*Table 1 Sequential MI models compared.*

| | | Slopes | | |
|---|---|---|---|---|
| | | Common | On/Off | Pattern |
| | Common | CICS | - | - |
| **Intercept** | On/Off | OICS | OIOS | - |
| | Pattern | PICS | PIOS | PIPS |

### 3.1.3 Sequential Multiple Imputation using previous residuals

To compare with our proposed MI models, we also consider both models outlined in Roger 2017. These models condition on previous residual values $R_j = Y_j - \hat{\mu}_j$ where $\hat{\mu}_j$ is the estimated mean at the $j$-th timepoint. Parameterizing the imputation model on these residuals leads to the intercept parameters coinciding with the estimated marginal means for each timepoint.

**On/Off Intercepts with Common Slopes using Residuals (OICS-R).**

$$Y_j = Intercept + D_j + R_0 + R_1 + \cdots + R_{j-1}$$

The OICS-R model adjusts the marginal means for on and off treatment but makes no distinction between on and off treatment when conditioning on the residuals. This is similar in structure to the OICS model we propose, but it can be shown these models are not equal (see appendix 1).

**Pattern Intercepts with Common Slopes using Residuals (PICS-R).**

$$Y_j = Intercept + P_j + R_0 + R_1 + \cdots + R_{j-1}$$

The PICS-R model adjusts the marginal means for each treatment discontinuation pattern possible up to timepoint $j$ but makes no distinction between on and off treatment when conditioning on the residuals. It can be shown that this model is a reparameterization of the PICS model we propose (see appendix 1).

# 4 Simulation Study

The goal of the simulation study was to assess the bias and precision of the estimated treatment effects from each model and also to investigate the convergence and stability of the models. The design of the simulated trials matched the parallel group structure of the proposed Phase III trial that motivated this work. Each simulated trial included 375 patients per group and aimed to detect a clinically meaningful $FEV_1$ difference of 100mL between active and control.

For each of the simulated trials, we considered 3 treatment discontinuation mechanisms, 4 pairs of discontinuation rates, 3 study withdrawal patterns and 2 off-treatment expected outcomes trajectories. This resulted in 72 discontinuation and withdrawal combinations with 1000 trials created for each (summarised in Figure 1). The large factorial design of the simulation was intended to "stress test" the imputation models with respect to convergence and the accuracy of the imputations and estimates. The following parts of this section outline how the data was generated and provide detail on the different scenarios for treatment discontinuation, study withdrawal and off-treatment expected outcomes. The final part of this section gives information on the analyses and performance measures. Data generation, imputation and analysis was performed using SAS [10].

*Figure 1 A schematic showing the full factorial nature of the simulation study with 72 Scenarios for each of the 1000 simulated studies.*

| Treatment discontinuation mechanism | | Treatment discontinuation rates in each arm | | Balance of 50% study withdrawal | | Off-treatment expected outcomes |
|---|---|---|---|---|---|---|
| $\begin{pmatrix} \text{DAR} \\ \text{DNAR1} \\ \text{DNAR2} \end{pmatrix}$ | $\times$ | $\begin{pmatrix} 10\%\ \text{Control}: 10\%\ \text{Active} \\ 10\%\ \text{Control}: 20\%\ \text{Active} \\ 20\%\ \text{Control}: 20\%\ \text{Active} \\ 50\%\ \text{Control}: 50\%\ \text{Active} \end{pmatrix}$ | $\times$ | $\begin{pmatrix} \text{Balanced} \\ \text{More Early} \\ \text{More Late} \end{pmatrix}$ | $\times$ | $\begin{pmatrix} \text{Return to Baseline} \\ \text{Same as Active} \end{pmatrix}$ |

## 4.1 Data Generation

To create outcome data for virtual patients, we generated a vector of correlated patient-level data comprising potential on-treatment $FEV_1$ data for all timepoints *and* potential off-treatment $FEV_1$ data for all timepoints, using a multivariate normal distribution. At each post-baseline timepoint, some patients were selected to discontinue treatment based on the discontinuation mechanism and rate of the specific scenario (see Section 4.2), with subsequent values using the generated off-treatment data. Some of the discontinued subjects were then selected to be withdrawn from the trial according to the specific withdrawal for that timepoint (see Section 4.3) with further outcomes set to missing.

The expected values of $FEV_1$ (in mL) for baseline and on-treatment outcomes in the control group were set as $[2140,\ 2470,\ 2520,\ 2540]$ and the expected values for the on-treatment effects (differences between Active and Control) were set to $[0,\ 100,\ 100,\ 100]$ for timepoints 0 (baseline), 1, 2 and 3, respectively. Two general scenarios were considered for the off treatment expected values (see Section 4.4). Full details of the data generation process are included in appendix 2.

## 4.2 Treatment Discontinuation Mechanism and Proportion

To investigate complex treatment discontinuation situations, three mechanisms were considered:

1. Discontinuation at random (DAR) based on lack of efficacy. This mechanism selected which patients discontinued after each timepoint using a propensity score linked to each patient's previous on-treatment $FEV_1$ value and selected the lowest (worst) ranked patients to discontinue.
2. Discontinuation not at random (DNAR) based on lack of efficacy. This mechanism selected which patients to discontinue after each timepoint using a propensity score linked to each patient's next potential on-treatment $FEV_1$ value and selected the lowest (worst) ranked patients to discontinue. This mechanism is labelled DNAR1.
3. A two-process DNAR based on a lack of tolerability in early visits and lack of efficacy in late visits. This mechanism was designed to add another discontinuation reason such as safety related problems for subjects who are 'more sensitive' to the drug and therefore experience both a higher response and lower tolerability early in the study. This mechanism selected which patients to discontinue after each timepoint using the same propensity score used in the DNAR1 mechanism but selected patients with the highest (best) ranked values at the first post-baseline timepoint and the lowest (worst) ranked values after further timepoints. This mechanism is labelled DNAR2.

For each treatment discontinuation mechanism above, four treatment discontinuation rates were considered for control and active groups corresponding to the percentage of patients that had discontinued treatment by the final timepoint:

1. 10% Control and 10% Active.
2. 10% Control and 20% Active.
3. 20% Control and 20% Active.
4. 50% Control and 50% Active.

The treatment discontinuations were spread over the post-baseline assessments $Y_1$, $Y_2$ and $Y_3$ in the ratio 5:3:2, creating a larger proportion of patients discontinuing treatment at earlier timepoints. As an example, a rate of 50% treatment discontinuation would be allocated as (25%,15%,10%) for times 1, 2 and 3 respectively.

## 4.3 Study withdrawal rate and balance

The process for simulating which of the patients who discontinued treatment then withdrew from the trial (creating missing outcomes) was simplified to be missing completely at random (MCAR). The process was designed as a single assessment at the point of treatment discontinuation, where the patient either withdrew and provided no off-treatment data or remained in the trial providing off treatment outcomes until completion. The proportion of discontinued patients simulated to withdraw from the study by the final timepoint was fixed at 50%, but the balance of where the withdrawals occurred was varied across the post-baseline assessments with scenarios of:

1. "Balanced" with 50% withdrawal in all patients who discontinued, regardless of their time of discontinuing, creating similar proportions of observed and missing off-treatment data at each timepoint.
2. "More early" with 80% withdrawal in patients who discontinued at timepoint 1 and 20% withdrawal in patients who discontinued at timepoints 2 and 3, creating smaller proportions of observed off-treatment data relative to missing off-treatment data at early timepoints, but larger proportions at later timepoints.
3. "More late" with 20% withdrawal in patients who discontinued at timepoint 1 and 80% withdrawal in patients who discontinued at timepoints 2 and 3, creating larger proportions of observed off-treatment data relative to missing off-treatment data at early timepoints, but smaller proportions at later timepoints.

The above withdrawal scenarios were designed to investigate if the proportion of observed and missing off-treatment data at different timepoints impacts the robustness of sequential imputation and estimation at the final timepoint.

## 4.4 Expected values for off-treatment outcomes

Two scenarios for the off-treatment outcomes were considered:

1. "Return to baseline". Under this assumption, patients who discontinue treatment would receive no further treatment and their mean $FEV_1$ values would revert to baseline levels.
2. "Same as active". Based on existing licensed medicines, it was considered possible for discontinuing patients to be offered further treatment using two separate components of the triple combination therapy being investigated, which could result in similar $FEV_1$ values compared with on-treatment patients receiving Active treatment.

## 4.5 Data analysis

We analysed the full simulated on- and off-treatment data (before setting outcomes to missing) using the simple ANCOVA model defined in section 3 for change from baseline $FEV_1$ at the final timepoint with model terms for baseline $FEV_1$ and treatment. Estimated marginal means (least-squared means in SAS terminology) were produced for the two arm-specific parts of the estimand. After setting generated post-withdrawal outcomes to missing, the MMRM analysis was fitted using restricted maximum likelihood with an unstructured variance-covariance matrix common to the two treatment arms. For the MI analyses, the imputation models were fitted each using 25 imputations for computational feasibility. The analysis model for each imputed data set was the same ANCOVA model used for the full data.

## 4.6 Performance measures

The performance measures used for each model were the percentage of simulated data sets for which the model converged and produced stable estimates, and the bias, 95% confidence interval (CI) halfwidth and coverage of the estimated treatment effects (Active vs Control) and the estimated group means. The bias and CI coverage were calculated using the analytical expected values (calculations included in appendix 2). Monte Carlo (MC) errors were also calculated.

# 5 Results

## 5.1 Convergence and MC Error

There were no problems fitting the ANCOVA analysis on the full data or the MMRM model. All MI models except PIPS (the most complex) fitted successfully in all scenarios. The PIPS model had convergence issues in all scenarios with 10% treatment discontinuation and also scenarios with 20% treatment discontinuation and "more early" study withdrawal. The convergence issues were due to insufficient off-treatment data in some discontinuation patterns.

In general, these fitting problems suggest that the use of the PIPS model could easily lead to convergence problems if used in the planned respiratory trial and so this model was discarded. No further results or comparisons for PIPS are presented in this work. For all other models and scenarios, the MC standard error for treatment effects were below 1.7ml (1.7% of the effect size).

## 5.2 Bias, Halfwidth and Coverage

To help understand the results for the treatment effects across all scenarios we present two heatmaps that summarise the bias, change in 95% CI halfwidth and 95% CI coverage. The first heatmap displays all the “Return to Baseline” scenarios (Figure 2) and the second displays all the “Same as Active” scenarios (Figure 3). Detailed plots of the bias, 95% CI halfwidth and coverage for each scenario are presented in the Supplementary Information (SI): SI figures 1-3 (treatment effects) and SI figures 3-6 (group means).

## 5.1 Bias, Halfwidth and Coverage

To help understand the results for the treatment effects across all scenarios we present two heatmaps that summarise the Bias, Change in 95% CI Halfwidth and 95% CI Coverage. The first heatmap displays all the “Return to Baseline” scenarios (Figure 2) and the second displays all the “Same as Active” scenarios (Figure 3). Detailed plots of the bias, 95% CI halfwidth and coverage for each scenario are presented in SI figures 1-3 (treatment effects) SI figures 3-6 (group means). Any differences we discuss in the heatmaps can be observed in the SI figures and are not due to MC error. We discuss the models in four categories - the full data ANCOVA (FULL), the models that rely on the common MAR assumption (MMRM and CICS), the models that have on- and off-treatment intercepts (OICS, OICS-R and OIOS), and the models with pattern-based intercepts (PICS, PICS-R and PIOS). We also consider similarities and differences between the corresponding scenarios for the two off-treatment expected outcome trajectories.

**Full data ANCOVA Model**

There was negligible bias for the treatment effects from the full data ANCOVA model which indicates that on average our simulated data matches the analytically calculated true expected values for the treatment policy in each scenario.

The CI halfwidths from the estimated treatment effects increased as the discontinuation rate increased, with larger increases seen in the “Return to Baseline” scenarios compared to the corresponding “Same as Active” scenarios (SI Figures 2). These provide a useful reference for comparing the other models against.

The full data analysis produced overcovered CI intervals for most scenarios (Figures 2 and 3). This is primarily due to the model assumptions of normal residuals and the fact that the outcomes are actually from a bimodal mixture distribution of on- and off-treatment. However, as the other models we consider also rely on assumptions of normality, these coverage values provide a useful reference for direct comparisons.

**Models with a common MAR assumption**

The MMRM and CICS models produced biased treatment effects in all scenarios. The bias increased as the treatment discontinuation rate increased and was magnified further in scenarios with unequal discontinuation rates. The largest bias for the treatment effects was approximately 30ml (30% of the target effect of 100ml) and occurred in the “Return to Baseline” scenario with DNAR1 mechanism in combination with “10% Control 20% Active” treatment discontinuation rates and “More Late” study withdrawal (Figure 2: ID 18).

The bias in these models relates directly to the common MAR assumption – specifically that missing outcomes would be similar to an aggregate of all the observed on- and off-treatment outcomes at that timepoint. In all cases except for the patients in the Active group in the “Same as Active” scenarios (SI Figure 4.10), the missing off-treatment values are systematically different from the observed on-treatment values and not making a distinction results in MNAR data and bias.

The increase in bias across the discontinuation rates is driven by more discontinuations leading to more study withdrawal (which we fixed at 50% of treatment discontinuations) resulting in more MNAR data. Additionally, in the “Return to Baseline” scenarios (which have large differences between on- and off-treatment) the treatment effect bias is compounded further by unbalanced treatment discontinuation which creates different amounts of MNAR induced bias for each group estimate (See Figure 2).

In some scenarios the MMRM and CICS models produced CI halfwidths for the treatment effects that were smaller than the full data. This occurred in the “Return to Baseline” scenarios with 10% and 20% treatment discontinuation rates (Figure 2: IDs 1-9, 13-21 and 25-33) but not for 50% treatment discontinuation rates nor any of the “Same as Active” scenarios. The underestimation of variability also relates to the common MAR assumption that missing outcomes would be similar to all the observed on- and off-treatment data. In reality the missing values were only similar to patients that had discontinued treatment. The scenarios that produced the underestimation in variability had quite different on- and off-treatment expected outcomes coupled with unequal discontinuation rates in the two arms, and not making a distinction between them implicitly assumes the data came from the whole range of observed values, rather than just the off-treatment values which then artificially reduces the overall variability estimates. This reduction appears to be outweighed or masked by the increased uncertainty due to larger numbers of study withdrawals seen in the cases with 50% treatment discontinuation and in general for the “Same as Active” scenarios, where off-treatment outcomes are similar to the on-treatment outcomes for patients in the active group.

The CI coverage of the treatment effects for both the MMRM and CICS models were lower in all scenarios than the full data ANCOVA intervals. In general, the scenarios with the lowest coverages matched the situations with the largest bias implying the bias is the main cause for lack of coverage.

**Models with On- and Off-treatment Intercepts**

The OICS, OICS-R and OIOS models showed negligible bias for the treatment effects in all scenarios with "Balanced" study withdrawal. However, in "More Early" or "More Late" study withdrawal scenarios, the OICS and OIOS models were biased with the direction linked to unbalancing. Again, the bias increased with discontinuation rate and was magnified further by unequal discontinuation rates and DNAR mechanisms. In the scenarios with bias, the magnitude for OICS and OIOS was approximately the same with the maximum bias of 16ml and 15ml respectively (16% and 15% of the target effect).

The OICS-R model also had biased treatment effects for scenarios with "More Early" or "More Late" study withdrawals but only in the cases with unequal treatment discontinuation rates and DNAR mechanisms (See Figure 2: ID 18,30) (Figure 3: ID 18,29,30). The bias was consistently smaller than the OICS and OIOS models with a maximum absolute bias of 4ml (4% of the target effect).

The unbalanced study withdrawal complicates the missing data mechanism making it depend on the time of treatment discontinuation. This implies the missingness is also related to the pattern of treatment discontinuation and explains why we see bias in OICS, OICS-R and OIOS models in these scenarios as they have no pattern components.

The CI halfwidths for the treatment effects in the OICS, OIOS and OICS-R models were similar to or greater than the full data ANCOVA, MMRM and CICS models in all scenarios. All three models had similar CI halfwidths for scenarios with "Balanced" study withdrawal, but for "More Early" study withdrawal the OICS and OIOS halfwidths were larger than the OICS-R and for "More Late" withdrawal they were lower than OICS-R. The maximum change in halfwidths relative to the FULL model for the OICS, OICS-R and OIOS models were 32%, 14% and 33% respectively occurring in the "Same as Active" scenario with DNAR2 mechanism and 50% treatment discontinuation (Figure 3: ID 35).

The CI coverage for the treatment effects were largely similar between the OICS, OICS-R and OIOS models except in the "Same as Active" scenarios with 50% discontinuation rate and "More Early" study withdrawal, where there was drop in coverage in all three models, but a larger drop in the OICS and OIOS models. All three models had consistently lower coverage than the full data ANCOVA intervals with the lowest coverage of approximately 90% in the OICS and OIOS models.

Comparing the performance of the OICS and OICS-R models it seems that the use of the residuals in the OICS-R model leads to smaller bias in all cases. The maximum bias for the treatment effects was also considerably smaller (16ml vs. 4ml). The use of the residuals also appears to limit the increase in variability considerably in the majority of scenarios (but not all). The maximum change in CI halfwidth was also considerably smaller (32% vs. 14%). The OICS-R model also appears to be unbiased in some situations with equal discontinuation rates where the missingness is related to discontinuation pattern and we see bias in OICS. Comparing the structure of both (see Appendix 1) it is clear that the use of residuals for conditioning in OICS-R implies that the regression terms account for the on- or off-treatment status of the past values. This is not the case for the OICS model where the direct conditioning on the values makes no distinction between the status of the past data. This difference implies the OICS-R has more structural flexibility to condition on the past history and may explain why the model outperforms the simpler OICS model in these scenarios. It should be noted that although OICS-R was unbiased for treatment effects in these cases, it is clear that the group means were biased (SI Figures 4.1-4.12). However, with approximately the same treatment discontinuation in both groups and fixed study withdrawal, the difference between the groups is maintained. If there was unequal discontinuation or withdrawal rates in each group, this model is also likely to show bias.

**Models with Pattern based Intercepts**

There was negligible bias for the treatment effects in the PICS, PICS-R and PIOS models in all scenarios. The lack of bias for the simpler pattern models (PICS and PICS-R) suggests the inclusion of pattern-based intercepts is sufficient to deal with the most complex scenarios we generated including time dependent missingness from unbalanced study withdrawal and DNAR mechanisms. One reason for this sufficiency is likely to be that our data generation model used the same variance-covariance values for on-treatment and off-treatment outcomes. If we had created data with different variances for on-and off-treatment, the PIOS model may have performed better the PICS and PICS-R.

All three models had similar CI halfwidths except scenarios with “More Early” study withdrawal where the PIOS model halfwidths were slightly higher. The maximum change in halfwidth occurred in the “Same as Active” scenario with DAR mechanism, 50% discontinuation and “More Early” withdrawal and was approximately 43% (Figure 3: ID 11). As with the on- and off-treatment intercept models, the increase in CI halfwidths was largest for cases with “More Early” study withdrawal and smallest for “More Late” study withdrawal.

The increase in variability seen in PICS, PICS-R and PIOS is driven by the increased model complexity. The use of parameters to distinguish between discontinuation patterns results in a considerable reduction of information contributing to the estimation of each parameter and is the main driver behind the substantial increases to the variability.

The CI coverage for the treatment effects for the PICS, PICS-R and PIOS models were more comparable to the full ANCOVA intervals in the majority of scenarios except cases with 50% treatment discontinuation where there was a drop in coverage which was magnified further by DNAR mechanisms and unbalanced study withdrawal. The lowest coverage for all three models was approximately 90% (Figure 3: ID 24).

Finally, the bias, halfwidth and coverage results for PICS and PICS-R also provides confirmation by simulation that these models are essentially the same.

**Comparing Off-treatment Scenarios**

Comparing the models that showed bias (MMRM, CICS, OICS, OIOS and OICS-R) for each "Return to Baseline" and corresponding "Same as Active" scenarios we see similar levels of bias in the treatment effects when the treatment discontinuation rates are equal, but unequal discontinuation creates much larger bias for the "Return to Baseline" scenarios. Looking at the bias for the corresponding group means for each of these cases (SI Figures 4.1-4.12) shows there is larger bias in all the "Return to Baseline" scenarios compared to the corresponding "Same as Active". This difference is largely driven by the fact the "Return to Baseline" off-treatment data are different to both Control and Active on-treatment data, whereas the "Same as Active" scenario has off-treatment data matching the on-treatment data in the Active group and therefore the common MAR assumption holds in this group. The similar bias seen in the treatment effects for equal treatment discontinuation rates is driven largely by the choice of our simulated data. The bias in the group means and treatment effects in these scenarios are considered in more detail in Appendix 3.

# 6 Discussion

This work assessed a number of different sequential MI models for the estimation of treatment effects where the intercurrent event of treatment discontinuation is handled using the treatment policy strategy. The inspiration came from phase 3 design work for a respiratory drug where there were concerns about the accuracy of simple MMRM or MI models due to their reliance on a *common* MAR assumption across on- and off-treatment outcomes. We suggested a range of MI models to investigate the accuracy of performing imputation conditioning on previous values and treatment-discontinuation status or pattern. We evaluated the models for feasibility as estimation methods by comparing them to an ANCOVA analysis of the full data and a traditional MMRM using all available data.

The problems faced in fitting the PIPS model suggest it may have limited utility in many real trial applications. It may still be useful for very large trials, but the ease of fitting and good performance of the other MI models considered here suggests there is limited need for that level of complexity. Our work shows that accurate estimation is possible with some of the models we investigated. However, if they are to be used for the estimation of primary estimands in real clinical trials further work would be advisable to investigate type 1 error and efficiency at the design stage for the specific study. In addition, although we stress-tested the MI models by creating scenarios with small amounts of observed off-treatment data at particular timepoints, we ensured there would be 50% collection of off-treatment data at the final timepoint. Also, our simulation study only considered a simplified withdrawal set-up that ensured off-treatment outcomes existed for any situation where missing data needed imputing. In reality this may not be the case for all trials and further work could look at how likely these models are to run into fitting problems if patients can leave the trial at any timepoint, as this may lead to smaller amounts of off-treatment data collected and consequently the imputation models becoming non-estimable. We also ensured any discontinuation and withdrawal was monotone, a simplification that is unlikely to occur in practice and can make sequential MI more difficult because sequential imputation of missing data at a given visit does not take into account the observed data in future visits. Further work to investigate the impact of intermittent missingness on sequential MI and also comparing it to multivariate imputation would be interesting.

A major point for consideration is that failure to collect data off-treatment may have a serious impact on reliable estimation of the estimand targeted here (regardless of the estimation method considered). Our simulation results show that the uncertainly in the group means and treatment effects increases considerably with increasing amounts of missing off-treatment data. Investigating this further (Appendix 3) suggests that the relationship depends on the overall amount of missing off-treatment data *and* the relative proportion of observed and missing off-treatment data. It is possible that a change of mindset and educational work is needed to increase awareness among study investigators and participants about the importance of collecting data after discontinuation of treatment.

In an extreme case where little to no off-treatment data was collected, the only feasible models considered in this work would be the common MAR-based methods, however the resulting estimates and effects would really be based on data collected on-treatment. Such estimates would be more aligned with an estimand that uses a hypothetical strategy to deal with treatment discontinuation. Another option could be reference-based imputation methods [6,9]. This alternative also targets a treatment policy estimand and would be appropriate in settings where subjects in the active treatment arm who discontinue treatment receive post-discontinuation medication comparable to that in the Control arm. A detailed comparison of off-treatment MI with reference-based MI would also be an interesting extension to this work.

Another implication from these results is that a major factor in the size of bias when estimating a treatment policy estimand is the difference between on- and off-treatment outcomes. The large biases seen in our "Return to baseline" scenarios result from baseline measures that were very different to the outcome behaviour of both the Active and Control treated subjects. Although this extreme case is plausible in some trials that use an active comparator or expect some placebo effect, a situation where control and off-treatment outcomes are similar may be more common – this type of data structure is sometimes described as a "jump to reference" setting. The data generated in our "Same as Active" scenarios are mirror images of this and therefore we would expect bias to be of similar magnitude for the corresponding treatment discontinuation and study withdrawal rates.

# 7 Conclusions

In general, this work has demonstrated that treatment discontinuation complicates the missing data mechanism associated with study withdrawal and can make the common MAR assumption invalid. It also demonstrated that imbalances of treatment discontinuation rates across the treatment groups, imbalances of study withdrawal rates across time, DNAR mechanisms as well as different outcome behaviour on and off treatment will all create problems for estimating the effects of a treatment policy. In real trials the majority of missing outcomes will be from patients discontinuing treatment and then withdrawing from the study. It is also likely that the discontinuation rates (and therefore the missing rates) will differ between the trial groups as larger numbers may discontinue from control groups for lack of efficacy or larger numbers may discontinue from active groups for tolerability issues. The time of discontinuation is also likely to affect whether a patient remains in the trial until the final timepoint – e.g. it seems unlikely that patient discontinuing treatment two months into a two-year trial will have the same chance of being observed at the final timepoint compared to a patient discontinuing treatment one month before the trial ends. Given these likely scenarios, estimating an estimand with treatment policy handling of treatment discontinuation needs careful thought and is unlikely to be straight-forward in many cases.

No model we considered for estimating these effects seems to be optimal in all the scenarios we investigated. Given the large biases seen in the common-MAR-based (MMRM and CICS) models when there are only moderate amounts of unbalanced treatment discontinuation, we conclude these are poor choices for estimating treatment policy-based effects. An exception to this could be when the behaviour on- and off-treatment is expected to be similar or when there are low numbers of discontinuations and/or little missing data. The proposed alternative sequential imputation models that distinguish between on- and off-treatment data are clearly better but involve a trade-off between bias and variability. The pattern-based intercept models (PICS and PICS-R) appear to be the best choice for reducing bias, but in some settings lead to sizable increases in variability. They also have the potential for estimation problems in real trials where lack of off-treatment data may make specific discontinuation patterns inestimable. It is clear from the on- and off-treatment intercept models OICS, OICS-R and OIOS, that regression on the residuals is optimal and the OICS-R model can offer a good compromise by conceding small levels of bias in some settings as a trade-off for much less variance inflation. Logically, as it has fewer parameters than the pattern-based models it is also less likely to suffer from estimation issues related to lack of collected off-treatment data.

A pragmatic estimation strategy could be the pre-specification of a hierarchy of these models based on criteria related to successful fitting. As an example, the hierarchy could start with a pattern-based MI approach such as PICS or PICS-R, and if this failed the acceptance criteria then an intercept model such as OICS-R could be used. Finally, as a last resort if all off-treatment models failed the acceptance criteria, the hierarchy could default back to the CICS or MMRM approaches.

# 8 Acknowledgements

Ian White was supported by the Medical Research Council Programme MC_UU_00004/07.

The authors would like to thank James H. Roger, James Bell and Michael O'Kelly for useful discussions.

Figure 2 Summary of simulation results for treatment effect: Return to Baseline Scenarios. The bias is calculated using analytical expected values, change in CI halfwidth is relative to the full simulated data (FULL Results), and both the CI halfwidths and coverages are for 95% intervals.

Scenario: Return to Baseline

| Mechanism | Discontinuation Rate | Withdrawal Type | ID | Bias (ml) | | | | | | | | | Change in CI Halfwidth (%) | | | | | | | | | CI Coverage (%) | | | | | | | | |
|---|---|---|---|---|---|---|---|---|---|---|---|---|---|---|---|---|---|---|---|---|---|---|---|---|---|---|---|---|---|---|
| | | | | FULL | MMRM | CICS | OICS | OICS-R | OIOS | PICS | PICS-R | PIOS | FULL | MMRM | CICS | OICS | OICS-R | OIOS | PICS | PICS-R | PIOS | FULL | MMRM | CICS | OICS | OICS-R | OIOS | PICS | PICS-R | PIOS |
| DAR | 10% Control: 10% Active | Balanced | 1 | 0 | 4 | 4 | 0 | 0 | 0 | 0 | 0 | 0 | 0 | -2 | -1 | 2 | 2 | 3 | 2 | 2 | 3 | 96 | 96 | 96 | 96 | 96 | 96 | 96 | 96 | 96 |
| | | More Early | 2 | 0 | 4 | 4 | -3 | 0 | -2 | 0 | 0 | -1 | 0 | -1 | -1 | 8 | 2 | 9 | 8 | 8 | 12 | 96 | 96 | 96 | 96 | 96 | 96 | 96 | 95 | 96 |
| | | More Late | 3 | 0 | 4 | 4 | 2 | 1 | 2 | 1 | 1 | 1 | 0 | -2 | -2 | -1 | 2 | 0 | 2 | 2 | 3 | 96 | 96 | 95 | 96 | 96 | 96 | 96 | 96 | 96 |
| | 10% Control: 20% Active | Balanced | 4 | 0 | 24 | 24 | 0 | 0 | 0 | 0 | 0 | 0 | 0 | -2 | -1 | 3 | 3 | 4 | 3 | 3 | 4 | 96 | 91 | 91 | 96 | 95 | 96 | 96 | 96 | 96 |
| | | More Early | 5 | 0 | 25 | 25 | -14 | 1 | -13 | 1 | 1 | 1 | 0 | -1 | 0 | 12 | 3 | 12 | 11 | 11 | 14 | 96 | 90 | 90 | 95 | 95 | 95 | 96 | 96 | 97 |
| | | More Late | 6 | 0 | 28 | 28 | 14 | 0 | 14 | 0 | 0 | 0 | 0 | -3 | -3 | -1 | 3 | 0 | 4 | 4 | 5 | 96 | 89 | 89 | 94 | 95 | 94 | 95 | 94 | 95 |
| | 20% Control: 20% Active | Balanced | 7 | 0 | 9 | 9 | 0 | 0 | 0 | 0 | 0 | 0 | 0 | -1 | -1 | 4 | 4 | 4 | 4 | 4 | 5 | 97 | 95 | 95 | 95 | 95 | 95 | 95 | 95 | 96 |
| | | More Early | 8 | 0 | 9 | 9 | -4 | 1 | -4 | 1 | 1 | 1 | 0 | 0 | 0 | 14 | 4 | 14 | 14 | 14 | 16 | 97 | 95 | 95 | 96 | 96 | 96 | 95 | 95 | 95 |
| | | More Late | 9 | 0 | 9 | 9 | 5 | 0 | 5 | 0 | 0 | 0 | 0 | -3 | -3 | -1 | 4 | 0 | 5 | 5 | 6 | 97 | 95 | 95 | 96 | 96 | 96 | 95 | 95 | 95 |
| | 50% Control: 50% Active | Balanced | 10 | 0 | 18 | 17 | -1 | -1 | -1 | -1 | -1 | -1 | 0 | 5 | 6 | 10 | 10 | 10 | 10 | 10 | 11 | 98 | 96 | 96 | 96 | 96 | 96 | 96 | 96 | 96 |
| | | More Early | 11 | 0 | 16 | 15 | -13 | -1 | -12 | -1 | 0 | 0 | 0 | 9 | 11 | 29 | 11 | 28 | 33 | 31 | 34 | 98 | 97 | 97 | 94 | 96 | 94 | 95 | 94 | 95 |
| | | More Late | 12 | 0 | 21 | 20 | 12 | 0 | 11 | 0 | 0 | 0 | 0 | 0 | 0 | 2 | 9 | 3 | 11 | 11 | 11 | 98 | 95 | 95 | 95 | 95 | 96 | 94 | 94 | 94 |
| DNAR1 | 10% Control: 10% Active | Balanced | 13 | 0 | 4 | 4 | 0 | 0 | 0 | 0 | 0 | 0 | 0 | -2 | -2 | 3 | 2 | 3 | 2 | 2 | 3 | 97 | 96 | 95 | 95 | 95 | 95 | 95 | 95 | 95 |
| | | More Early | 14 | 0 | 4 | 4 | -3 | 0 | -3 | 0 | 0 | 0 | 0 | -1 | -1 | 9 | 2 | 9 | 8 | 8 | 11 | 97 | 96 | 96 | 96 | 96 | 96 | 96 | 95 | 96 |
| | | More Late | 15 | 0 | 4 | 4 | 2 | 0 | 1 | 0 | 0 | 0 | 0 | -2 | -2 | -1 | 2 | 0 | 2 | 2 | 3 | 97 | 96 | 96 | 95 | 96 | 95 | 96 | 96 | 96 |
| | 10% Control: 20% Active | Balanced | 16 | 0 | 27 | 26 | -1 | -1 | -1 | -1 | -1 | -1 | 0 | -2 | -2 | 3 | 3 | 4 | 3 | 3 | 4 | 98 | 90 | 90 | 97 | 96 | 96 | 96 | 96 | 96 |
| | | More Early | 17 | 0 | 28 | 28 | -16 | 1 | -15 | -1 | -1 | -1 | 0 | -1 | -1 | 12 | 2 | 12 | 11 | 11 | 14 | 98 | 89 | 89 | 94 | 97 | 95 | 96 | 97 | 97 |
| | | More Late | 18 | 0 | 30 | 30 | 15 | -3 | 14 | -1 | -1 | -1 | 0 | -3 | -3 | -1 | 3 | 0 | 4 | 3 | 4 | 98 | 88 | 88 | 95 | 96 | 95 | 96 | 96 | 96 |
| | 20% Control: 20% Active | Balanced | 19 | 0 | 8 | 8 | 0 | 0 | 0 | 0 | 0 | 0 | 0 | -2 | -2 | 4 | 3 | 4 | 4 | 4 | 4 | 98 | 96 | 96 | 97 | 96 | 96 | 97 | 96 | 97 |
| | | More Early | 20 | 0 | 8 | 8 | -5 | -1 | -4 | 0 | 0 | 0 | 0 | -1 | -1 | 15 | 3 | 14 | 13 | 13 | 16 | 98 | 96 | 96 | 95 | 96 | 95 | 95 | 95 | 95 |
| | | More Late | 21 | 0 | 9 | 9 | 5 | 1 | 5 | 0 | 0 | 0 | 0 | -3 | -3 | -1 | 4 | 0 | 5 | 5 | 6 | 98 | 96 | 96 | 96 | 97 | 96 | 96 | 96 | 96 |
| | 50% Control: 50% Active | Balanced | 22 | 0 | 18 | 18 | -1 | -1 | -1 | -1 | -1 | -1 | 0 | 4 | 5 | 10 | 9 | 10 | 9 | 9 | 10 | 97 | 96 | 96 | 96 | 96 | 96 | 96 | 96 | 96 |
| | | More Early | 23 | 0 | 17 | 16 | -13 | 0 | -11 | -1 | -1 | -1 | 0 | 8 | 10 | 29 | 9 | 27 | 31 | 30 | 33 | 97 | 96 | 96 | 95 | 96 | 95 | 95 | 95 | 94 |
| | | More Late | 24 | 0 | 21 | 21 | 11 | -1 | 11 | 0 | -1 | -1 | 0 | 0 | 1 | 2 | 10 | 3 | 10 | 10 | 11 | 97 | 94 | 94 | 96 | 96 | 96 | 94 | 94 | 95 |
| DNAR2 | 10% Control: 10% Active | Balanced | 25 | 0 | 4 | 4 | 0 | 0 | 0 | 0 | 0 | 0 | 0 | -1 | -1 | 2 | 2 | 3 | 2 | 2 | 3 | 96 | 96 | 96 | 96 | 96 | 96 | 97 | 97 | 97 |
| | | More Early | 26 | 0 | 4 | 4 | -2 | 0 | -2 | 1 | 1 | 1 | 0 | 0 | 0 | 8 | 3 | 10 | 8 | 8 | 12 | 96 | 95 | 95 | 95 | 96 | 96 | 96 | 96 | 96 |
| | | More Late | 27 | 0 | 4 | 4 | 2 | 0 | 2 | 0 | 0 | 0 | 0 | -2 | -2 | -1 | 1 | 0 | 2 | 2 | 3 | 96 | 96 | 96 | 96 | 96 | 96 | 96 | 96 | 96 |
| | 10% Control: 20% Active | Balanced | 28 | 1 | 24 | 24 | 1 | 1 | 1 | 1 | 1 | 1 | 0 | -1 | -1 | 3 | 3 | 4 | 3 | 3 | 4 | 97 | 90 | 90 | 95 | 95 | 96 | 95 | 95 | 96 |
| | | More Early | 29 | 1 | 23 | 23 | -13 | -2 | -13 | 1 | 1 | 1 | 0 | 0 | 0 | 12 | 4 | 13 | 11 | 11 | 14 | 97 | 91 | 91 | 94 | 96 | 94 | 96 | 95 | 96 |
| | | More Late | 30 | 1 | 29 | 29 | 16 | 4 | 15 | 1 | 1 | 1 | 0 | -3 | -3 | -1 | 2 | 0 | 4 | 4 | 5 | 97 | 87 | 87 | 93 | 95 | 93 | 95 | 96 | 96 |
| | 20% Control: 20% Active | Balanced | 31 | 0 | 8 | 8 | -1 | 0 | -1 | -1 | 0 | 0 | 0 | -1 | -1 | 4 | 3 | 5 | 4 | 4 | 5 | 96 | 95 | 95 | 96 | 96 | 96 | 96 | 96 | 96 |
| | | More Early | 32 | 0 | 8 | 8 | -6 | 0 | -6 | -1 | -1 | -1 | 0 | 1 | 1 | 15 | 5 | 16 | 14 | 14 | 17 | 96 | 95 | 95 | 95 | 96 | 95 | 95 | 94 | 95 |
| | | More Late | 33 | 0 | 9 | 9 | 4 | -1 | 4 | 0 | -1 | -1 | 0 | -4 | -3 | -1 | 3 | 0 | 5 | 5 | 6 | 96 | 95 | 95 | 96 | 96 | 96 | 95 | 95 | 95 |
| | 50% Control: 50% Active | Balanced | 34 | 0 | 19 | 18 | 0 | 0 | 0 | 0 | 0 | 0 | 0 | 5 | 6 | 10 | 9 | 12 | 10 | 10 | 11 | 97 | 95 | 95 | 94 | 94 | 95 | 94 | 94 | 94 |
| | | More Early | 35 | 0 | 17 | 17 | -12 | 0 | -12 | 0 | 0 | 0 | 0 | 10 | 11 | 29 | 12 | 31 | 33 | 32 | 35 | 97 | 96 | 96 | 94 | 95 | 94 | 95 | 94 | 95 |
| | | More Late | 36 | 0 | 21 | 21 | 12 | 0 | 12 | 0 | 0 | 0 | 0 | -1 | 0 | 2 | 8 | 3 | 11 | 11 | 12 | 97 | 93 | 93 | 95 | 95 | 95 | 93 | 93 | 93 |

Bias colour scale: -35, 0, 35

Change in CI Halfwidth colour scale: -25, 0, 25, 50

CI Coverage colour scale: 85, 90, 95, 100

Figure 3 Summary of simulation results for treatment effect: Same as Active Scenarios. The bias is calculated using analytical expected values, change in CI halfwidth is relative to the full simulated data (FULL Results), and both the CI halfwidths and coverages are for 95% intervals.

Scenario: Same as Active

| | | | | Bias (ml) | | | | | | | | | Change in CI Halfwidth (%) | | | | | | | | | CI Coverage (%) | | | | | | | | |
|---|---|---|---|---|---|---|---|---|---|---|---|---|---|---|---|---|---|---|---|---|---|---|---|---|---|---|---|---|---|---|
| Mechanism | Discontinuation Rate | Withdrawal Type | ID | FULL | MMRM | CICS | OICS | OICS-R | OIOS | PICS | PICS-R | PIOS | FULL | MMRM | CICS | OICS | OICS-R | OIOS | PICS | PICS-R | PIOS | FULL | MMRM | CICS | OICS | OICS-R | OIOS | PICS | PICS-R | PIOS |
| DAR | 10% Control: 10% Active | Balanced | 1 | 0 | 4 | 4 | 0 | 0 | 0 | 0 | 0 | 0 | 0 | 0 | 1 | 2 | 2 | 3 | 3 | 3 | 3 | 95 | 95 | 95 | 95 | 95 | 95 | 95 | 95 | 95 |
| | | More Early | 2 | 0 | 4 | 4 | -3 | 0 | -3 | 0 | 0 | 0 | 0 | 1 | 1 | 7 | 2 | 8 | 8 | 9 | 13 | 95 | 95 | 95 | 94 | 95 | 94 | 95 | 95 | 95 |
| | | More Late | 3 | 0 | 4 | 4 | 2 | 0 | 2 | 0 | 0 | 0 | 0 | 0 | 0 | 1 | 2 | 1 | 2 | 2 | 3 | 95 | 95 | 95 | 95 | 95 | 95 | 95 | 95 | 95 |
| | 10% Control: 20% Active | Balanced | 4 | 0 | 4 | 4 | 0 | 0 | 0 | 0 | 0 | 0 | 0 | 1 | 1 | 3 | 3 | 4 | 4 | 3 | 4 | 96 | 95 | 95 | 94 | 95 | 95 | 95 | 95 | 95 |
| | | More Early | 5 | 0 | 4 | 4 | -3 | 0 | -3 | -1 | -1 | -1 | 0 | 2 | 2 | 10 | 3 | 11 | 13 | 12 | 17 | 96 | 95 | 95 | 94 | 94 | 93 | 93 | 94 | 95 |
| | | More Late | 6 | 0 | 4 | 4 | 2 | 1 | 2 | 1 | 1 | 1 | 0 | 0 | 0 | 1 | 3 | 2 | 4 | 4 | 5 | 96 | 96 | 96 | 96 | 95 | 97 | 94 | 94 | 94 |
| | 20% Control: 20% Active | Balanced | 7 | 0 | 9 | 9 | 0 | 0 | 0 | 0 | 0 | 0 | 0 | 1 | 2 | 4 | 4 | 5 | 5 | 5 | 5 | 96 | 94 | 94 | 94 | 94 | 94 | 94 | 94 | 94 |
| | | More Early | 8 | 0 | 9 | 8 | -6 | 0 | -5 | 0 | 0 | 0 | 0 | 2 | 3 | 13 | 5 | 14 | 17 | 16 | 19 | 96 | 95 | 94 | 93 | 94 | 94 | 94 | 94 | 94 |
| | | More Late | 9 | 0 | 9 | 9 | 5 | 0 | 5 | 0 | 0 | 0 | 0 | 0 | 0 | 2 | 4 | 3 | 6 | 6 | 7 | 96 | 94 | 94 | 95 | 94 | 95 | 94 | 94 | 93 |
| | 50% Control: 50% Active | Balanced | 10 | 0 | 18 | 17 | -1 | -1 | -1 | -1 | -1 | -1 | 0 | 6 | 7 | 12 | 12 | 13 | 13 | 12 | 13 | 95 | 93 | 93 | 93 | 94 | 94 | 94 | 93 | 94 |
| | | More Early | 11 | 0 | 17 | 16 | -15 | 0 | -15 | 0 | 0 | 0 | 0 | 9 | 11 | 32 | 13 | 32 | 40 | 38 | 43 | 95 | 94 | 94 | 91 | 94 | 91 | 94 | 93 | 95 |
| | | More Late | 12 | 0 | 21 | 20 | 12 | 0 | 13 | 1 | 1 | 1 | 0 | 2 | 3 | 6 | 11 | 6 | 14 | 13 | 15 | 95 | 92 | 92 | 94 | 94 | 93 | 93 | 93 | 92 |
| DNAR1 | 10% Control: 10% Active | Balanced | 13 | 0 | 4 | 4 | 0 | 0 | 0 | 0 | 0 | 0 | 0 | 0 | 1 | 2 | 2 | 3 | 3 | 3 | 3 | 95 | 95 | 95 | 94 | 94 | 95 | 95 | 95 | 95 |
| | | More Early | 14 | 0 | 5 | 5 | -3 | 0 | -3 | 0 | 0 | 0 | 0 | 1 | 1 | 7 | 2 | 7 | 8 | 8 | 13 | 95 | 95 | 95 | 94 | 94 | 95 | 95 | 94 | 96 |
| | | More Late | 15 | 0 | 4 | 4 | 2 | 0 | 2 | 0 | 0 | 0 | 0 | 0 | 0 | 1 | 2 | 1 | 2 | 2 | 4 | 95 | 95 | 95 | 95 | 95 | 95 | 95 | 95 | 95 |
| | 10% Control: 20% Active | Balanced | 16 | 0 | 7 | 7 | 0 | 0 | 0 | 0 | 0 | 0 | 0 | 1 | 1 | 3 | 3 | 4 | 4 | 4 | 4 | 95 | 94 | 94 | 94 | 93 | 94 | 94 | 94 | 94 |
| | | More Early | 17 | 0 | 8 | 8 | -3 | 2 | -3 | 0 | -1 | 0 | 0 | 2 | 2 | 10 | 3 | 11 | 13 | 12 | 16 | 95 | 95 | 95 | 93 | 94 | 94 | 95 | 95 | 95 |
| | | More Late | 18 | 0 | 6 | 6 | 2 | -3 | 2 | 0 | 0 | -1 | 0 | 0 | 0 | 1 | 3 | 2 | 4 | 4 | 5 | 95 | 94 | 94 | 94 | 94 | 94 | 94 | 93 | 93 |
| | 20% Control: 20% Active | Balanced | 19 | 0 | 9 | 8 | 0 | 0 | 0 | 0 | 0 | 0 | 0 | 1 | 2 | 4 | 4 | 5 | 5 | 5 | 5 | 95 | 94 | 94 | 93 | 93 | 93 | 93 | 93 | 94 |
| | | More Early | 20 | 0 | 9 | 9 | -4 | 1 | -4 | 0 | 0 | 0 | 0 | 2 | 3 | 13 | 5 | 14 | 16 | 16 | 20 | 95 | 95 | 95 | 93 | 94 | 93 | 94 | 94 | 95 |
| | | More Late | 21 | 0 | 9 | 9 | 5 | 0 | 5 | 1 | 0 | 0 | 0 | 0 | 0 | 2 | 5 | 3 | 6 | 6 | 7 | 95 | 94 | 94 | 94 | 94 | 94 | 93 | 93 | 93 |
| | 50% Control: 50% Active | Balanced | 22 | 0 | 19 | 19 | 0 | 0 | 0 | 0 | 0 | 0 | 0 | 6 | 7 | 12 | 12 | 13 | 12 | 12 | 14 | 95 | 93 | 93 | 93 | 92 | 93 | 93 | 92 | 93 |
| | | More Early | 23 | 0 | 18 | 17 | -14 | 0 | -14 | 0 | 0 | -1 | 0 | 9 | 10 | 31 | 13 | 32 | 40 | 38 | 42 | 95 | 92 | 92 | 90 | 93 | 90 | 91 | 92 | 92 |
| | | More Late | 24 | 0 | 21 | 21 | 12 | -1 | 12 | 0 | 0 | 0 | 0 | 2 | 3 | 6 | 12 | 6 | 14 | 13 | 14 | 95 | 93 | 92 | 94 | 93 | 94 | 92 | 91 | 90 |
| DNAR2 | 10% Control: 10% Active | Balanced | 25 | 0 | 4 | 4 | 0 | 0 | 0 | 0 | 0 | 0 | 0 | 0 | 0 | 2 | 2 | 3 | 3 | 2 | 3 | 95 | 95 | 95 | 95 | 95 | 95 | 95 | 95 | 95 |
| | | More Early | 26 | 0 | 4 | 4 | -3 | 0 | -3 | -1 | 0 | 0 | 0 | 1 | 1 | 7 | 2 | 8 | 8 | 8 | 14 | 95 | 94 | 95 | 94 | 94 | 94 | 94 | 94 | 95 |
| | | More Late | 27 | 0 | 4 | 4 | 2 | 0 | 2 | 0 | 0 | 0 | 0 | 0 | 0 | 1 | 2 | 1 | 2 | 2 | 3 | 95 | 95 | 95 | 95 | 95 | 95 | 95 | 95 | 95 |
| | 10% Control: 20% Active | Balanced | 28 | 0 | 3 | 3 | -1 | 0 | 0 | 0 | 0 | 0 | 0 | 1 | 1 | 3 | 3 | 4 | 4 | 4 | 4 | 96 | 95 | 95 | 95 | 95 | 95 | 95 | 95 | 95 |
| | | More Early | 29 | 0 | 1 | 1 | -4 | -3 | -3 | -1 | -1 | -1 | 0 | 1 | 2 | 10 | 3 | 11 | 13 | 12 | 16 | 96 | 94 | 95 | 93 | 94 | 94 | 94 | 94 | 95 |
| | | More Late | 30 | 0 | 4 | 4 | 2 | 3 | 2 | 0 | 0 | 0 | 0 | 0 | 0 | 1 | 3 | 2 | 4 | 4 | 5 | 96 | 95 | 95 | 95 | 94 | 95 | 94 | 94 | 94 |
| | 20% Control: 20% Active | Balanced | 31 | 0 | 8 | 8 | -1 | -1 | -1 | -1 | -1 | -1 | 0 | 1 | 1 | 4 | 4 | 5 | 5 | 5 | 5 | 96 | 94 | 95 | 94 | 94 | 94 | 94 | 94 | 94 |
| | | More Early | 32 | 0 | 8 | 8 | -6 | 0 | -5 | 0 | 0 | 0 | 0 | 2 | 2 | 13 | 5 | 14 | 17 | 16 | 19 | 96 | 95 | 95 | 92 | 94 | 93 | 94 | 94 | 94 |
| | | More Late | 33 | 0 | 9 | 9 | 5 | 1 | 5 | 1 | 1 | 1 | 0 | 0 | 0 | 2 | 4 | 3 | 6 | 6 | 7 | 96 | 94 | 94 | 94 | 93 | 94 | 93 | 92 | 93 |
| | 50% Control: 50% Active | Balanced | 34 | 0 | 18 | 18 | 0 | 0 | 0 | 0 | 0 | 0 | 0 | 6 | 6 | 12 | 12 | 13 | 12 | 12 | 13 | 95 | 92 | 92 | 93 | 93 | 93 | 93 | 93 | 93 |
| | | More Early | 35 | 0 | 16 | 16 | -16 | -2 | -15 | -2 | -2 | -3 | 0 | 9 | 10 | 32 | 14 | 33 | 40 | 38 | 42 | 95 | 93 | 93 | 90 | 93 | 91 | 93 | 92 | 93 |
| | | More Late | 36 | 0 | 21 | 20 | 12 | 0 | 11 | -1 | -1 | -1 | 0 | 2 | 3 | 6 | 11 | 7 | 14 | 13 | 15 | 95 | 92 | 92 | 93 | 92 | 93 | 91 | 91 | 91 |

Colour scales: Bias -35 to 0 to 35; Change in CI Halfwidth -25, 0, 25, 50; CI Coverage 85, 90, 95, 100.

# Appendix 1: MI Models

## 1 MI model descriptions

### CICS: Common Intercept Common Slope

*Pseudo code in paper*

$$Y_j = Intercept + Y_0 + Y_1 + \cdots + Y_{j-1}$$

*Model Parameterization*

$$Y_1 = \alpha_{10} + \beta_{10} \cdot Y_0$$

$$Y_2 = \alpha_{20} + \beta_{20} \cdot Y_0 + \beta_{21} \cdot Y_1$$

$$Y_3 = \alpha_{30} + \beta_{30} \cdot Y_0 + \beta_{31} \cdot Y_1 + \beta_{32} \cdot Y_2$$

*Data patterns contributing to each parameter*

B = Baseline, O = On-treatment, X = Off-treatment.

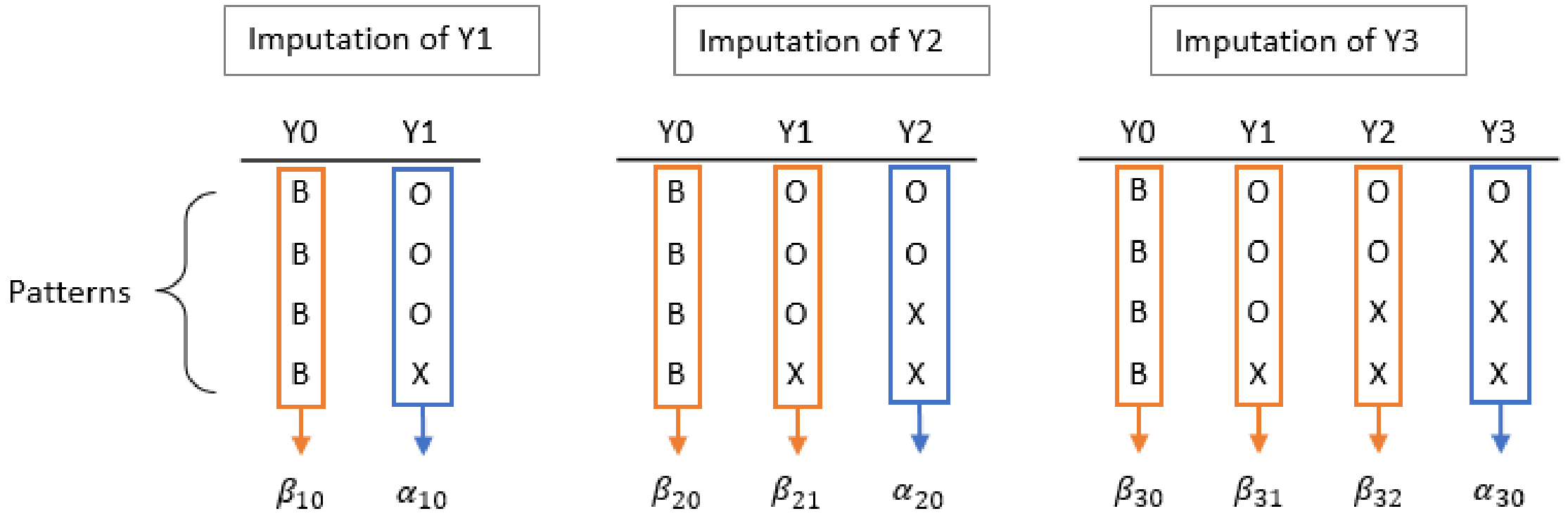

*Example SAS Code*

```
proc mi data     = <dataset>
        out      = <imputed>
        nimpute  = 100
        seed     = 12345;
  by TREAT;
  var Y0 Y1 Y2 Y3;
  monotone reg ( Y1 = Y0);
  monotone reg ( Y2 = Y0 Y1);
  monotone reg ( Y3 = Y0 Y1 Y2);
run;
```

# OICS: On/Off Intercept Common Slope

*Pseudo code in paper*

$$Y_j = Intercept + D_j + Y_0 + Y_1 + \cdots + Y_{j-1}$$

*Model Parameterization*

$$Y_1 = \alpha_{10} + D_1 \cdot \alpha_{11} + \beta_{10} \cdot Y_0$$

$$Y_2 = \alpha_{20} + D_2 \cdot \alpha_{21} + \beta_{20} \cdot Y_0 + \beta_{21} \cdot Y_1$$

$$Y_3 = \alpha_{30} + D_3 \cdot \alpha_{31} + \beta_{30} \cdot Y_0 + \beta_{31} \cdot Y_1 + \beta_{32} \cdot Y_2$$

*Data patterns contributing to each parameter*

B = Baseline, O = On-treatment, X = Off-treatment.

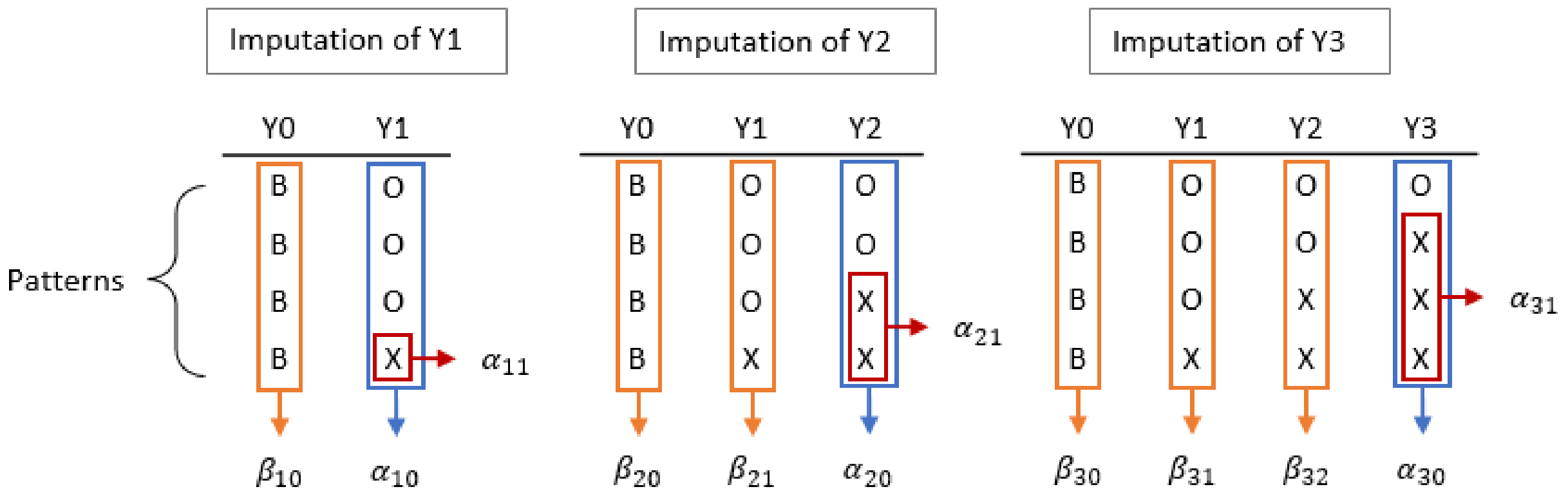

*Example SAS Code*

```
proc mi data     = <dataset>
        out      = <imputed>
        nimpute  = 100
        seed     = 12345;
  by TREAT;
  class D1 D2 D3;
  var D1 D2 D3 Y0 Y1 Y2 Y3;
  monotone reg ( Y1 = D1 Y0);
  monotone reg ( Y2 = D2 Y0 Y1);
  monotone reg ( Y3 = D3 Y0 Y1 Y2);
run;
```

# PICS: Pattern Intercept Common Slope

*Pseudo code in paper*

$$Y_j = Intercept + P_j + Y_0 + Y_1 + \cdots + Y_{j-1}$$

*Model Parameterization*

$$Y_1 = \alpha_{10} + P_{11} \cdot \alpha_{11} + \beta_{10} \cdot Y_0$$

$$Y_2 = \alpha_{20} + P_{21} \cdot \alpha_{21} + P_{22} \cdot \alpha_{22} + \beta_{20} \cdot Y_0 + \beta_{21} \cdot Y_1$$

$$Y_3 = \alpha_{30} + P_{31} \cdot \alpha_{31} + P_{32} \cdot \alpha_{32} + P_{33} \cdot \alpha_{33} + \beta_{30} \cdot Y_0 + \beta_{31} \cdot Y_1 + \beta_{32} \cdot Y_2$$

*Data patterns contributing to each parameter*

B = Baseline, O = On-treatment, X = Off-treatment.

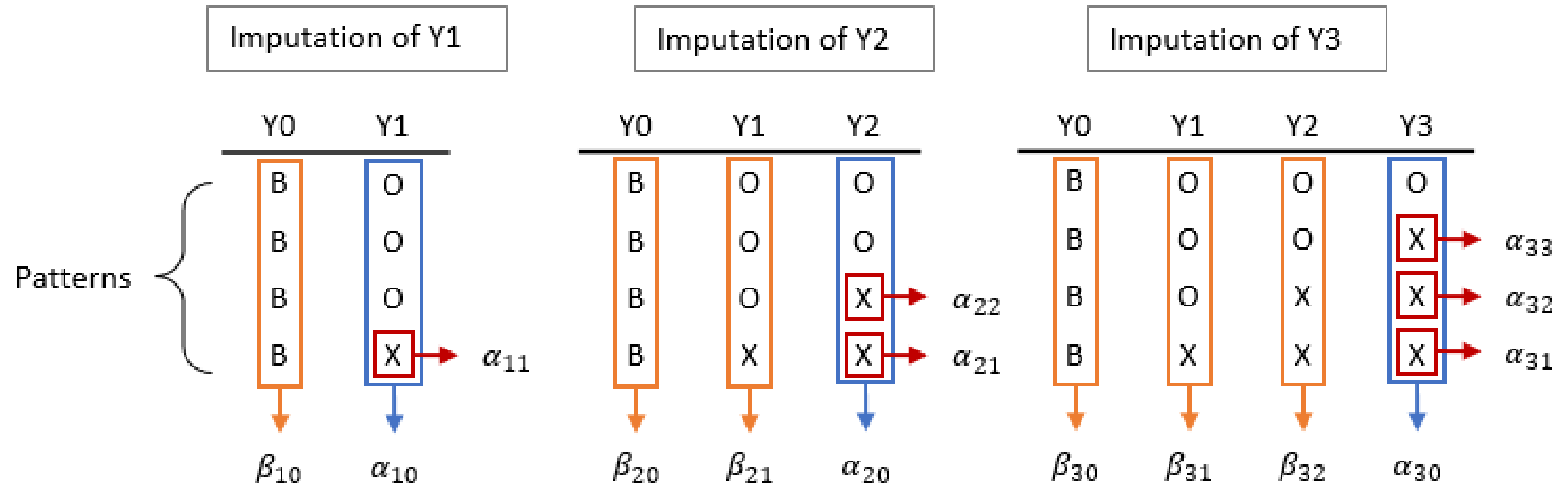

*Example SAS Code*

```
proc mi data      = <dataset>
        out       = <imputed>
        nimpute   = 100
        seed      = 12345;
  by TREAT;
  class P1 P2 P3;
  var P1 P2 P3 Y0 Y1 Y2 Y3;
  monotone reg ( Y1 = P1 Y0);
  monotone reg ( Y2 = P2 Y0 Y1);
  monotone reg ( Y3 = P3 Y0 Y1 Y2);
run;
```

# OIOS: On/Off Intercept On/Off Slope

*Pseudo code in paper*

$$Y_j = Intercept + D_j + Y_0 + Y_1 + \cdots + Y_{j-1} + D_j * Y_1 + \cdots + D_j * Y_j$$

*Model Parameterization*

$$Y_1 = \alpha_{10} + D_1 \cdot \alpha_{11} + \beta_{10} \cdot Y_0$$

$$Y_2 = \alpha_{20} + D_2 \cdot \alpha_{21} + \beta_{20} \cdot Y_0 + \left(\beta_{21} + D_2 \cdot \beta_{21}^{(1)}\right) \cdot Y_1$$

$$Y_3 = \alpha_{30} + D_3 \cdot \alpha_{31} + \beta_{30} \cdot Y_0 + \left(\beta_{31} + D_3 \cdot \beta_{31}^{(1)}\right) \cdot Y_1 + \left(\beta_{32} + D_3 \cdot \beta_{32}^{(1)}\right) \cdot Y_2$$

*Data patterns contributing to each parameter*

B = Baseline, O = On-treatment, X = Off-treatment.

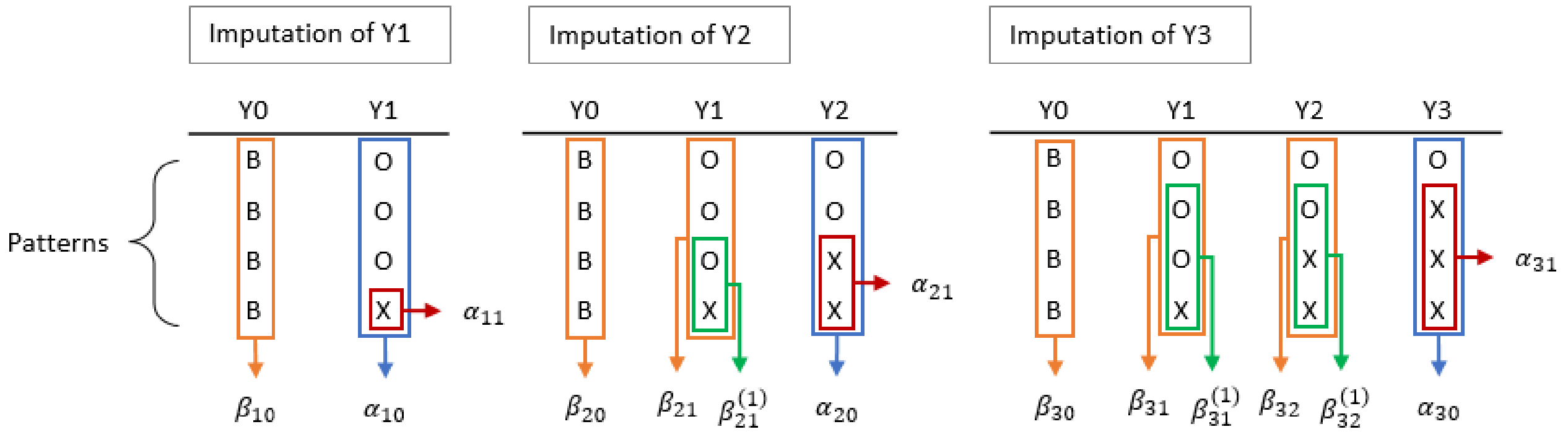

*Example SAS Code*

```
proc mi data    = <dataset>
        out     = <imputed>
        nimpute = 100
        seed    = 12345;
  by TREAT;
  class D1 D2 D3;
  var D1 D2 D3 Y0 Y1 Y2 Y3;
  monotone reg ( Y1 = D1 Y0);
  monotone reg ( Y2 = D2 Y0 Y1 D2*Y1);
  monotone reg ( Y3 = D3 Y0 Y1 D3*Y1 Y2 D3*Y2);
run;
```

# PIOS: Pattern Intercept On/Off Slope

*Pseudo code in paper*

$$Y_j = Intercept + P_j + Y_0 + Y_1 + \cdots + Y_{j-1} + D_1 * Y_1 + \cdots + D_j * Y_j$$

*Model Parameterization*

$$Y_1 = \alpha_{10} + P_{11} \cdot \alpha_{11} + \beta_{10} \cdot Y_0$$

$$Y_2 = \alpha_{20} + P_{21} \cdot \alpha_{21} + P_{22} \cdot \alpha_{22} + \beta_{20} \cdot Y_0 + \left(\beta_{21} + D_1 \cdot \beta_{21}^{(1)}\right) \cdot Y_1$$

$$Y_3 = \alpha_{30} + P_{31} \cdot \alpha_{31} + P_{32} \cdot \alpha_{32} + P_{33} \cdot \alpha_{33} + \beta_{30} \cdot Y_0 + \left(\beta_{31} + D_2 \cdot \beta_{31}^{(1)}\right) \cdot Y_1 + \left(\beta_{32} + D_2 \cdot \beta_{32}^{(1)}\right) \cdot Y_2$$

*Data patterns contributing to each parameter*

B = Baseline, O = On-treatment, X = Off-treatment.

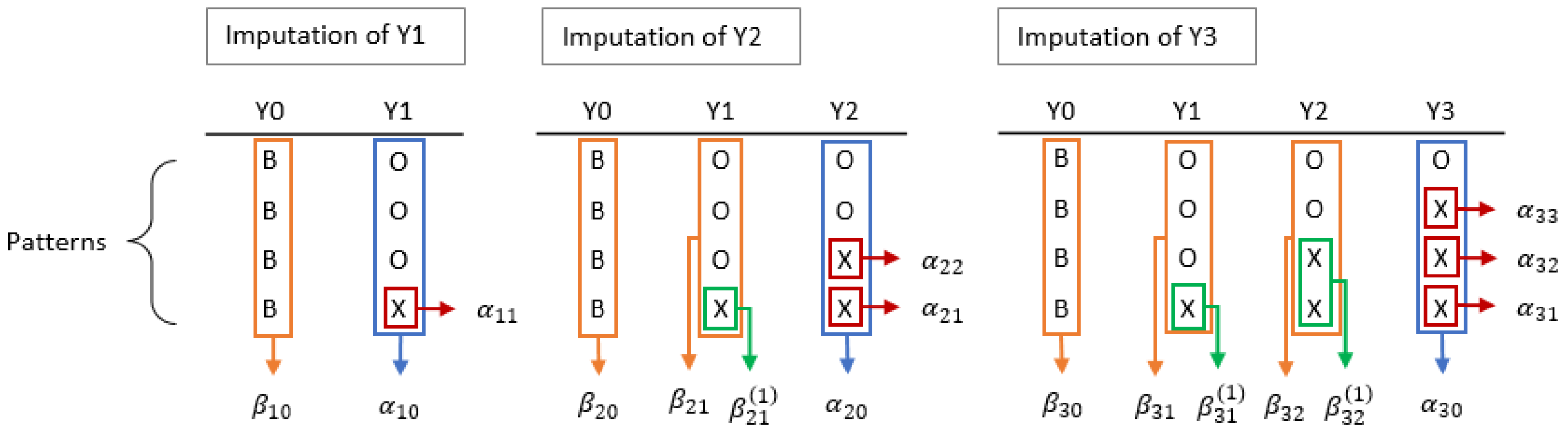

*Example SAS Code*

```
proc mi data      = <dataset>
        out       = <imputed>
        nimpute   = 100
        seed      = 12345;
  where D3 = 1;
  by TREAT;
  class P1 P2 P3 D1 D2;
  var P1 P2 P3 D1 D2 Y0 Y1 Y2 Y3;
  monotone reg ( Y1 = P1 Y0);
  monotone reg ( Y2 = P2 Y0 Y1 D1*Y1);
  monotone reg ( Y3 = P3 Y0 Y1 D1*Y1 Y2 D2*Y2);
run;
```

# PIOS: Pattern Intercept Pattern Slope

*Pseudo code in paper*

$$Y_j = P_J * (Intercept + Y_0 + Y_1 + \cdots + Y_{j-1})$$

*Model Parameterization*

$$Y_1 = P_{30}\left(\alpha_{10} + \beta_{10}^{(0)} \cdot Y_0\right) + P_{31}\left(\alpha_{11} + \beta_{10}^{(1)} \cdot Y_0\right) + P_{32}\left(\alpha_{12} + \beta_{10}^{(2)} \cdot Y_0\right) + P_{33}\left(\alpha_{13} + \beta_{10}^{(3)} \cdot Y_0\right)$$

$$Y_2 = P_{30}\left(\alpha_{20} + \beta_{20}^{(0)} \cdot Y_0 + \beta_{21}^{(1)} \cdot Y_1\right) + P_{31}\left(\alpha_{21} + \beta_{20}^{(1)} \cdot Y_0 + \beta_{21}^{(1)} \cdot Y_1\right) + P_{32}\left(\alpha_{22} + \beta_{20}^{(2)} \cdot Y_0 + \beta_{21}^{(2)} \cdot Y_1\right) + P_{33}\left(\alpha_{23} + \beta_{20}^{(3)} \cdot Y_0 + \beta_{21}^{(3)} \cdot Y_1\right)$$

$$Y_3 = P_{30}\left(\alpha_{30} + \beta_{30}^{(0)} \cdot Y_0 + \beta_{31}^{(0)} \cdot Y_1 + \beta_{32}^{(0)} \cdot Y_2\right) + P_{31}\left(\alpha_{31} + \beta_{30}^{(1)} \cdot Y_0 + \beta_{31}^{(1)} \cdot Y_1 + \beta_{32}^{(1)} \cdot Y_2\right) + P_{32}\left(\alpha_{32} + \beta_{30}^{(2)} \cdot Y_0 + \beta_{31}^{(2)} \cdot Y_1 + \beta_{32}^{(2)} \cdot Y_2\right) + P_{33}\left(\alpha_{33} + \beta_{30}^{(3)} \cdot Y_0 + \beta_{31}^{(3)} \cdot Y_1 + \beta_{32}^{(3)} \cdot Y_2\right)$$

*Data patterns contributing to each parameter*

B = Baseline, O = On-treatment, X = Off-treatment.

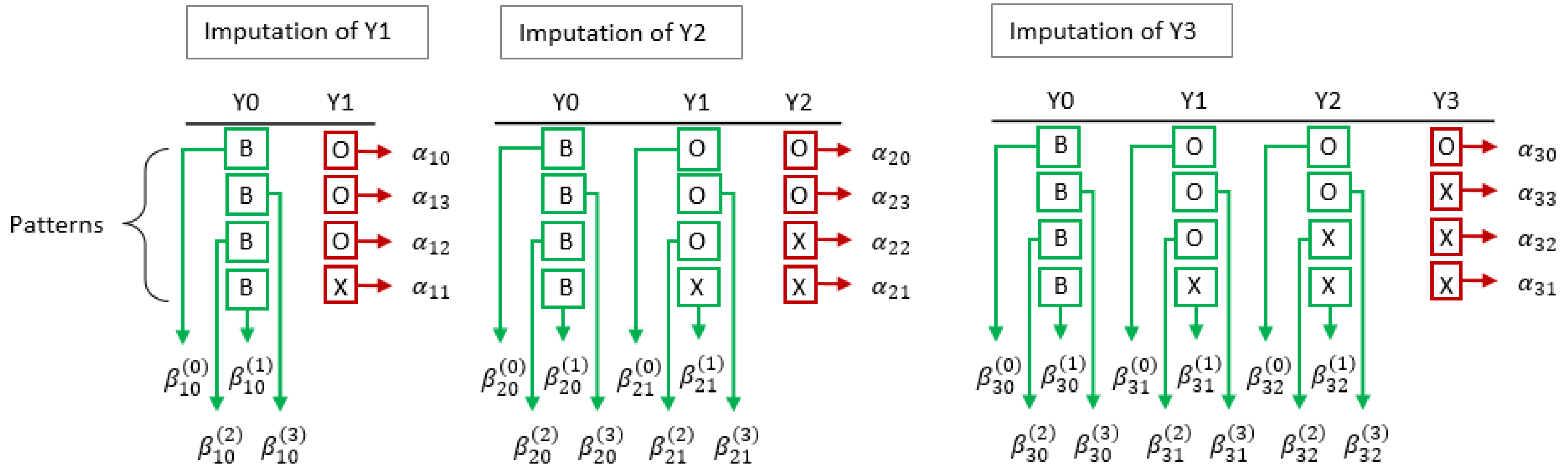

*Example SAS Code*

```
proc mi data     = <dataset>
        out      = <imputed>
        nimpute  = 100
        seed     = 12345;
  by TREAT P3;
  var Y0 Y1 Y2 Y3;
  monotone reg ( Y1 = Y0);
  monotone reg ( Y2 = Y0 Y1);
  monotone reg ( Y3 = Y0 Y1 Y2);
run;
```

# OICS-R: On/Off Intercept Common Slope on Residuals

*Pseudo code in paper*

$$Y_j = \mu_j + D_j + R_0 + R_1 + \cdots + R_{j-1}$$

*Model Parameterization*

$$Y_1 = \mu_1 + D_1 \cdot \delta_1 + \beta_{10} \cdot R_0$$

$$Y_2 = \mu_2 + D_2 \cdot \delta_2 + \beta_{20} \cdot R_0 + \beta_{21} \cdot R_1$$

$$Y_3 = \mu_3 + D_3 \cdot \delta_3 + \beta_{30} \cdot R_0 + \beta_{31} \cdot R_1 + \beta_{32} \cdot R_2$$

*Data patterns contributing to each parameter*

B = Baseline, O = On-treatment, X = Off-treatment.

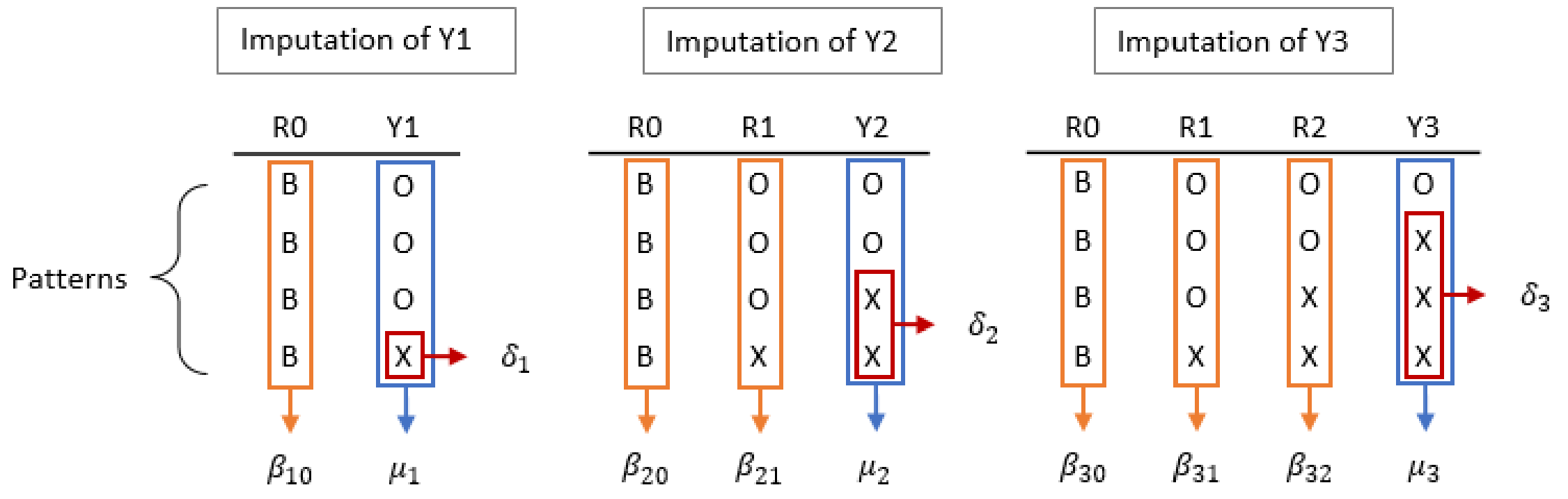

*Example SAS Code*

```
%mistep(data=<dataset>, out=imputed0, by=TREAT
       ,response=Y0, class=sim, model=sim, suffix=0
       ,nimpute=100, seed=1234500);

%mistep(data=imputed0, out=imputed1, by=TREAT
       ,response=Y1, class=D1, model=%str(D1 R0)
       ,nimpute=1, seed=1234501);

%mistep(data=imputed1, out=imputed2, by=TREAT
       ,response=Y2, class=D2, model=%str(D2 R0 R1)
       ,nimpute=1, seed=1234502);

%mistep(data=imputed2, out=imputed3, by=TREAT
       ,response=Y3, class=D3, model=%str(D3 R0 R1 R2)
       ,nimpute=1, seed=1234503);
```

# PICS-R: Pattern Intercept Common Slope on Residuals

*Pseudo code in paper*

$$Y_j = \mu_j + P_j + R_0 + R_1 + \cdots + R_{j-1}$$

*Model Parameterization*

$$Y_1 = \mu_1 + P_{11} \cdot \delta_{11} + \beta_{10} \cdot R_0$$

$$Y_2 = \mu_2 + P_{21} \cdot \delta_{21} + P_{22} \cdot \delta_{22} + \beta_{20} \cdot R_0 + \beta_{21} \cdot R_1$$

$$Y_3 = \mu_3 + P_{31} \cdot \delta_{31} + P_{32} \cdot \delta_{32} + P_{33} \cdot \delta_{33} + \beta_{30} \cdot R_0 + \beta_{31} \cdot R_1 + \beta_{32} \cdot R_2$$

*Data patterns contributing to each parameter*

B = Baseline, O = On-treatment, X = Off-treatment.

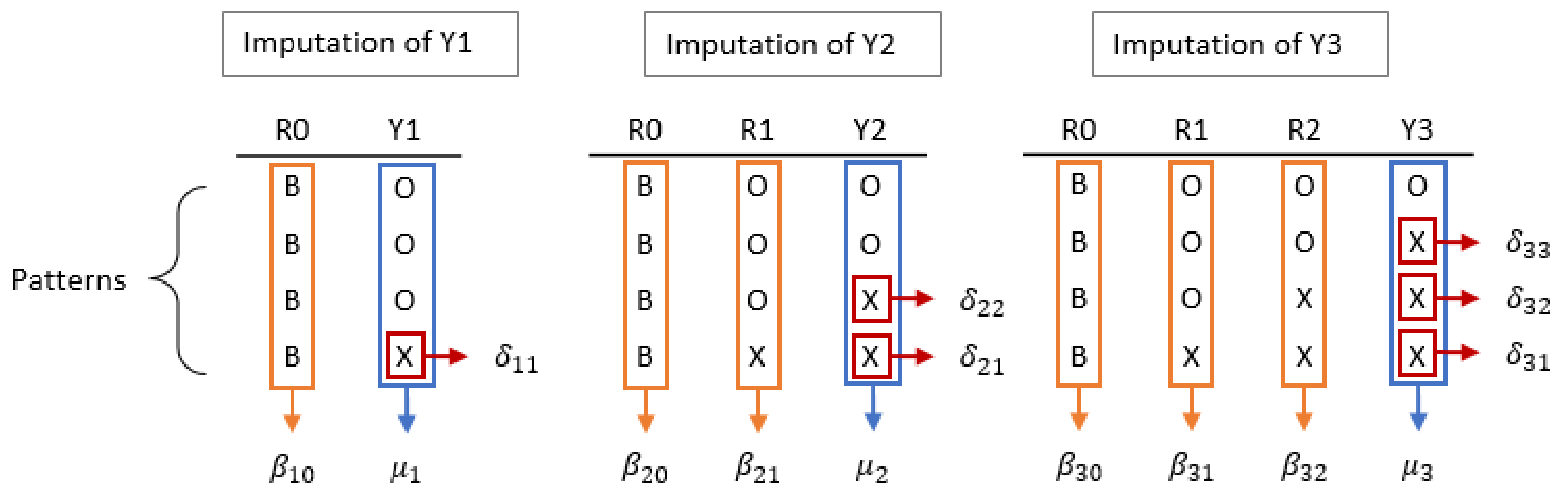

*Example SAS Code*

```
%mistep(data=<dataset>, out=imputed0, by=TREAT
        ,response=Y0, class=sim, model=sim, suffix=0
        ,nimpute=100, seed=1234500);

%mistep(data=imputed0, out=imputed1, by=TREAT
        ,response=Y1, class=P1, model=%str(P1 R0)
        ,nimpute=1, seed=1234501);

%mistep(data=imputed1, out=imputed2, by=TREAT
        ,response=Y2, class=P2, model=%str(P2 R0 R1)
        ,nimpute=1, seed=1234502);

%mistep(data=imputed2, out=imputed3, by=TREAT
        ,response=Y3, class=P3, model=%str(P3 R0 R1 R2)
        ,nimpute=1, seed=1234503);
```

## 2 Comparison of OICS and OICS-R

The following notes illustrate that OICS and OICS-R may appear to be the same but are in fact different imputation models. We show this by comparing the expected values used for imputing the different discontinuation patterns. The OICS and OICS-R models are defined as:

| OICS | OICS-R |
|---|---|
| $Y_1 = \alpha_{10} + D_1 \cdot \alpha_{11} + \beta_{10} \cdot Y_0$ | $Y_1 = \mu_1 + D_1 \cdot \delta_1 + \beta_{10} \cdot R_0$ |
| $Y_2 = \alpha_{20} + D_2 \cdot \alpha_{21} + \beta_{20} \cdot Y_0 + \beta_{21} \cdot Y_1$ | $Y_2 = \mu_2 + D_2 \cdot \delta_2 + \beta_{20} \cdot R_0 + \beta_{21} \cdot R_1$ |
| $Y_3 = \alpha_{30} + D_3 \cdot \alpha_{31} + \beta_{30} \cdot Y_0 + \beta_{31} \cdot Y_1 + \beta_{32} \cdot Y_2$ | $Y_3 = \mu_3 + D_3 \cdot \delta_3 + \beta_{30} \cdot R_0 + \beta_{31} \cdot R_1 + \beta_{32} \cdot R_2$ |

Letting $\mu_0$ be the baseline expected value and $\mu_j^P$ be the expected value for time point $j$ and pattern $P$ which we specify as combinations of "O" and "X" to represent on- and off-treatment at prior timepoints, the OICS and OICS-R expected values for each pattern are:

| Mean | OICS | OICS-R |
|---|---|---|
| $\mu_1^O$ | $\alpha_{10} + \beta_{10} \cdot \mu_0$ | $\mu_1$ |
| $\mu_1^X$ | $\alpha_{10} + \alpha_{11} + \beta_{10} \cdot \mu_0$ | $\mu_1 + \delta_1$ |
| $\mu_2^{OO}$ | $\alpha_{20} + \beta_{20} \cdot \mu_0 + \beta_{21}(\alpha_{10} + \beta_{10} \cdot \mu_0)$ | $\mu_2$ |
| $\mu_2^{OX}$ | $\alpha_{20} + \alpha_{21} + \beta_{20} \cdot \mu_0 + \beta_{21}(\alpha_{10} + \beta_{10} \cdot \mu_0)$ | $\mu_2 + \delta_2$ |
| $\mu_2^{XX}$ | $\alpha_{20} + \alpha_{21} + \beta_{20} \cdot \mu_0 + \beta_{21}(\alpha_{10} + \alpha_{11} + \beta_{10} \cdot \mu_0)$ | $\mu_2 + \delta_2$ |
| $\mu_3^{OOO}$ | $\alpha_{30} + \beta_{30} \cdot \mu_0 + \beta_{31}(\alpha_{10} + \beta_{10} \cdot \mu_0) + \beta_{32}\left(\alpha_{20} + \beta_{20} \cdot \mu_0 + \beta_{21}(\alpha_{10} + \beta_{10} \cdot \mu_0)\right)$ | $\mu_3$ |
| $\mu_3^{OOX}$ | $\alpha_{30} + \alpha_{31} + \beta_{30} \cdot \mu_0 + \beta_{31}(\alpha_{10} + \beta_{10} \cdot \mu_0) + \beta_{32}\left(\alpha_{20} + \beta_{20} \cdot \mu_0 + \beta_{21}(\alpha_{10} + \beta_{10} \cdot \mu_0)\right)$ | $\mu_3 + \delta_3$ |
| $\mu_3^{OXX}$ | $\alpha_{30} + \alpha_{31} + \beta_{30} \cdot \mu_0 + \beta_{31}(\alpha_{10} + \beta_{10} \cdot \mu_0) + \beta_{32}\left(\alpha_{20} + \alpha_{21} + \beta_{20} \cdot \mu_0 + \beta_{21}(\alpha_{10} + \beta_{10} \cdot \mu_0)\right)$ | $\mu_3 + \delta_3$ |
| $\mu_3^{XXX}$ | $\alpha_{30} + \alpha_{31} + \beta_{30} \cdot \mu_0 + \beta_{31}(\alpha_{10} + \alpha_{11} + \beta_{10} \cdot \mu_0) + \beta_{32}\left(\alpha_{20} + \alpha_{21} + \beta_{20} \cdot \mu_0 + \beta_{21}(\alpha_{10} + \alpha_{11} + \beta_{10} \cdot \mu_0)\right)$ | $\mu_3 + \delta_3$ |

The structure of the expected values for OICS-R simplify due to $E(R_j) = 0$ and so we can see the expected values used of imputation in OICS depend upon the pattern of discontinuation, whereas the OICS-R model uses the same expected value for imputation across patterns and therefore these models are not the same.

# 3 Comparison of PICS and PICS-R

The following notes illustrate that the PICS and PICS-R models are essentially the same model but parameterised differently. The PICS and PICS-R models are defined as:

| PICS | PICS-R |
|---|---|
| $Y_1 = \alpha_{10} + P_{11} \cdot \alpha_{11} + \beta_{10} \cdot Y_0$ | $Y_1 = \mu_1 + P_{11} \cdot \delta_{11} + \beta_{10} \cdot R_0$ |
| $Y_2 = \alpha_{20} + P_{21} \cdot \alpha_{21} + P_{22} \cdot \alpha_{22} + \beta_{20} \cdot Y_0 + \beta_{21} \cdot Y_1$ | $Y_2 = \mu_2 + P_{21} \cdot \delta_{21} + P_{22} \cdot \delta_{22} + \beta_{20} \cdot R_0 + \beta_{21} \cdot R_1$ |
| $Y_3 = \alpha_{30} + P_{31} \cdot \alpha_{31} + P_{32} \cdot \alpha_{32} + P_{33} \cdot \alpha_{33} + \beta_{30} \cdot Y_0 + \beta_{31} \cdot Y_1 + \beta_{32} \cdot Y_2$ | $Y_3 = \mu_3 + P_{31} \cdot \delta_{31} + P_{32} \cdot \delta_{32} + P_{33} \cdot \delta_{33} + \beta_{30} \cdot R_0 + \beta_{31} \cdot R_1 + \beta_{32} \cdot R_2$ |

Letting $\mu_0$ be the baseline expected value and $\mu_j^P$ be the pattern mean for time point $j$ and pattern $P$ which we specify as combinations of “O” and “X” to represent on- and off-treatment at prior timepoints, the PICS and PICS-R expected values for each pattern are:

| Mean | PICS | PICS-R |
|---|---|---|
| $\mu_1^{O}$ | $\alpha_{10} + \beta_{10} \cdot \mu_0$ | $\mu_1$ |
| $\mu_1^{X}$ | $\alpha_{10} + \alpha_{11} + \beta_{10} \cdot \mu_0$ | $\mu_1 + \delta_{11}$ |
| $\mu_2^{OO}$ | $\alpha_{20} + \beta_{20} \cdot \mu_0 + \beta_{21}(\alpha_{10} + \beta_{10} \cdot \mu_0)$ | $\mu_2$ |
| $\mu_2^{OX}$ | $\alpha_{20} + \alpha_{21} + \beta_{20} \cdot \mu_0 + \beta_{21}(\alpha_{10} + \beta_{10} \cdot \mu_0)$ | $\mu_2 + \delta_{21}$ |
| $\mu_2^{XX}$ | $\alpha_{20} + \alpha_{22} + \beta_{20} \cdot \mu_0 + \beta_{21}(\alpha_{10} + \alpha_{11} + \beta_{10} \cdot \mu_0)$ | $\mu_2 + \delta_{22}$ |
| $\mu_3^{OOO}$ | $\alpha_{30} + \beta_{30} \cdot \mu_0 + \beta_{31}(\alpha_{10} + \beta_{10} \cdot \mu_0) + \beta_{32}\big(\alpha_{20} + \beta_{20}\mu_0 + \beta_{21}(\alpha_{10} + \beta_{10} \cdot \mu_0)\big)$ | $\mu_3$ |
| $\mu_3^{OOX}$ | $\alpha_{30} + \alpha_{31} + \beta_{30} \cdot \mu_0 + \beta_{31}(\alpha_{10} + \beta_{10} \cdot \mu_0) + \beta_{32}\big(\alpha_{20} + \beta_{20}\mu_0 + \beta_{21}(\alpha_{10} + \beta_{10} \cdot \mu_0)\big)$ | $\mu_3 + \delta_{31}$ |
| $\mu_3^{OXX}$ | $\alpha_{30} + \alpha_{32} + \beta_{30} \cdot \mu_0 + \beta_{31}(\alpha_{10} + \beta_{10} \cdot \mu_0) + \beta_{32}\big(\alpha_{20} + \alpha_{21} + \beta_{20}\mu_0 + \beta_{21}(\alpha_{10} + \beta_{10} \cdot \mu_0)\big)$ | $\mu_3 + \delta_{32}$ |
| $\mu_3^{XXX}$ | $\alpha_{30} + \alpha_{33} + \beta_{30} \cdot \mu_0 + \beta_{31}(\alpha_{10} + \alpha_{11} + \beta_{10} \cdot \mu_0) + \beta_{32}\big(\alpha_{20} + \alpha_{22} + \beta_{20}\mu_0 + \beta_{21}(\alpha_{10} + \alpha_{11} + \beta_{10} \cdot \mu_0)\big)$ | $\mu_3 + \delta_{33}$ |

Equating the expected values between the models for each discontinuation pattern and working from timepoint one using substitution we can see:

Timepoint 1:

$\alpha_{10} = \mu_1 - \beta_{10} \cdot \mu_0$

$\alpha_{11} = \delta_{11}$

Timepoint 2:

$\alpha_{20} = \mu_2 - \beta_{20} \cdot \mu_0 - \beta_{21} \cdot \mu_1$

$\alpha_{21} = \delta_{21}$

$\alpha_{22} = \delta_{22} - \beta_{21} \cdot \delta_{11}$

Timepoint 3:

$\alpha_{30} = \mu_3 - \beta_{30} \cdot \mu_0 - \beta_{31} \cdot \mu_1 - \beta_{32} \cdot \mu_2$

$\alpha_{31} = \delta_{31}$

$\alpha_{32} = \delta_{32} - \beta_{32} \cdot \delta_{21}$

$\alpha_{33} = \delta_{33} - \beta_{31} \cdot \delta_{11} - \beta_{32} \cdot \delta_{22}$

So, the parameters of PICS are just linear combinations of the parameters in PICS-R.

# Appendix 2: Data Generation Model

## 1 Expected values of the treatment policy

For 10% Discontinuation (5%, 3%, 2%):

$$E[Y_0] = \mu_0^{on}$$
$$E[Y_1] = 0.95 \cdot \mu_1^{on} + 0.05 \cdot \mu_1^{off}$$
$$E[Y_2] = 0.92 \cdot \mu_2^{on} + 0.08 \cdot \mu_2^{off}$$
$$E[Y_3] = 0.90 \cdot \mu_3^{on} + 0.10 \cdot \mu_3^{off}$$

For 20% Discontinuation (10%, 6%, 4%):

$$E[Y_0] = \mu_0^{on}$$
$$E[Y_1] = 0.90 \cdot \mu_1^{on} + 0.10 \cdot \mu_1^{off}$$
$$E[Y_2] = 0.84 \cdot \mu_2^{on} + 0.16 \cdot \mu_2^{off}$$
$$E[Y_3] = 0.80 \cdot \mu_3^{on} + 0.20 \cdot \mu_3^{off}$$

For 50% Discontinuation (25%, 15%, 10%):

$$E[Y_0] = \mu_0^{on}$$
$$E[Y_1] = 0.75 \cdot \mu_1^{on} + 0.25 \cdot \mu_1^{off}$$
$$E[Y_2] = 0.60 \cdot \mu_2^{on} + 0.40 \cdot \mu_2^{off}$$
$$E[Y_3] = 0.50 \cdot \mu_3^{on} + 0.50 \cdot \mu_3^{off}$$

The exact values of $\mu_j^{on}$ and $\mu_j^{off}$ depend on the timepoint and off-treatment scenario.

## 2 Details of the DGM

### 2.1 Stage 1 - Generating On-treatment and Off-treatment Data

All multivariate normal on treatment outcomes are simulated based on the mean vector and variance-covariance matrix for the on-treatment control arm and then adjusted by the active treatment effect if the subject is in the Active group and a subject specific on-treatment heterogeneity is applied to each simulated value.

To generate the set of off treatment values, on-treatment values for the second timepoint onwards are adjusted according to the expected value for the off-treatment scenario and subject specific heterogeneity. For "return to baseline" scenarios the off-treatment values are adjusted to make the expected value equal to the baseline expected value for each treatment. In the "same as active" scenarios the off-treatment values are adjusted so that the expected value matches that of the Active, on treatment value for the same time point.

Let $\boldsymbol{Y}_C = (Y_{0C}, Y_{1C}, Y_{2C}, Y_{3C})'$ be the normal outcome vector for a control treatment with mean $\boldsymbol{\mu}_C$ and covariance matrix $\boldsymbol{\Sigma}_C$ i.e. $\boldsymbol{Y}_C \sim MVN(\boldsymbol{\mu}_C, \boldsymbol{\Sigma}_C)$.

Let $\boldsymbol{\Delta} = (\Delta_0, \Delta_1, \Delta_2, \Delta_3)'$ be the vector of expected differences for active treatment compared with control.

Also define $u \sim normal(0, \theta_{\text{on}}^2)$ and $v \sim normal(0, \theta_{\text{off}}^2)$ be the patient specific heterogeneity for on and off treatment outcomes respectively.

The patient on-treatment outcome vectors for group Z can be created as:

$$\boldsymbol{Y}_Z^{on} = \boldsymbol{Y}_C + I_{[Z=A]} \cdot \boldsymbol{\Delta} + \mathbf{1}_4 \cdot u$$

The patient off-treatment outcome vectors for group Z in the **return to baseline** scenarios can be created as:

$$\boldsymbol{Y}_Z^{off} = \boldsymbol{Y}_{Z,-1}^{on} - \boldsymbol{\mu}_{C,-1} - I_{[Z=A]} \cdot \boldsymbol{\Delta}_{-1} + \mathbf{1}_3 \cdot \mu_{0C} + \mathbf{1}_3 \cdot v$$

The patient off-treatment outcome vectors for group Z in the **same as active** scenarios can be created as:

$$\boldsymbol{Y}_Z^{off} = \boldsymbol{Y}_{Z,-1}^{on} + I_{[Z=C]} \cdot \boldsymbol{\Delta}_{-1} + \mathbf{1}_3 \cdot v$$

Where:

- $I_{[\cdot]}$ is an indicator function taking the value 1 when condition $[\cdot]$ is true and zero otherwise.
- $\boldsymbol{Y}_{Z,-1}^{on}$ is the vector of on-treatment outcomes for a patient in group Z with the first element removed.
- $\boldsymbol{\mu}_{C,-1}$ is the vector of expected values of the control treatment with the first element removed.
- $\boldsymbol{\Delta}_{-1}$ is the vector of expected values of active treatment effects with the first element removed.
- $\mu_{0C}$ is the expected value of the control treatment at timepoint 0 (baseline).
- $\mathbf{1}_n$ is an n-dimensional vector with each element being the number 1.

The multivariate normal distribution used to simulate $\boldsymbol{Y}_C \sim MVN(\boldsymbol{\mu}_C, \boldsymbol{\Sigma}_C)$ was:

$$\boldsymbol{Y}_C \sim MVN\left(\begin{bmatrix} 2.14 \\ 2.47 \\ 2.52 \\ 2.54 \end{bmatrix}, \begin{bmatrix} 0.45 & 0.46 & 0.46 & 0.47 \\ 0.46 & 0.66 & 0.62 & 0.63 \\ 0.46 & 0.62 & 0.65 & 0.63 \\ 0.47 & 0.63 & 0.63 & 0.68 \end{bmatrix}\right)$$

And the active treatment effects $\boldsymbol{\Delta}$ were:

$$\boldsymbol{\Delta} = [0 \quad 0.1 \quad 0.1 \quad 0.1]'$$

The on and off-treatment heterogeneity standard deviations $\theta_{on}$ and $\theta_{off}$ were both set to 0.3.

This creates a single set of On-treatment and Off-treatment values for each subject and each scenario and is based on a single draw from the multivariate normal distribution. This allows the off-treatment values to be the same outcomes generated for the on treatment but adjusted for the scenario of interest.

## 2.2 Stage 2 - Selecting patients to discontinue from treatment

Once both sets of On and Off treatment outcomes values have been generated, a discontinuation process is applied to the on-treatment data to determine if a subject discontinues. To flag the subjects that discontinue we used a heuristic selection process based on a propensity score.

For the $j$-th timepoint we defined a conditioning parameter $\omega_j$ using the standard deviation $\sigma_j$ as:

$$\omega_j = \frac{0.5}{\sigma_j}$$

We then use the conditioning parameter as the basis for a simple logistic selection model which for the DAR process is:

$$P\left(D_j^{DAR} = 1 \middle| Y_{j-1}\right) = logistic(\omega_j \cdot Y_{j-1})$$

And for the DNAR processes:

$$P\left(D_j^{DNAR} = 1 \middle| Y_j\right) = logistic(\omega_j \cdot Y_j)$$

To select the subject to discontinue we simulate $u \sim uniform(0,1)$, apply the inverse logistic transform (the logit) and subtract that value from both sides which creates a propensity like relative measure of the simulated on-treatment values.

$$\kappa_j^{DAR} = (\omega_j \cdot y_{j-1}) - logit(u)$$
$$\kappa_j^{DNAR} = (\omega_j \cdot y_j) - logit(u)$$

The measure $\kappa_j$ is then ranked and percentages above or below thresholds selected. This was to ensure exact proportions of subjects were selected for each simulated trial.

For the one process DAR and DNAR setting a percentage of the worst ranked subjects are selected to discontinue at each timepoint. The actual percentage thresholds for selection are based on the discontinuation rates for the particular scenario.

For the two process DNAR setting a percentage of the best ranked subjects are selected to discontinue at timepoint 1 and then a percentage of the worst ranked subjects are selected to discontinue at subsiquent timepoints. Again the actual percentage thresholds for selection are based on the discontinuation rates for the particular scenario.

## 2.3 Stage 3 – Selecting patients to withdraw from the trial

Once patients have been selected as discontinuing treatment a simple MCAR selection process is applied for selecting the patients that withdraw from the trial. We simulate $u \sim uniform(0,1)$ and select the patients with values of $u$ below the threshold for the timepoint at which they discontinued. Similar to discontinuation selection, the actual percentage thresholds for withdrawal selection are based on the withdrawal balance for the particular scenario (e.g The withdrawal scenario "More Late" would have thresholds of 20% at timepoint 1 and 80% at timepoints 2 or 3). If patients are selected to withdraw all subsequent values are set to missing, otherwise all their off-treatment values are kept.

# Appendix 3: Bias and Variance Discussion

## 1 Comparison of bias between off-treatment scenarios

To understand the bias for the treatment effects and group means for the "Return to Baseline" and "Same as Active" scenarios it is useful to consider a simplified setup the final timepoint. For each treatment group $Z \in (C, A)$, we define:

- $n_{1Z}$ as the number of patients completing the study on-treatment and $\mu_{1Z}$ as the expected change from baseline $FEV_1$ for these patients.
- $n_{2Z}$ as the number of patients completing the study but discontinued treatment and therefore provide off-treatment data at the final timepoint. Also define $\mu_{1Z}$ as the expected change from baseline from these patients.
- $n_{3Z}$ as the number of patients that discontinued treatment and then withdrew from the study creating missing data at the final timepoint.

In both "Return to Baseline" and "Same as Active" scenarios the missing data for patients that discontinued treatment were generated to be similar to the patients that discontinued treatment but completed the study (i.e. both would have the same expected value $\mu_{2Z}$). Both scenarios also assumed 50% of patients discontinuing treatment would then withdraw from the study to create missing data (i.e. $n_{2Z} = n_{3Z}$). This means the expected change from baseline in $FEV_1$ for the full treatment policy for each group Z would be:

$$E\left[Y_Z^{Full}\right] = \frac{n_{1Z}}{n_{1Z} + 2n_{2Z}} \cdot \mu_{1Z} + \frac{2n_{2Z}}{n_{1Z} + 2n_{2Z}} \cdot \mu_{2Z}$$

We use this expected value as a benchmark for comparing the relative bias in both off-treatment scenarios resulting from making a naïve "common MAR" assumption that the $n_{3Z}$ missing outcomes are expected to be an aggregate of all the observed on- and off-treatment outcomes.

Under the common MAR assumption, the expected value of change from baseline FEV1 would be:

$$E[Y_Z^{MAR}] = \frac{n_{1Z}}{n_{1Z} + n_{2Z}} \cdot \mu_{1Z} + \frac{n_{2Z}}{n_{1Z} + n_{2Z}} \cdot \mu_{2Z}$$

The bias from making this assumption would then be,

$$E[Y_Z^{MAR}] - E\left[Y_Z^{Full}\right] = \frac{n_{1Z}}{n_{1Z} + n_{2Z}} \cdot \mu_{1Z} + \frac{n_{2Z}}{n_{1Z} + n_{2Z}} \cdot \mu_{2Z} - \frac{n_{1Z}}{n_{1Z} + 2n_{2Z}} \cdot \mu_{1Z} - \frac{2n_{2Z}}{n_{1Z} + 2n_{2Z}} \cdot \mu_{2Z}$$

And so,

$$Bias_Z = \left(\frac{n_{1Z}}{n_{1Z} + n_{2Z}} - \frac{n_{1Z}}{n_{1Z} + 2n_{2Z}}\right) \cdot \mu_{1Z} + \left(\frac{n_{2Z}}{n_{1Z} + n_{2Z}} - \frac{2n_{2Z}}{n_{1Z} + 2n_{2Z}}\right) \cdot \mu_{2Z}$$

From this expression we see that the bias for treatment group $Z$ is related to the proportion of patients observed to be on- and off-treatment and how different they are from the proportions of on- and off-treatment with respect to the full data. It also shows the bias is related to the expected values of the on- and off-treatment outcomes. This explains the rise in bias in the group means and treatment effects for both off-treatment scenarios when the discontinuation rate increased and created more missing outcomes (as we fixed withdrawal to 50% of treatment discontinuations). Specifically the increased missingness created a larger disparity between the proportions of observed on- and off-treatment data compared to the full proportions. I

Rearranging the expression further allows us to explain why the bias for the "Return to Baseline" scenario is larger than the "Same as Active" scenario for a given discontinuation rate, mechanism, and study withdrawal type. Simplifying the expression, the bias for treatment group $Z$ becomes:

$$Bias_Z = \frac{n_{1Z} \cdot n_{2Z} \cdot \mu_{1Z} - n_{1Z} \cdot n_{2Z} \cdot \mu_{2Z}}{(n_{1Z} + n_{2Z})(n_{1Z} + 2n_{2Z})}$$

Using the on- and off-treatment expected values for the "Return to Baseline" and "Same as Active" scenario we see the bias as:

$$Bias_C^{RTB} = \frac{400 \cdot n_{1C} \cdot n_{2C} - 0 \cdot n_{1C} \cdot n_{2C}}{(n_{1C} + n_{2C})(n_{1C} + 2n_{2C})} = \frac{400 \cdot n_{1C} \cdot n_{2C}}{(n_{1C} + n_{2C})(n_{1C} + 2n_{2C})}$$

$$Bias_A^{RTB} = \frac{500 \cdot n_{1A} \cdot n_{2A} - 0 \cdot n_{1A} \cdot n_{2A}}{(n_{1A} + n_{2A})(n_{1A} + 2n_{2A})} = \frac{500 \cdot n_{1A} \cdot n_{2A}}{(n_{1A} + n_{2A})(n_{1A} + 2n_{2A})}$$

$$Bias_C^{SAA} = \frac{400 \cdot n_{1C} \cdot n_{2C} - 500 \cdot n_{1C} \cdot n_{2C}}{(n_{1C} + n_{2C})(n_{1C} + 2n_{2C})} = \frac{-100 \cdot n_{1C} \cdot n_{2C}}{(n_{1C} + n_{2C})(n_{1C} + 2n_{2C})}$$

$$Bias_A^{SAA} = \frac{500 \cdot n_{1A} \cdot n_{2A} - 500 \cdot n_{1A} \cdot n_{2A}}{(n_{1A} + n_{2A})(n_{1A} + 2n_{2A})} = 0$$

Looking at the corresponding bias values for each group it can be seen that the "Return to Baseline" scenario will have more bias than the corresponding "Same as Active" bias for a given discontinuation rate. It is also possible to see there is no bias for the "Same as Active" active group because the expected values before and after treatment discontinuation are the same.

Finally, we can also use these bias terms to understand the similar levels of bias seen in the treatment effects for both the "Return to Baseline" and "Same as Active" scenarios when they have equal treatment discontinuation rates and the same corresponding discontinuation mechanism, and study withdrawal type. Because equal patients were allocated to each group any scenarios with the same discontinuation rate, mechanism and withdrawal type have $n_{1C} = n_{1A}$, $n_{2C} = n_{2A}$ and $n_{3C} = n_{3A}$ (which we will denote, $n_1$, $n_2$ and $n_3$). In this setting the bias in the treatment effects for both scenarios have a shift of $\frac{100 \cdot n_1 \cdot n_2}{(n_1 + n_2)(n_1 + 2n_2)}$.

# 2 The impact of missing off-treatment data on variability

To provide some theoretical guidance on the relationship between the amount of missing off-treatment data and the variance inflation, we consider a simplified scenario of a single final timepoint with three potential groups of subjects within a given arm:

- Group 1 are those who complete the study on-treatment. Suppose there are $n_1$ subjects in this group. The observed average change from baseline in FEV1 at timepoint 3 is $\bar{y}_1$ with standard error $\sigma_1/\sqrt{n_1}$.
- Group 2 are those who complete the study, but experienced treatment discontinuation, so that they provide some data on-treatment and some data off-treatment. We assume there $n_2$ subjects in the group. The observed average change from baseline in FEV1 at timepoint 3 is $\bar{y}_2$ with S.E. $\sigma_2/\sqrt{n_2}$.
- Group 3 are those who also experienced treatment discontinuation but did not complete the study. This group provided some data on-treatment, discontinued from treatment, then (potentially) provided some data off-treatment and eventually withdrew from the study (perhaps at the same time they stopped treatment), and consequently have missing data. Assume there are $n_3$ subjects in this group. Their average change from baseline in FEV1 at timepoint 3 is not observed.

The total number of subjects is $n = n_1 + n_2 + n_3$.

Suppose for simplicity that $\sigma_1 = \sigma_2 = \sigma_3 = \sigma$, assumed known.

Under the assumption that those subjects in Group3 are like those in Group 2, the estimated value of the change from baseline in FEV1 at timepoint 3 would be

$$\frac{n_1\bar{y}_1 + (n_2+n_3)\bar{y}_2}{n_1+n_2+n_3}$$

Now, assuming independence between the means in groups 1 and 2, the variance of that estimator is

$$\frac{n_1^2}{(n_1+n_2+n_3)^2}Var(\bar{y}_1) + \frac{(n_2+n_3)^2}{(n_1+n_2+n_3)^2}Var(\bar{y}_2)$$

When we observe all the data there are $n_2 + n_3$ subjects with on- and off-treatment data, and therefore $Var(\bar{y}_2) = \frac{\sigma^2}{n_2+n_3}$. But in the presence of missing data, $Var(\bar{y}_2) = \frac{\sigma^2}{n_2}$ because there are only $n_2$ subjects with on- and off-treatment data.

When we observe all the data for all subjects, the above variance expression becomes

$$\frac{n_1^2}{(n_1+n_2+n_3)^2}\cdot\frac{\sigma^2}{n_1} + \frac{(n_2+n_3)^2}{(n_1+n_2+n_3)^2}\cdot\frac{\sigma^2}{(n_2+n_3)} = \frac{\sigma^2}{(n_1+n_2+n_3)^2}\cdot(n_1+n_2+n_3) = \frac{\sigma^2}{n}$$

When there are missing data, then the variance is

$$\frac{n_1^2}{(n_1+n_2+n_3)^2}\cdot\frac{\sigma^2}{n_1} + \frac{(n_2+n_3)^2}{(n_1+n_2+n_3)^2}\cdot\frac{\sigma^2}{n_2} = \frac{\sigma^2}{n^2}\cdot\left(n_1 + n_2\left(1+\frac{n_3}{n_2}\right)^2\right)$$

The relative increase in the variance of the estimator in the presence of missing data would therefore be

$$\frac{\frac{\sigma^2}{n^2}\cdot\left(n_1+n_2\left(1+\frac{n_3}{n_2}\right)^2\right)-\frac{\sigma^2}{n}}{\frac{\sigma^2}{n}}=\frac{1}{n}\left(n_1+n_2+2n_3+\frac{n_3^2}{n_2}\right)-1$$

$$=\frac{1}{n}\left(n_1+n_2+n_3+n_3\left(1+\frac{n_3}{n_2}\right)\right)-1=1+\frac{n_3}{n}\left(1+\frac{n_3}{n_2}\right)-1=\frac{n_3}{n}\left(1+\frac{n_3}{n_2}\right)$$

The above expression says that the relative increase in the variance of the estimator due to missing data becomes larger as the proportion of subjects with missing data $\frac{n_3}{n}$ increases, but also that the impact of a certain proportion of subjects with missing data $\frac{n_3}{n}$ will be amplified if the number (proportion) of subjects with missing data $n_3$ ($n_3/n$) is larger than the number (proportion) of subjects with off-treatment data $n_2$ ($n_2/n$).

Using the discontinuation and withdrawal scenarios from our simulation study, the increase in variance implied by this simplified set up would be:

- 10% Discontinuation and 50% Withdrawal → $0.05\cdot\left(1+\frac{0.05}{0.05}\right)=10\%$ increase in variance for group means.
- 20% Discontinuation and 50% Withdrawal → $0.10\cdot\left(1+\frac{0.10}{0.10}\right)=20\%$ increase in variance for the group means.
- 50% Discontinuation and 50% Withdrawal → $0.25\cdot\left(1+\frac{0.25}{0.25}\right)=50\%$ increase in variance for the group means.

Using these increases for the group means and making an independence assumption for comparing the groups, the variance inflation for the treatment effects in our discontinuation scenarios would be:

- 10% Control: 10% Active → 10% Increase.
- 10% Control: 20% Active → 15% Increase.
- 20% Control: 20% Active → 20% Increase.
- 50% Control: 50% Active → 50% Increase.

The increases in variance seen in our simulation study was less than the values in this simplified set up, with the maximum seen in the PIOS model with a 41% increase compared to the full ANCOVA in the "Same as Active" scenario with 50% Discontinuation, DNAR1 mechanism and "More Early" withdrawal. This suggests that using MI that conditions on previous values can be used to successfully reduce impact of missing off-treatment data on variance inflation.

**SI Figure 1.1**
**Bias for Treatment Effects - (Active vs. Control)**

**Disc. Scenario: Return To Baseline**
**Disc. Mechanism: DAR**

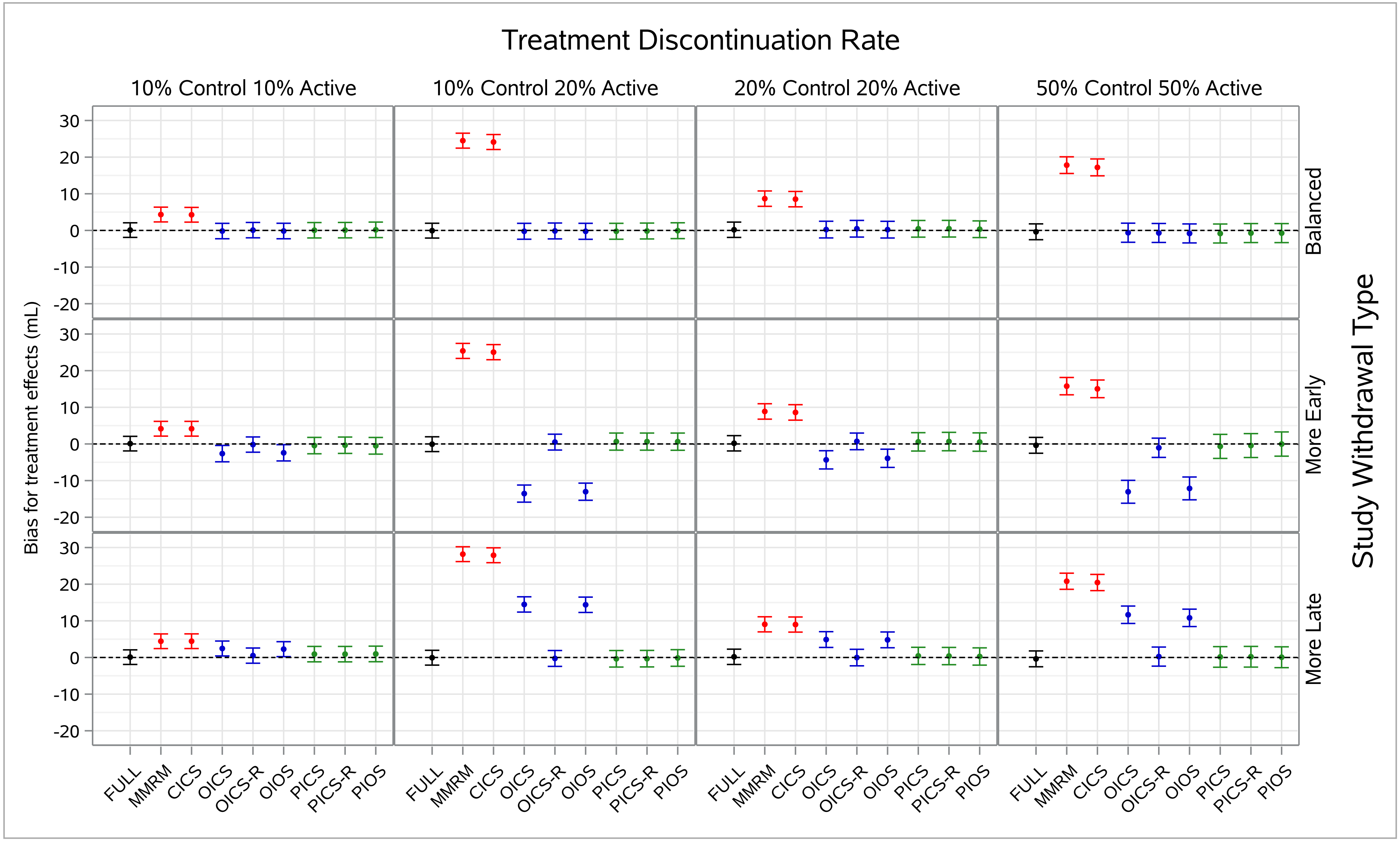

**SI Figure 1.2**
**Bias for Treatment Effects - (Active vs. Control)**

**Disc. Scenario: Return To Baseline**
**Disc. Mechanism: DNAR1**

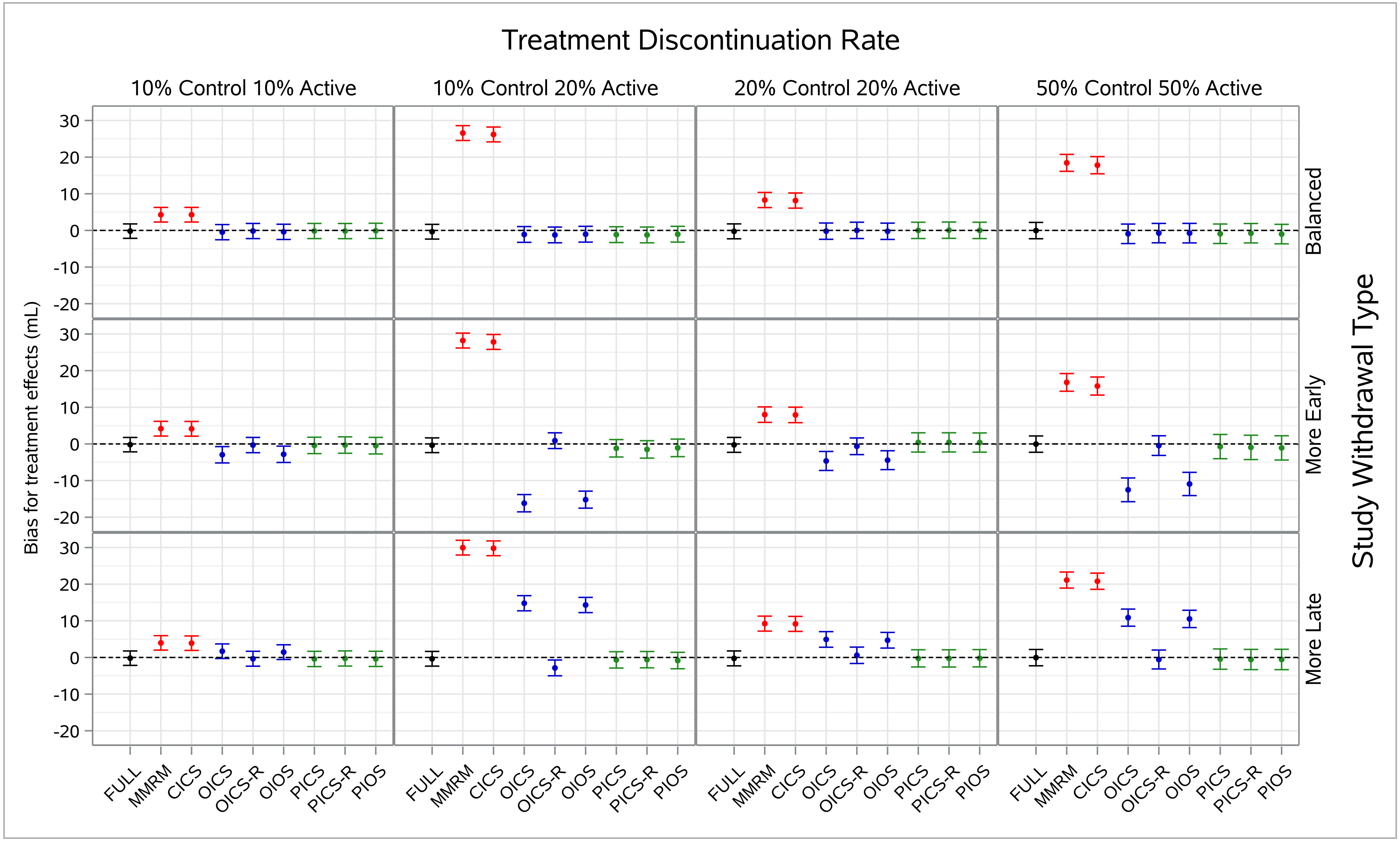

**SI Figure 1.3**
**Bias for Treatment Effects - (Active vs. Control)**

**Disc. Scenario: Return To Baseline**
**Disc. Mechanism: DNAR2**

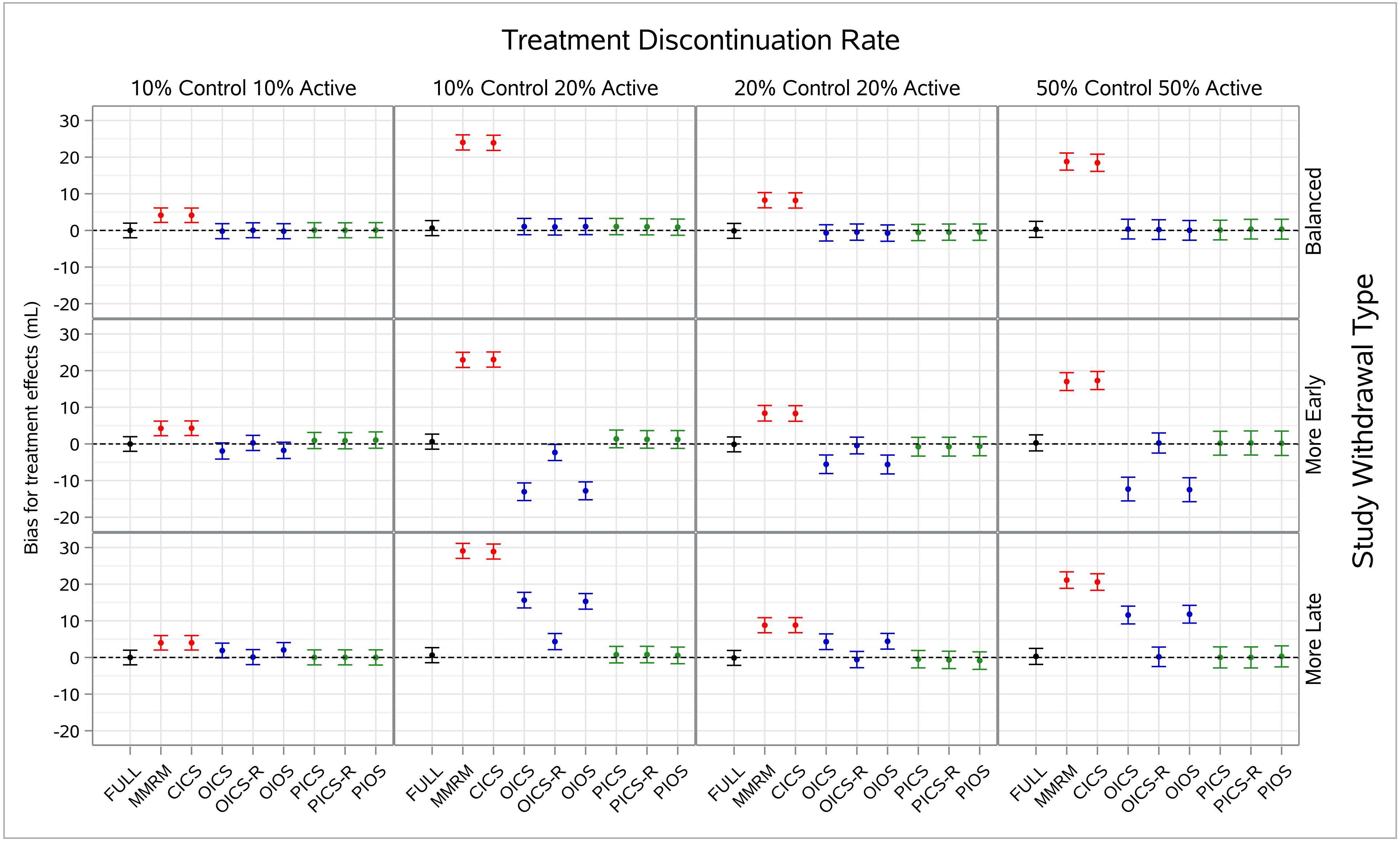

**SI Figure 1.4**
**Bias for Treatment Effects - (Active vs. Control)**

**Disc. Scenario: Same As Active**
**Disc. Mechanism: DAR**

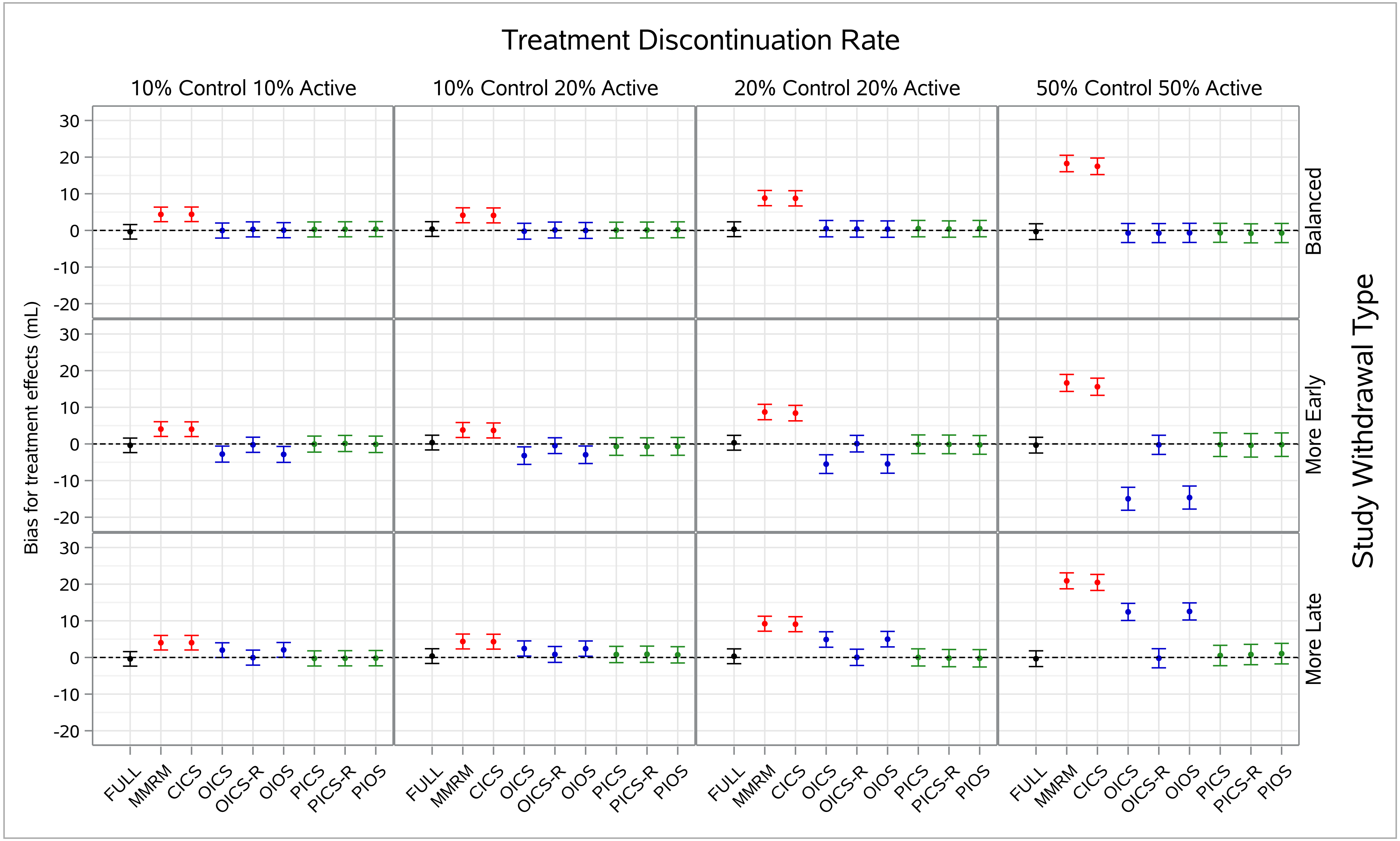

**SI Figure 1.5**
**Bias for Treatment Effects - (Active vs. Control)**

**Disc. Scenario: Same As Active**
**Disc. Mechanism: DNAR1**

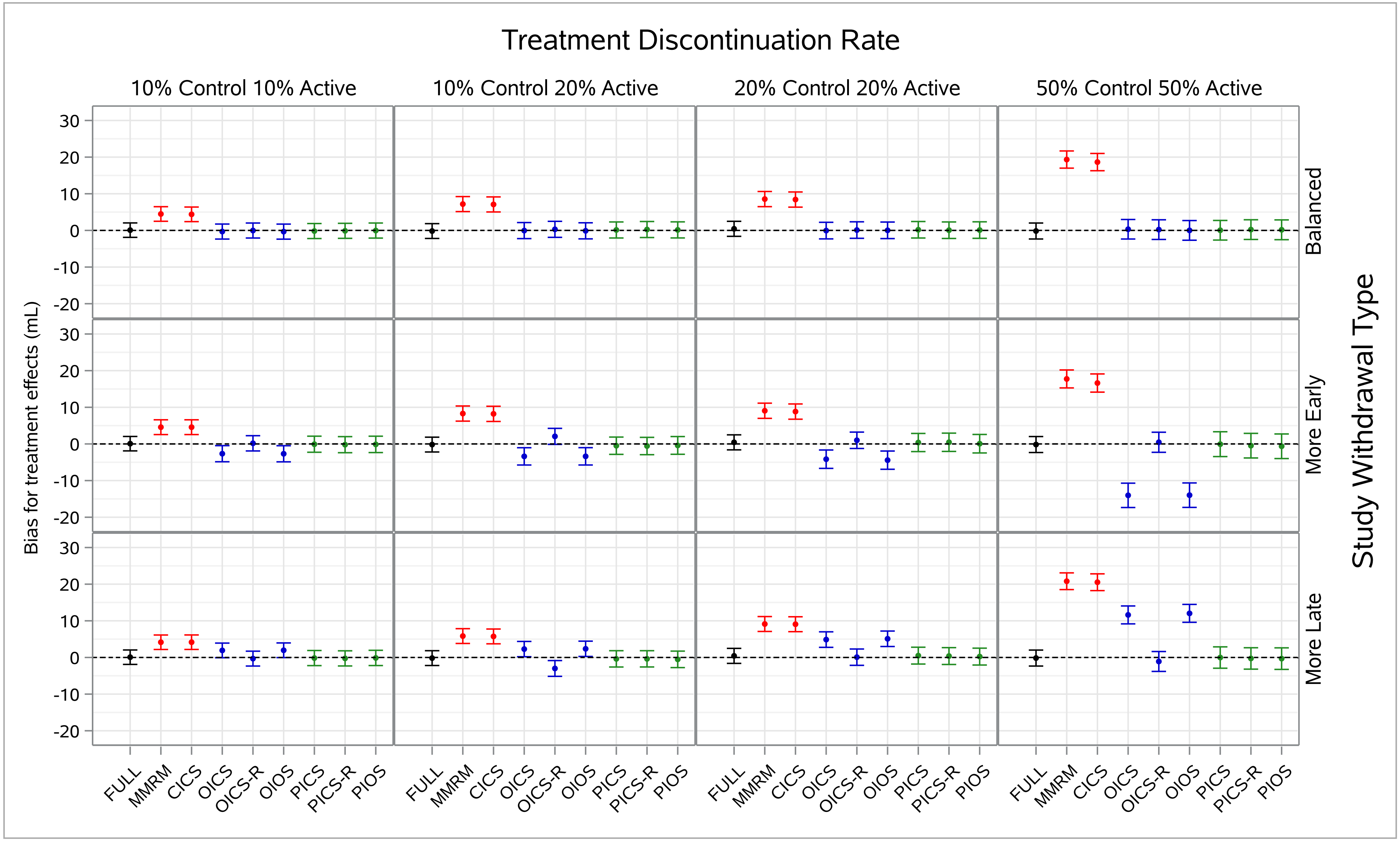

**SI Figure 1.6**
**Bias for Treatment Effects - (Active vs. Control)**

**Disc. Scenario: Same As Active**
**Disc. Mechanism: DNAR2**

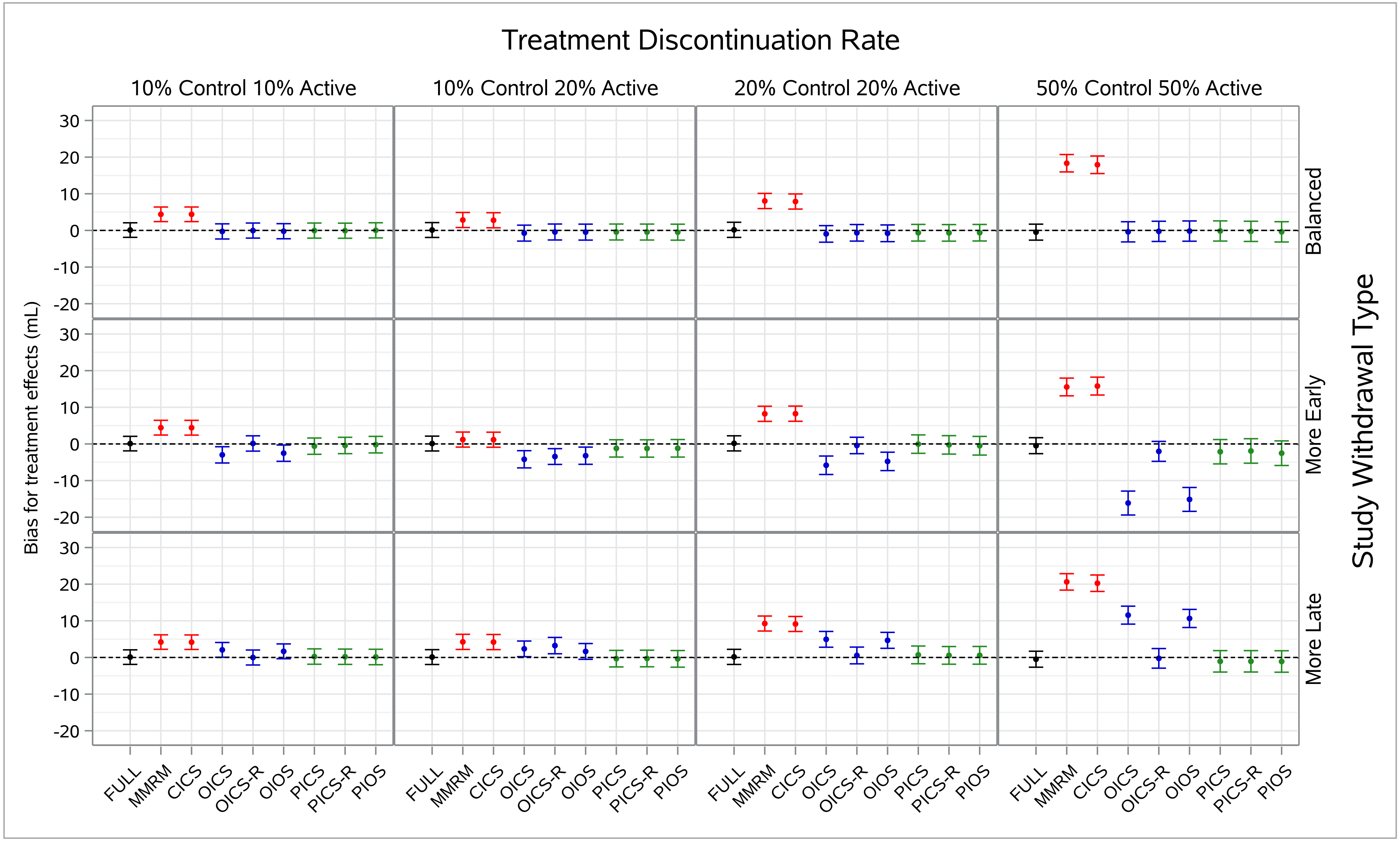

**SI Figure 2.1**
**Average 95% CI Halfwidth for Treatment Effects (Active vs. Control)**

**Disc. Scenario: Return To Baseline**
**Disc. Mechanism: DAR**

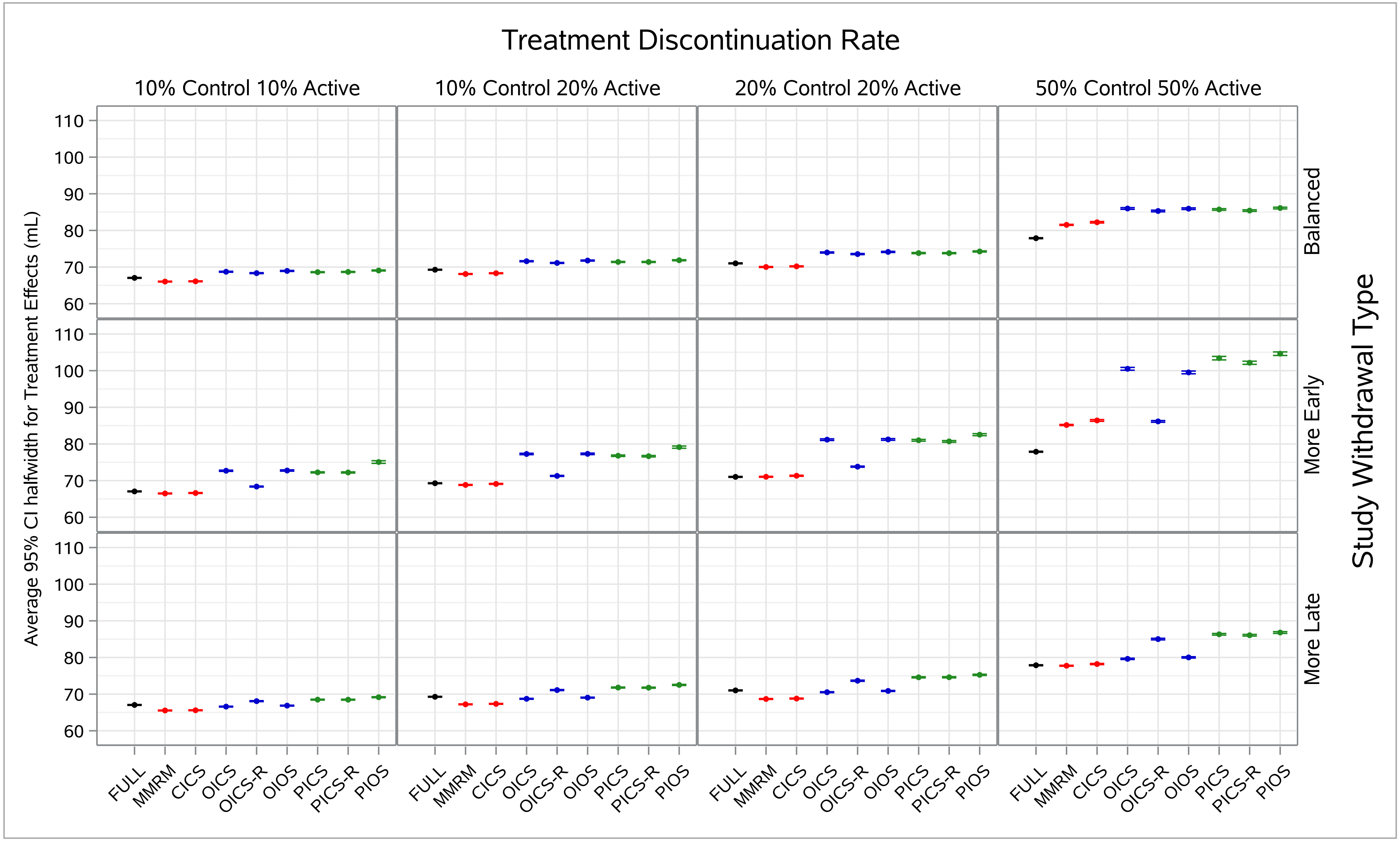

**SI Figure 2.2**
**Average 95% CI Halfwidth for Treatment Effects (Active vs. Control)**

**Disc. Scenario: Return To Baseline**
**Disc. Mechanism: DNAR1**

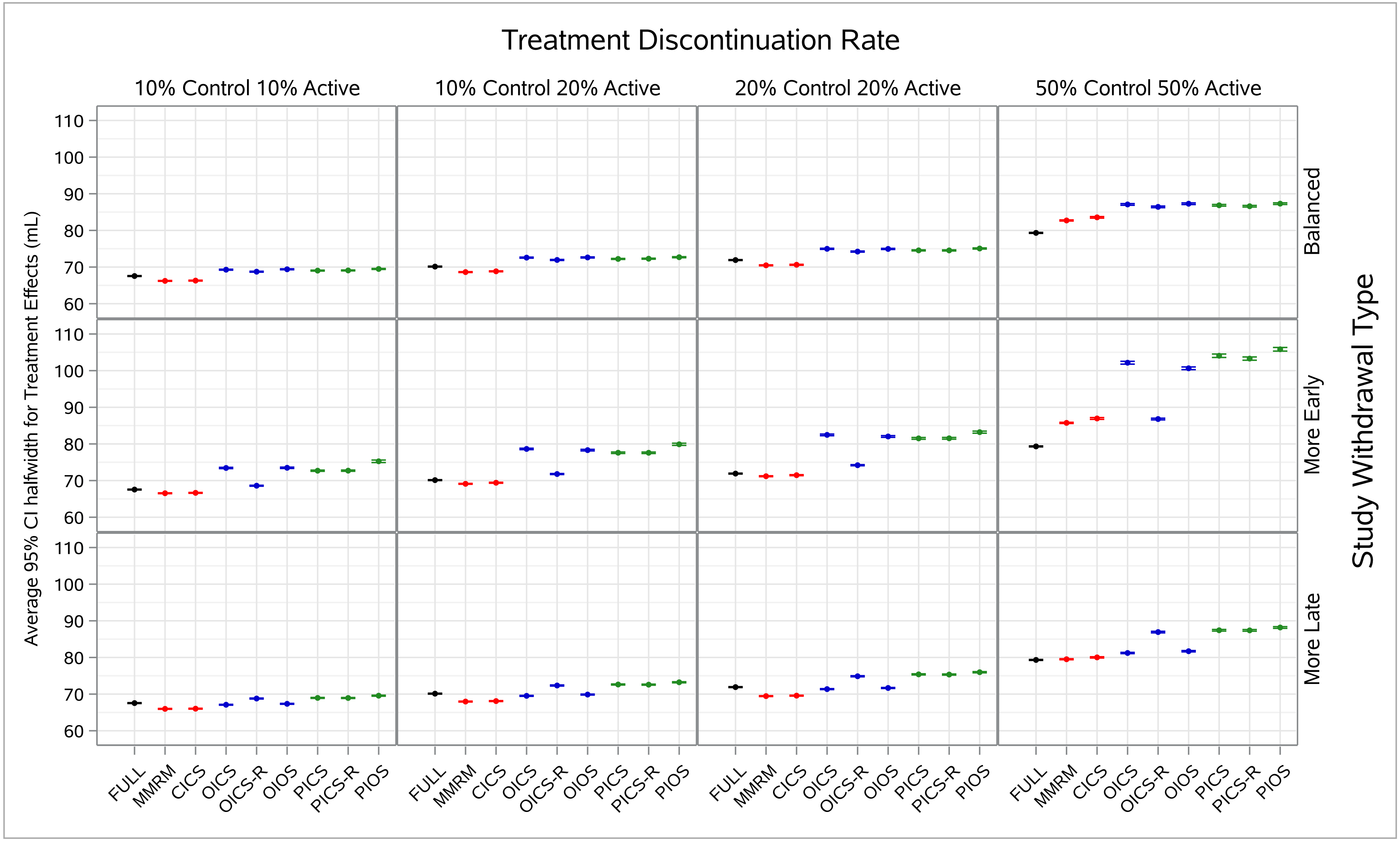

**SI Figure 2.3**
**Average 95% CI Halfwidth for Treatment Effects (Active vs. Control)**

**Disc. Scenario: Return To Baseline**
**Disc. Mechanism: DNAR2**

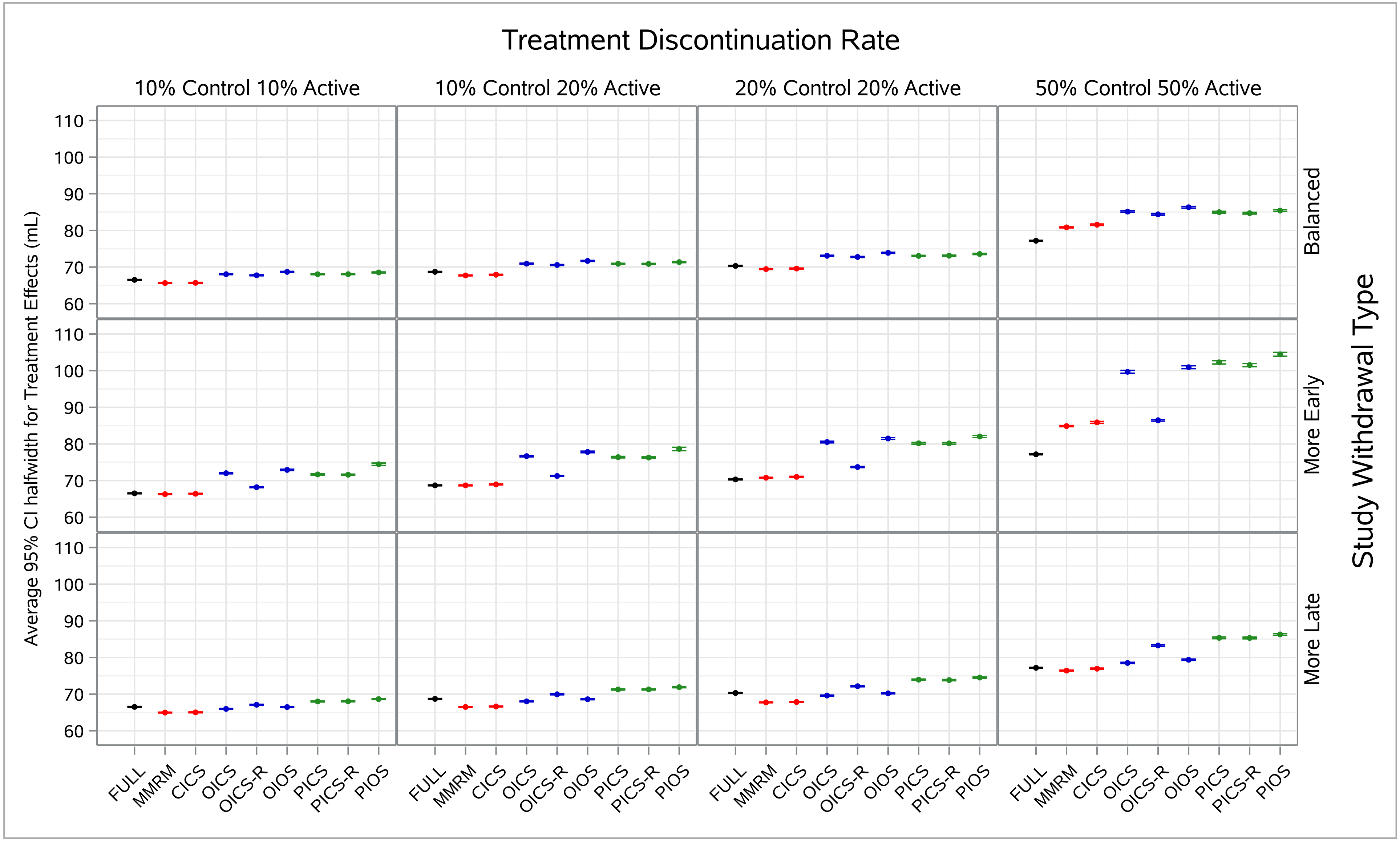

**SI Figure 2.4**
**Average 95% CI Halfwidth for Treatment Effects (Active vs. Control)**

**Disc. Scenario: Same As Active**
**Disc. Mechanism: DAR**

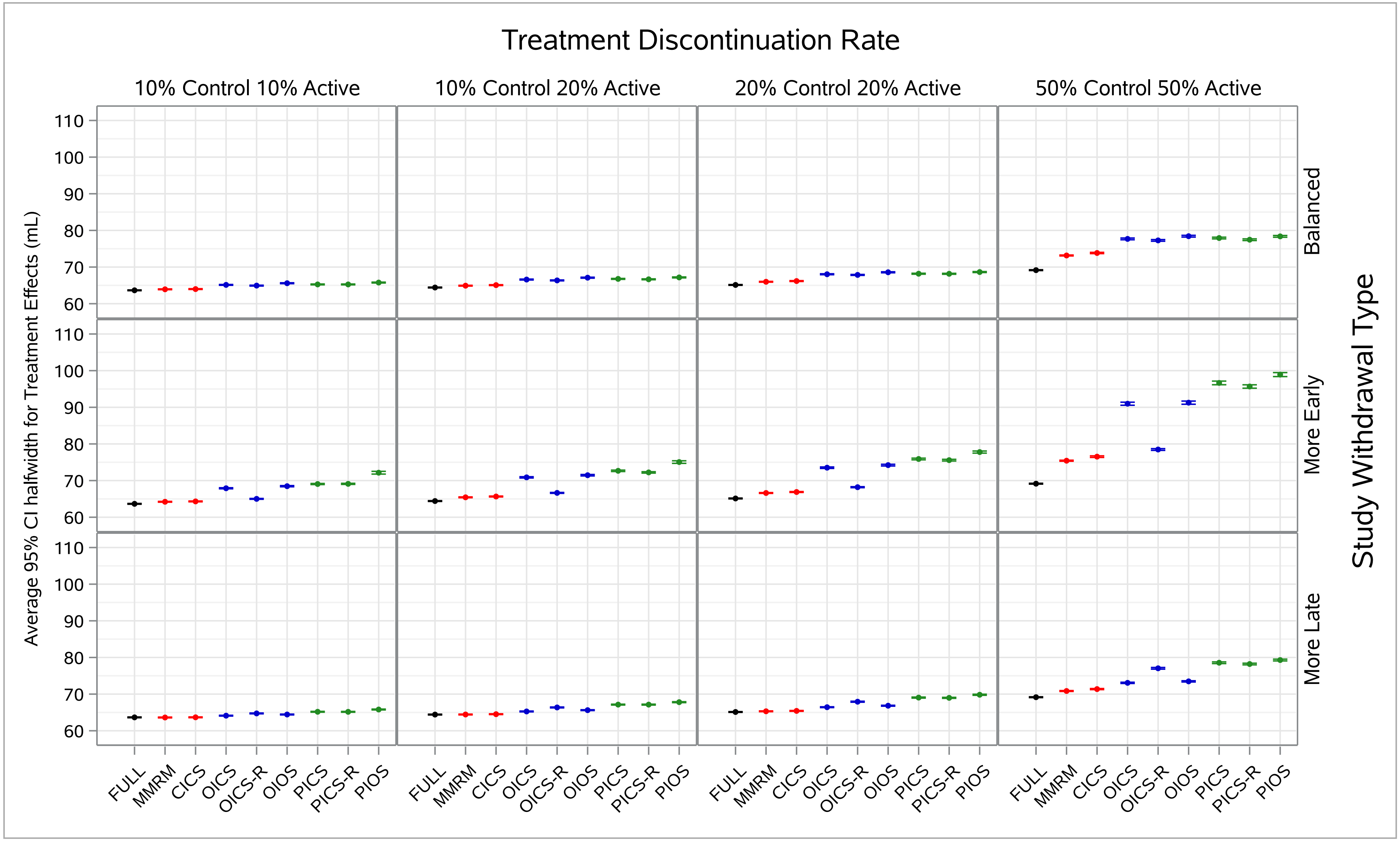

**SI Figure 2.5**
**Average 95% CI Halfwidth for Treatment Effects (Active vs. Control)**

**Disc. Scenario: Same As Active**
**Disc. Mechanism: DNAR1**

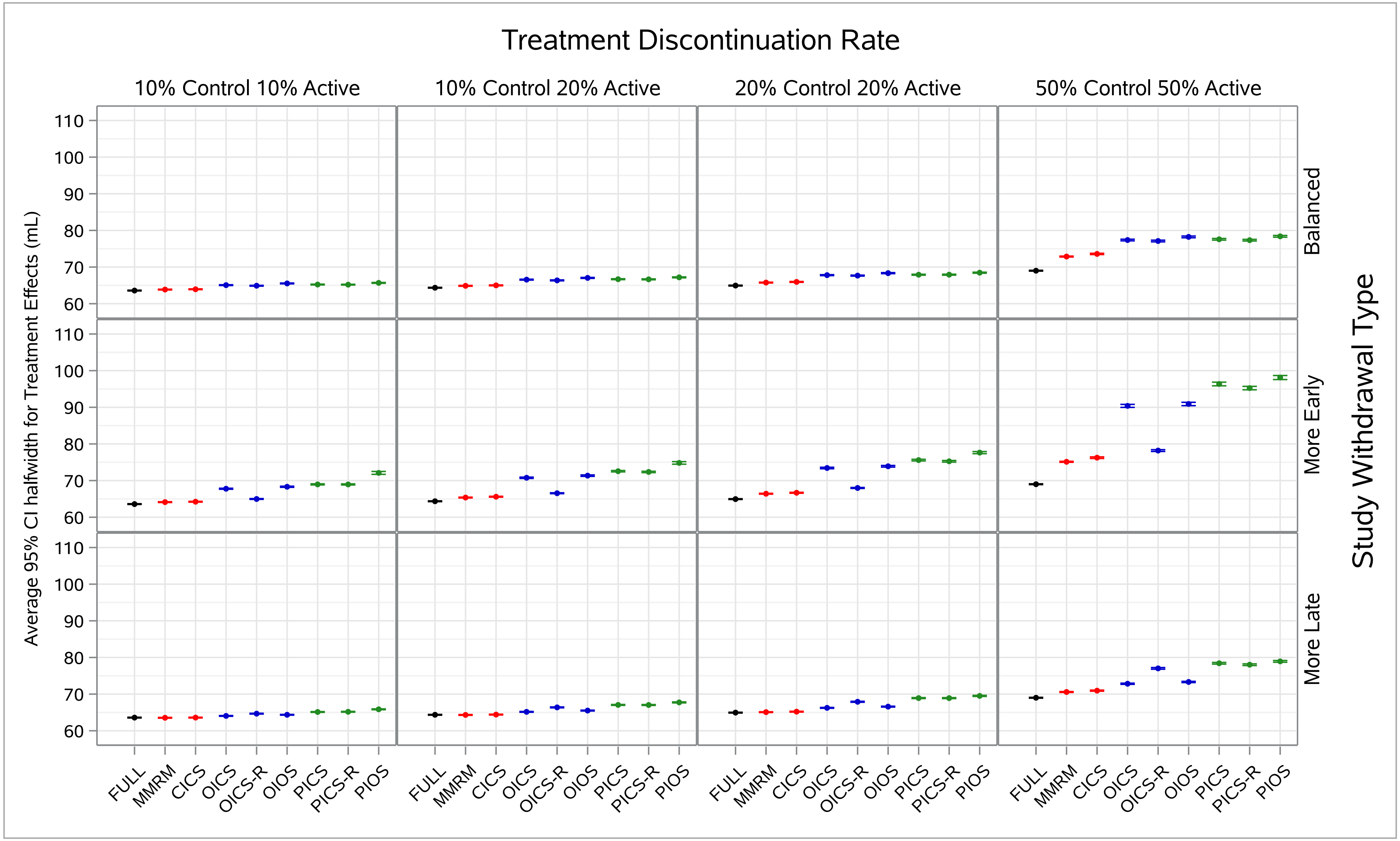

**SI Figure 2.6**
**Average 95% CI Halfwidth for Treatment Effects (Active vs. Control)**

**Disc. Scenario: Same As Active**
**Disc. Mechanism: DNAR2**

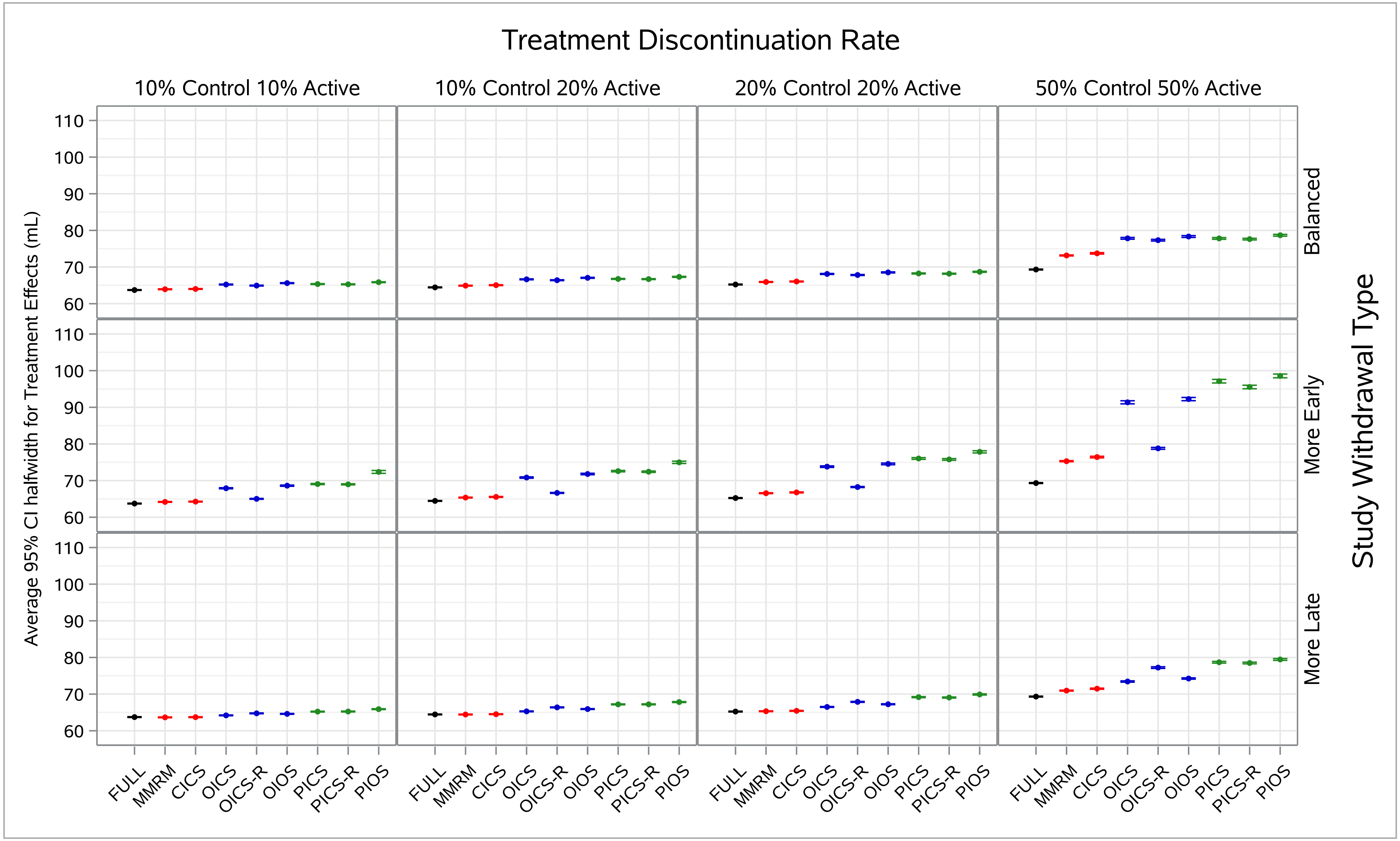

**SI Figure 3.1**

**95% CI Coverage for Treatment Estimates (Active vs. Control)**

**Disc. Scenario: Return To Baseline**

**Disc. Mechanism: DAR**

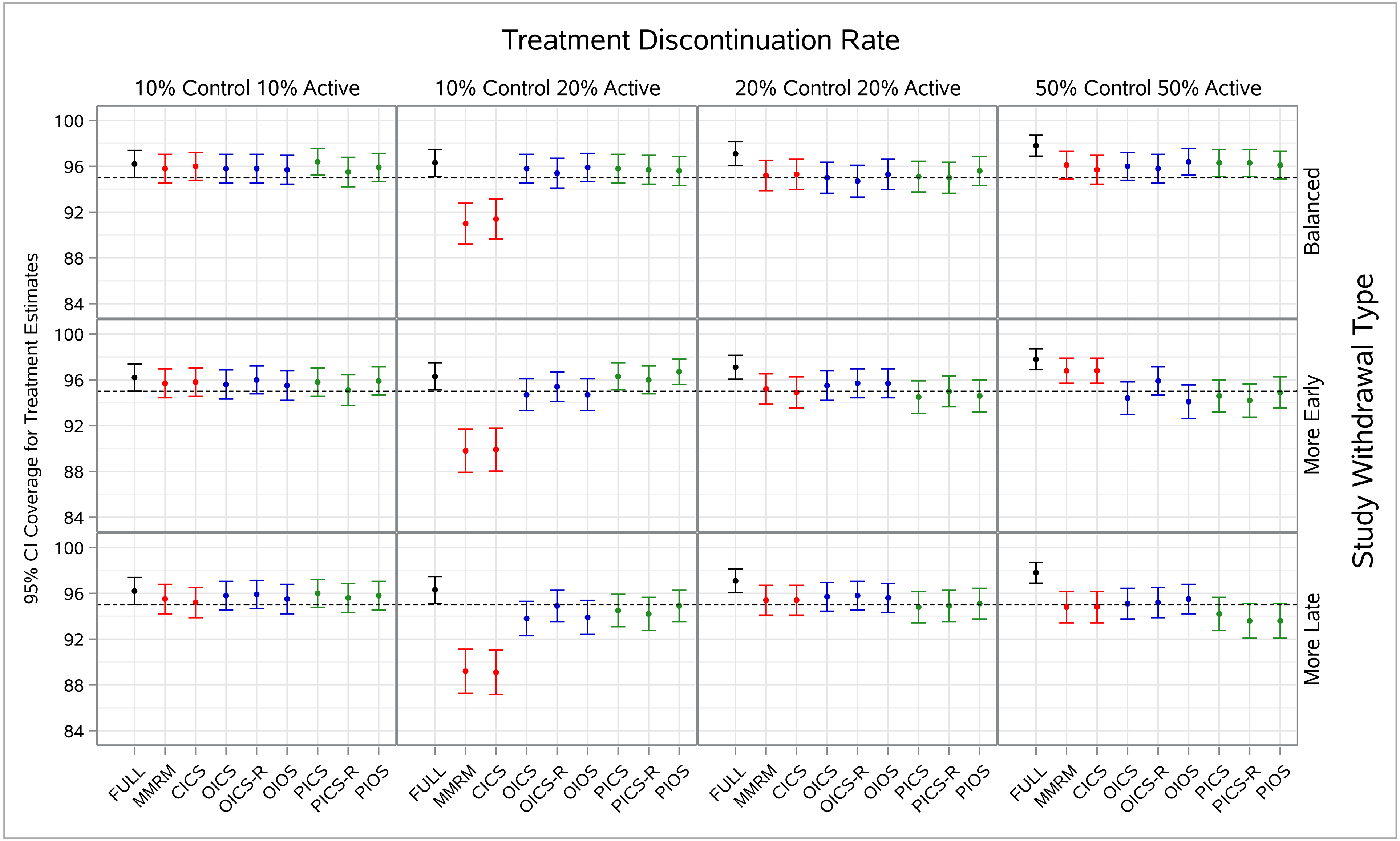

**SI Figure 3.2**
**95% CI Coverage for Treatment Estimates (Active vs. Control)**

**Disc. Scenario: Return To Baseline**
**Disc. Mechanism: DNAR1**

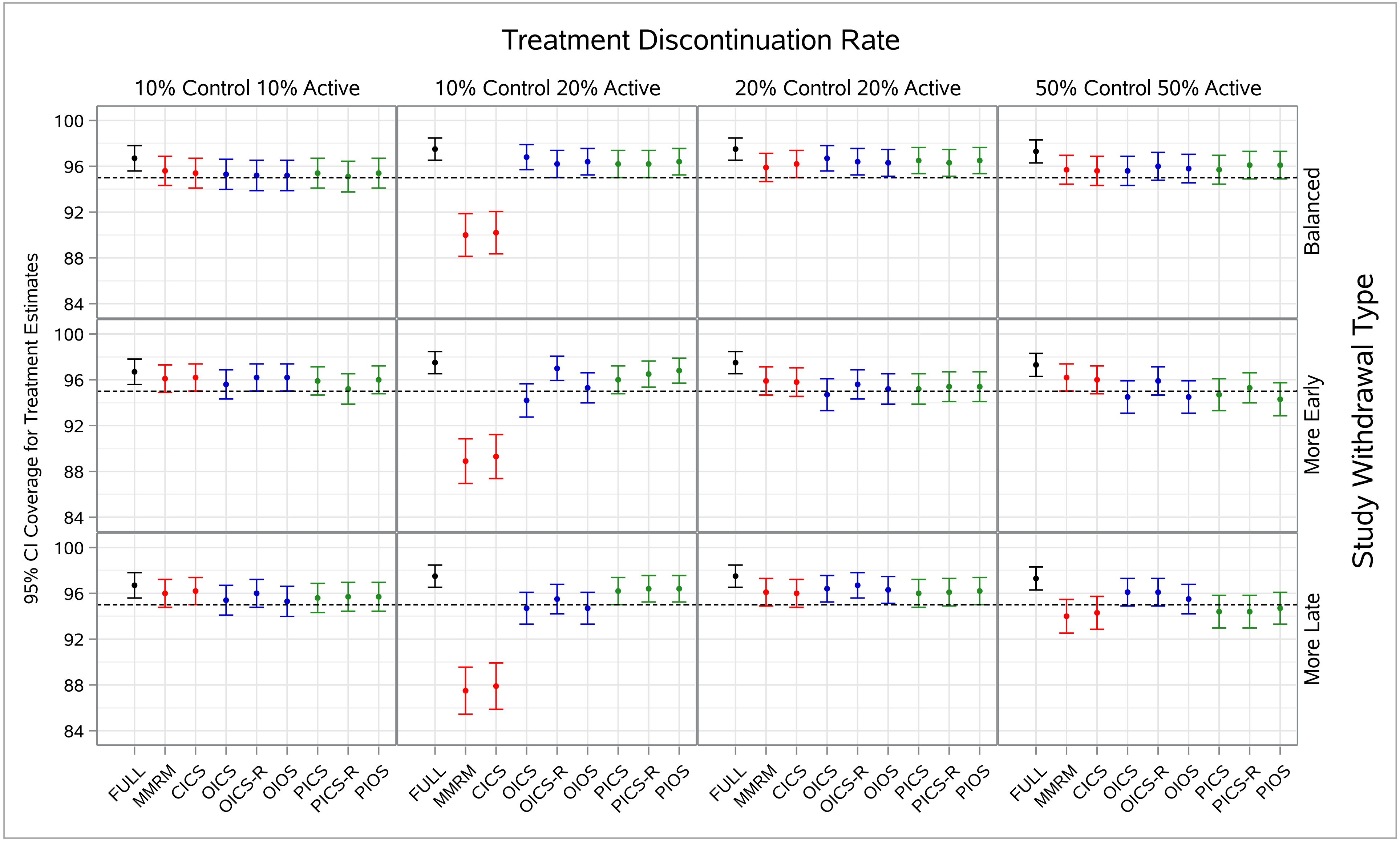

**SI Figure 3.3**
**95% CI Coverage for Treatment Estimates (Active vs. Control)**

**Disc. Scenario: Return To Baseline**
**Disc. Mechanism: DNAR2**

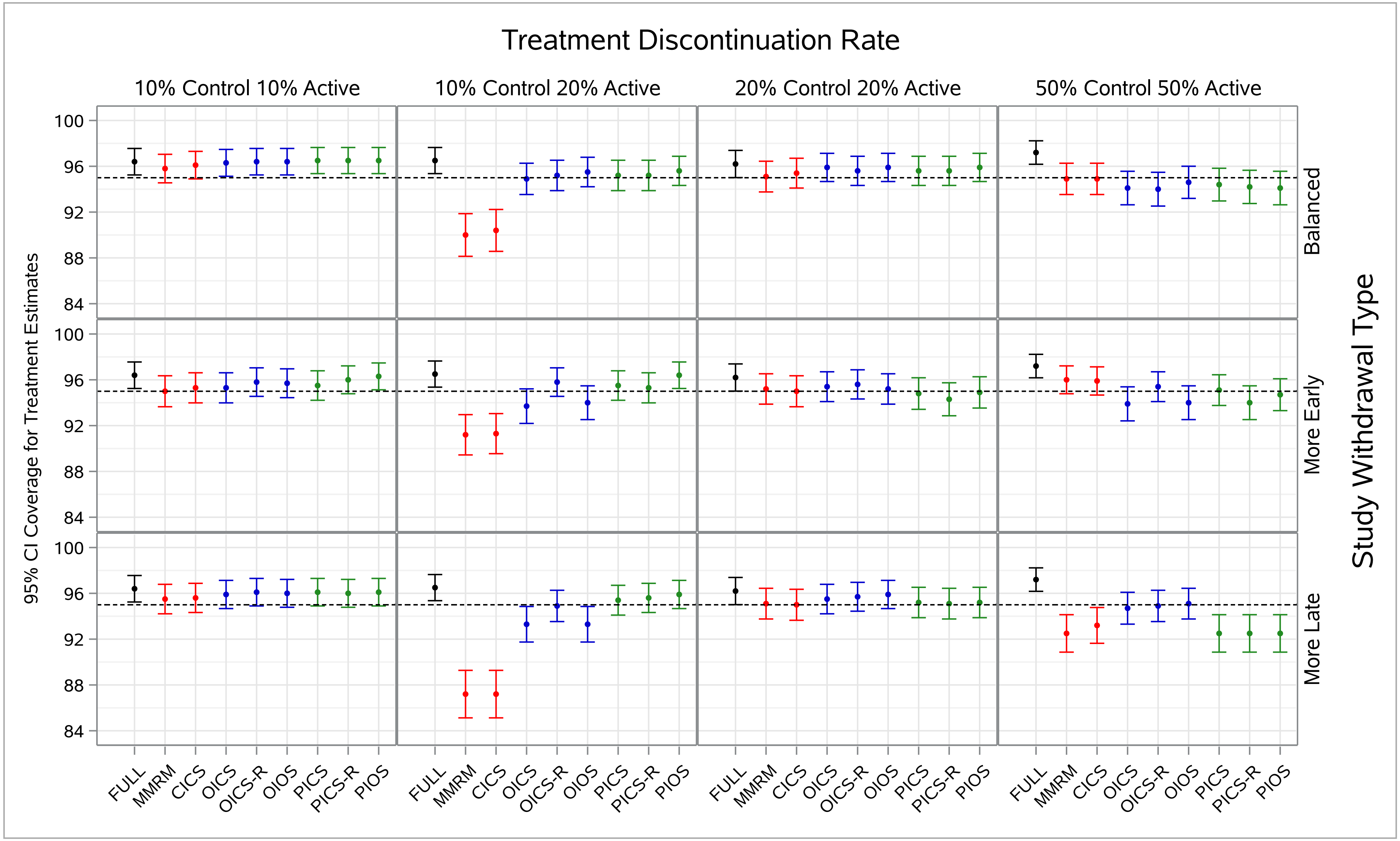

**SI Figure 3.4**
**95% CI Coverage for Treatment Estimates (Active vs. Control)**

**Disc. Scenario: Same As Active**
**Disc. Mechanism: DAR**

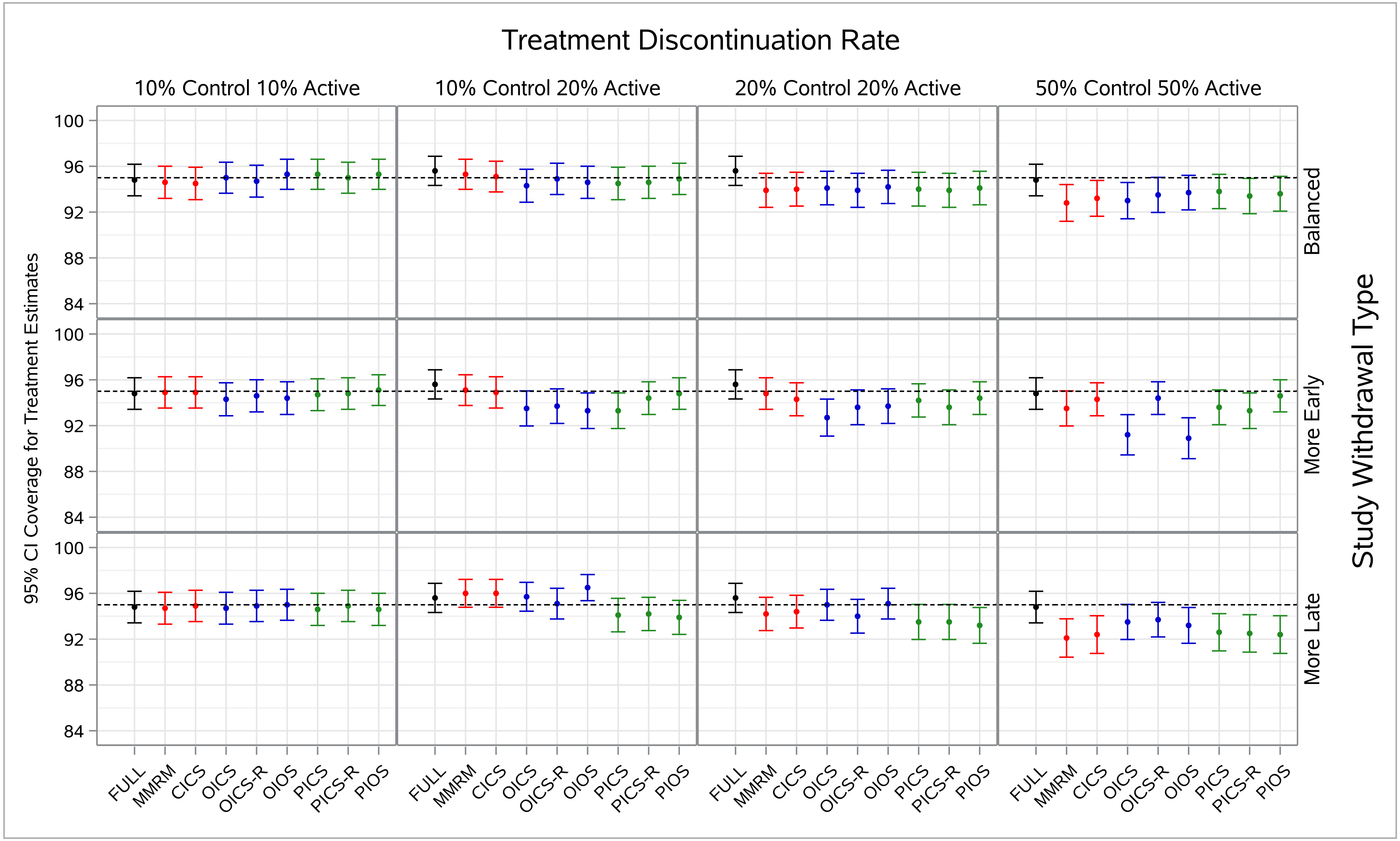

**SI Figure 3.5**
**95% CI Coverage for Treatment Estimates (Active vs. Control)**

**Disc. Scenario: Same As Active**
**Disc. Mechanism: DNAR1**

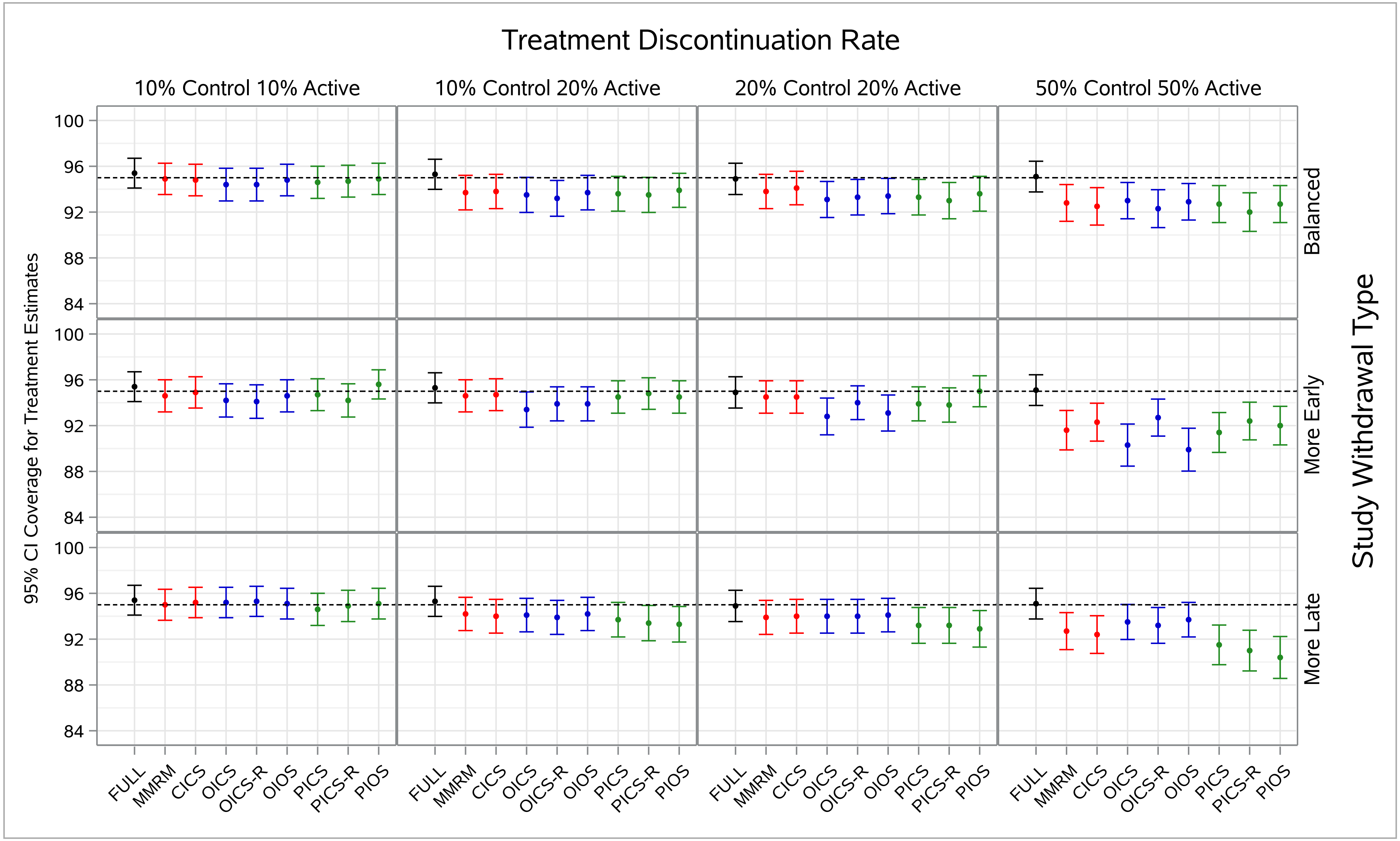

**SI Figure 3.6**
**95% CI Coverage for Treatment Estimates (Active vs. Control)**

**Disc. Scenario: Same As Active**
**Disc. Mechanism: DNAR2**

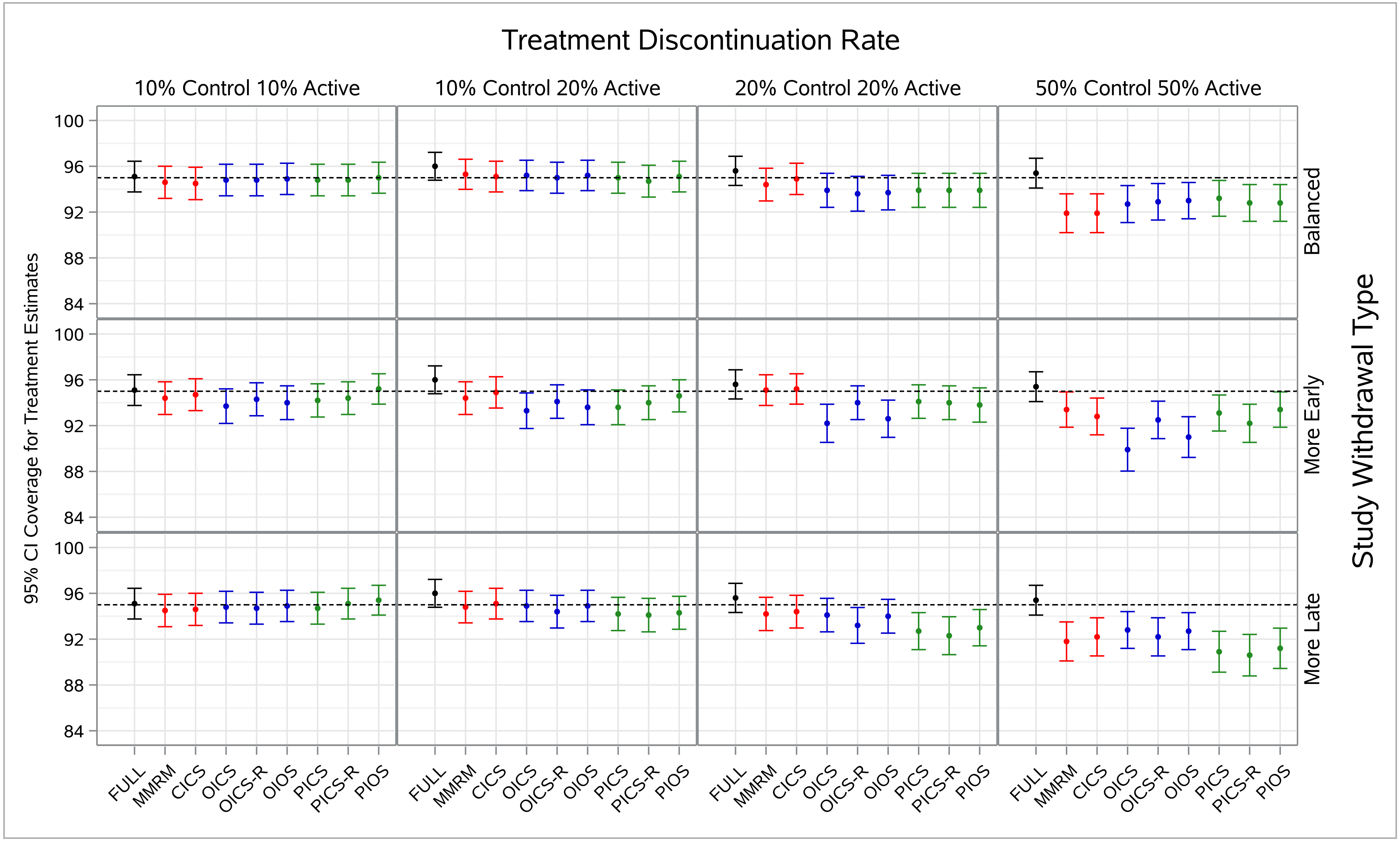

**SI Figure 4.1**
**Bias for Group Means - Control Group**

**Disc. Scenario: Return To Baseline**
**Disc. Mechanism: DAR**

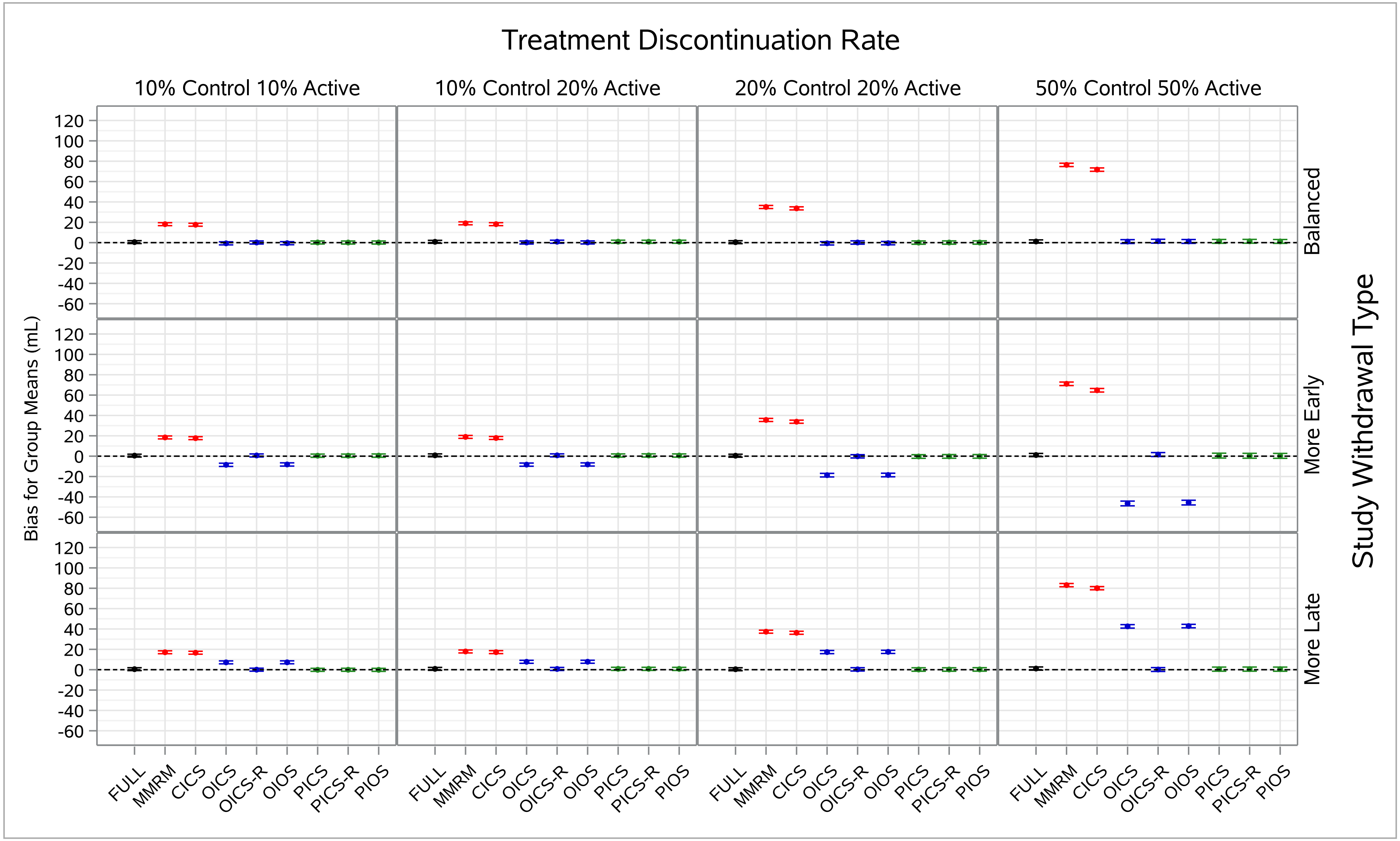

**SI Figure 4.2**
**Bias for Group Means - Control Group**

**Disc. Scenario: Return To Baseline**
**Disc. Mechanism: DNAR1**

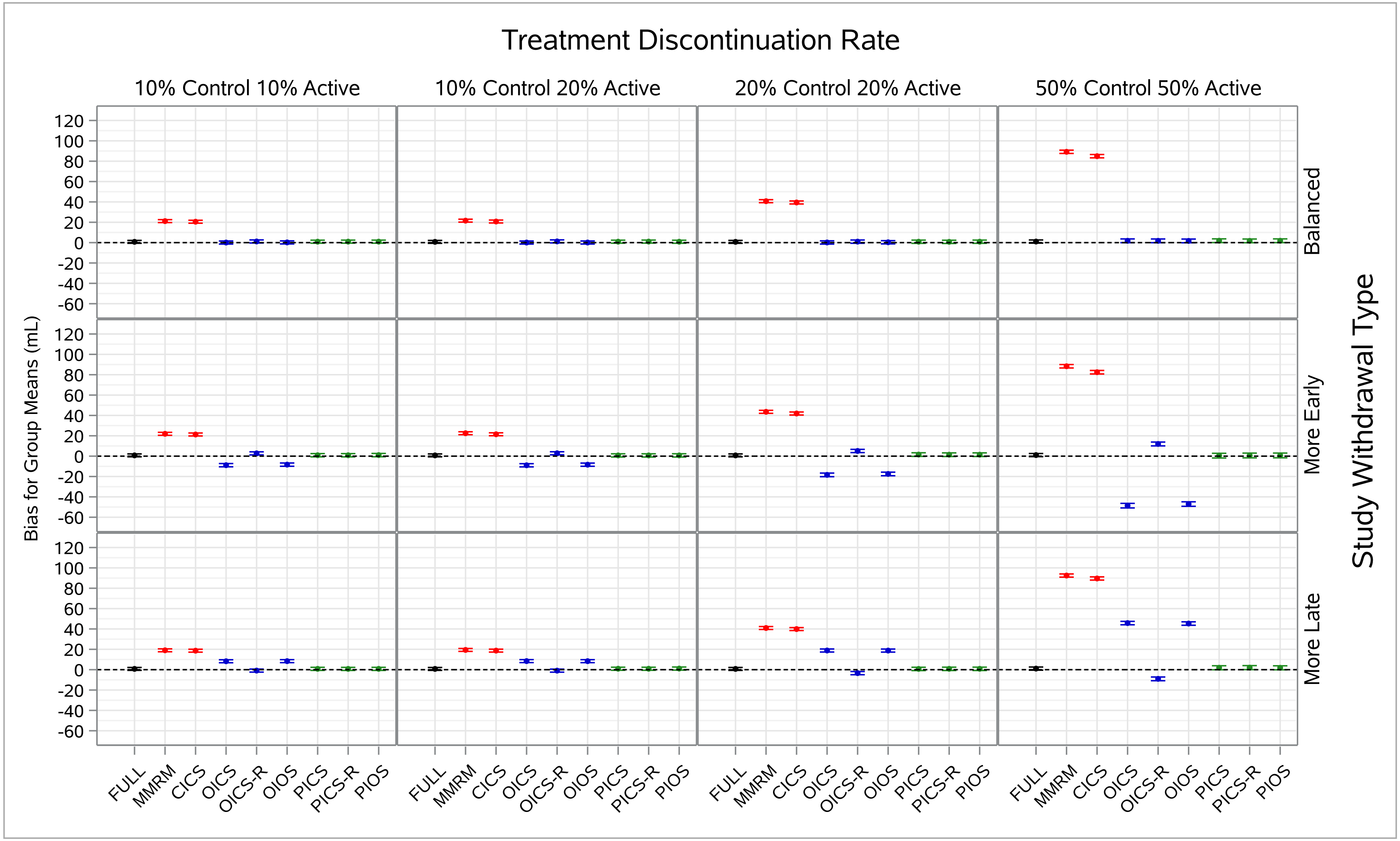

**SI Figure 4.3**
**Bias for Group Means - Control Group**

**Disc. Scenario: Return To Baseline**
**Disc. Mechanism: DNAR2**

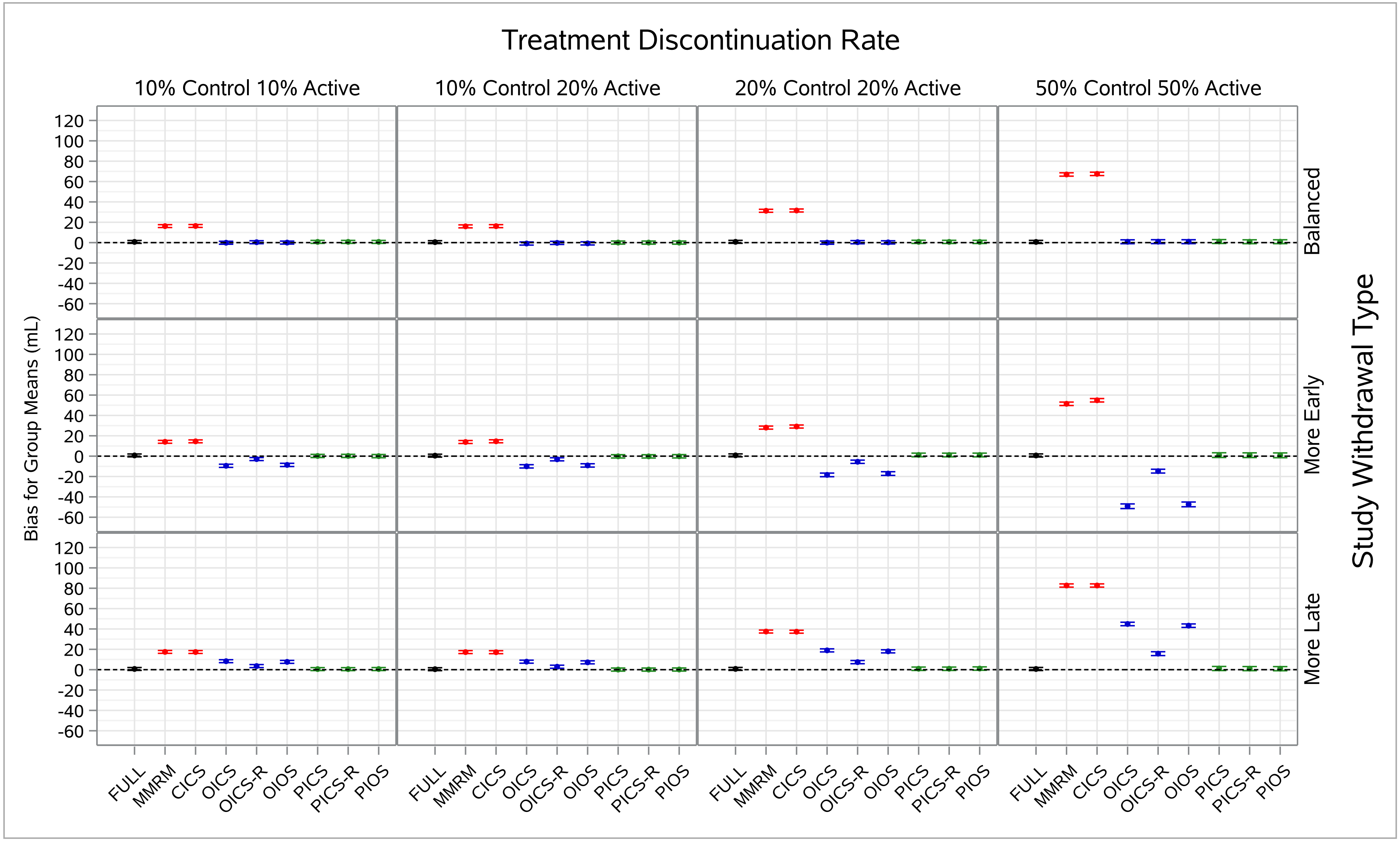

**SI Figure 4.4**
**Bias for Group Means - Active Group**

**Disc. Scenario: Return To Baseline**
**Disc. Mechanism: DAR**

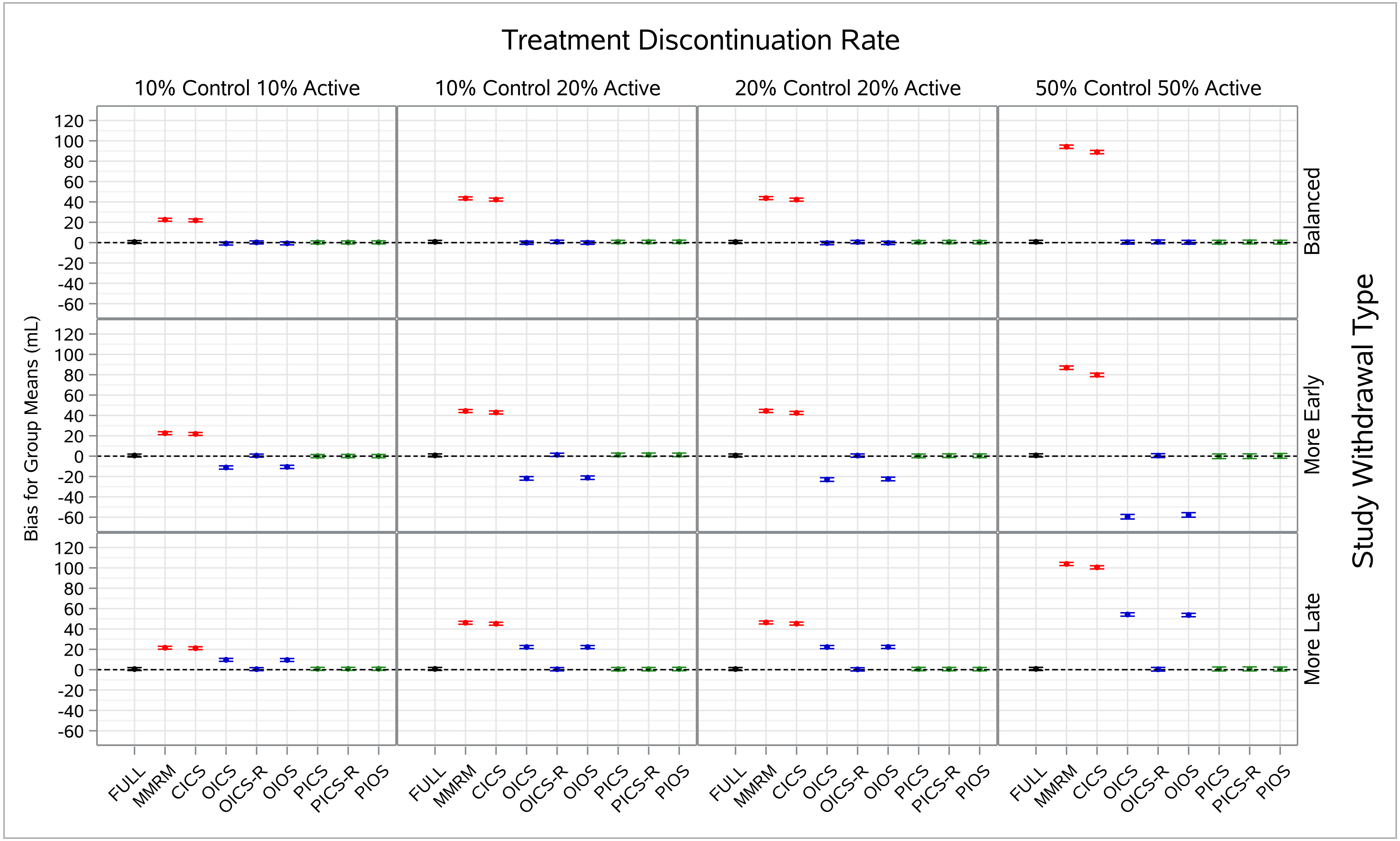

**SI Figure 4.5**
**Bias for Group Means - Active Group**

**Disc. Scenario: Return To Baseline**
**Disc. Mechanism: DNAR1**

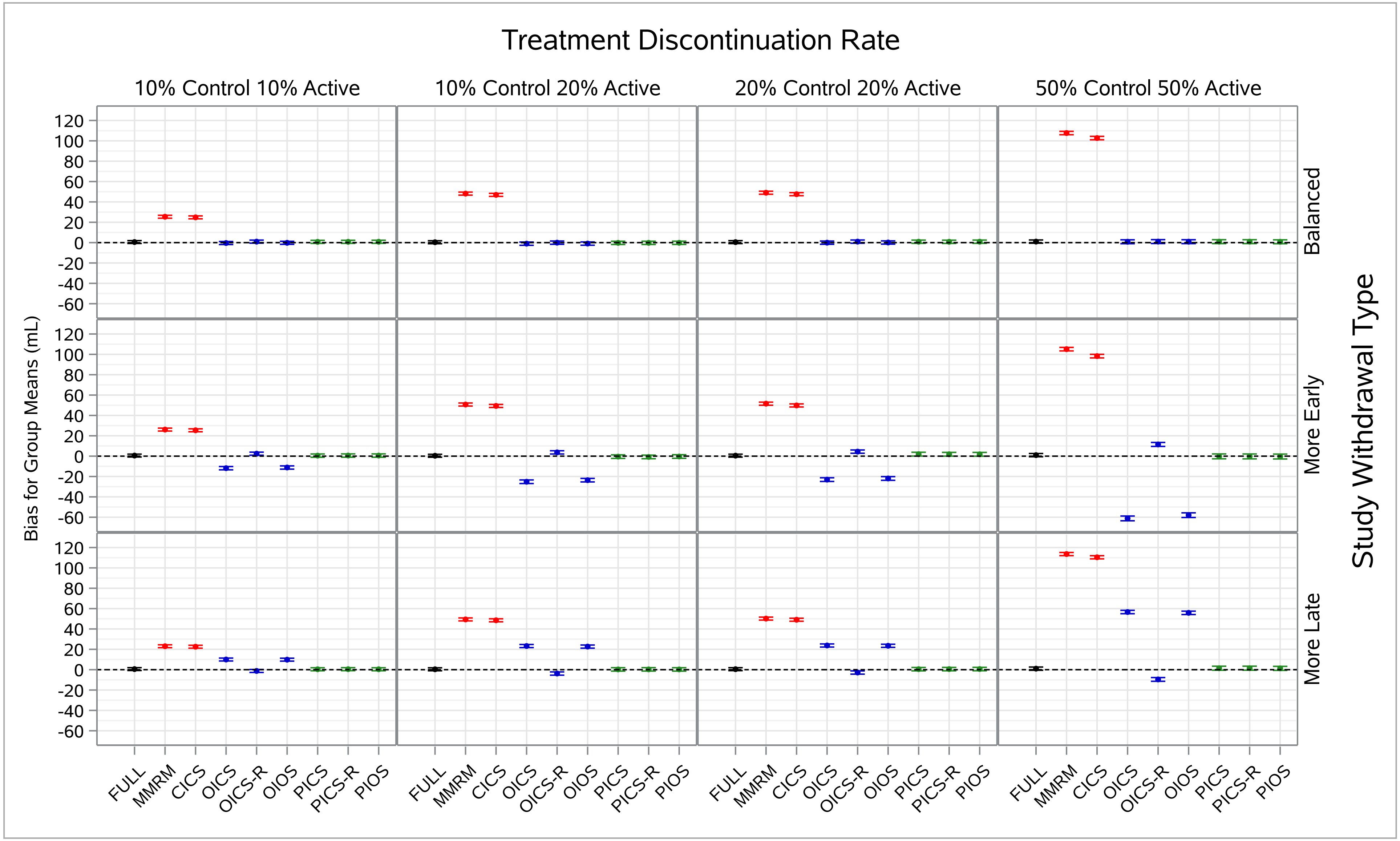

**SI Figure 4.6**
**Bias for Group Means - Active Group**

**Disc. Scenario: Return To Baseline**
**Disc. Mechanism: DNAR2**

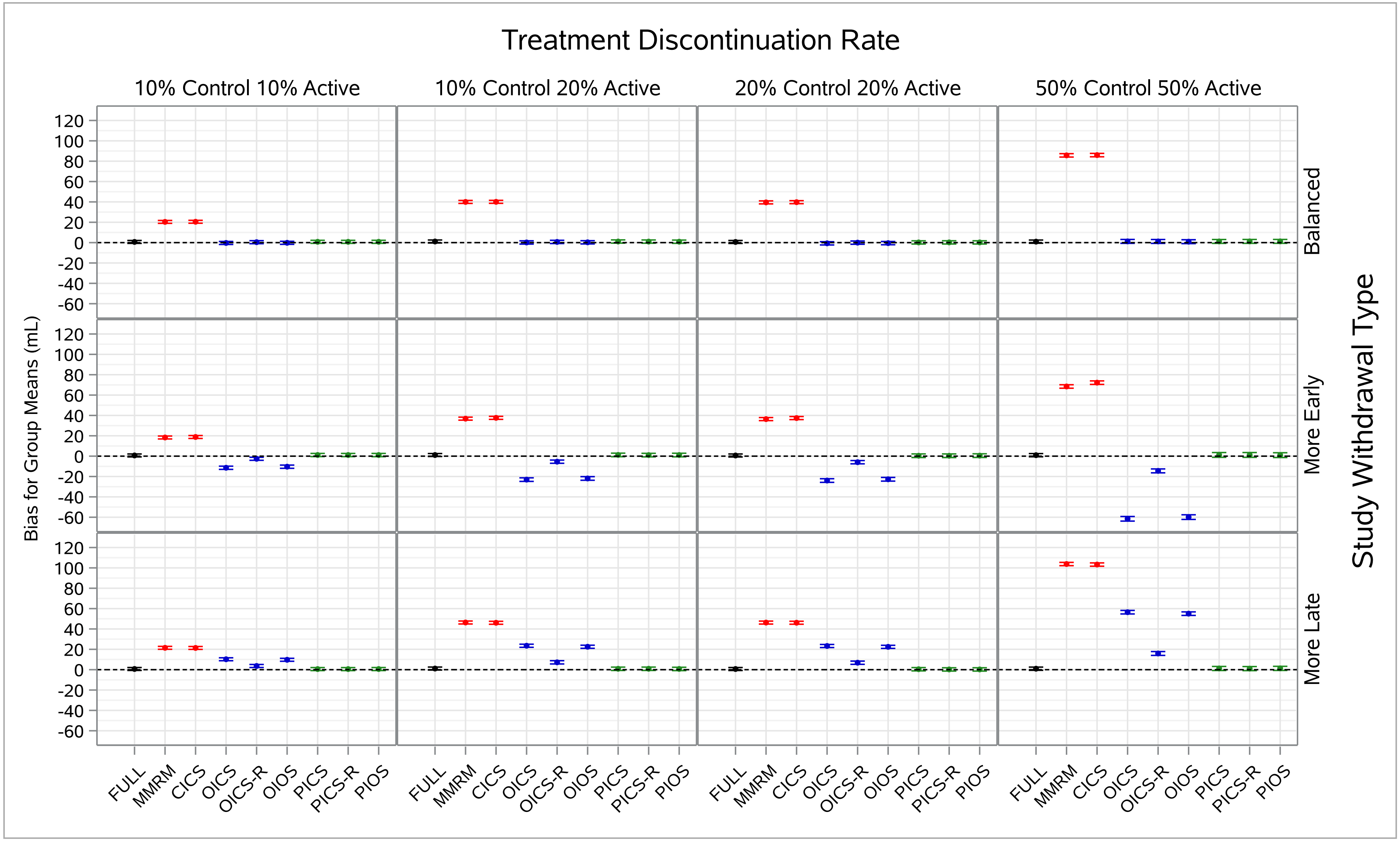

**SI Figure 4.7**
**Bias for Group Means - Control Group**

**Disc. Scenario: Same as Active**
**Disc. Mechanism: DAR**

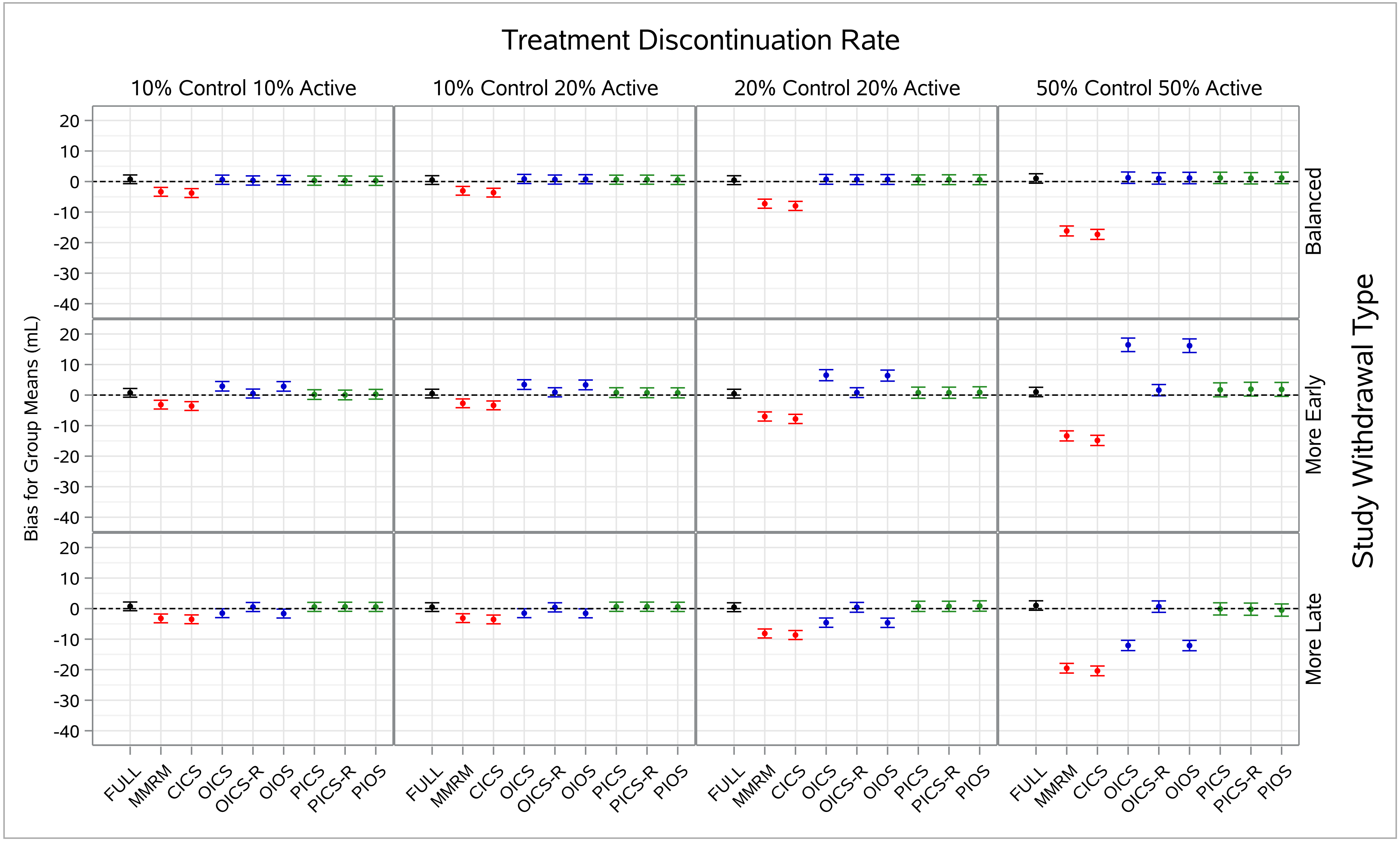

**SI Figure 4.8**
**Bias for Group Means - Control Group**

**Disc. Scenario: Same as Active**
**Disc. Mechanism: DNAR1**

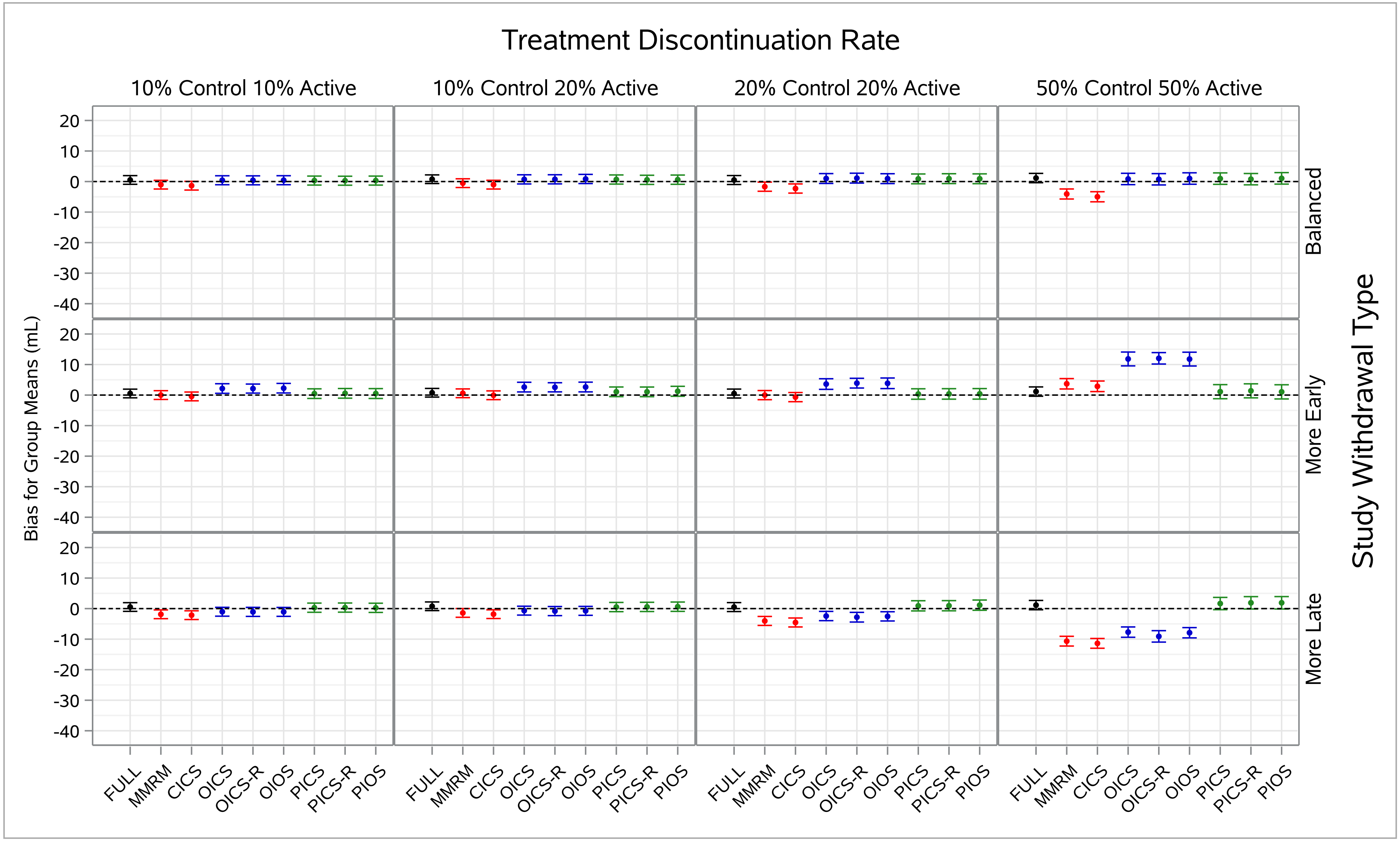

**SI Figure 4.9**
**Bias for Group Means - Control Group**

**Disc. Scenario: Same as Active**
**Disc. Mechanism: DNAR2**

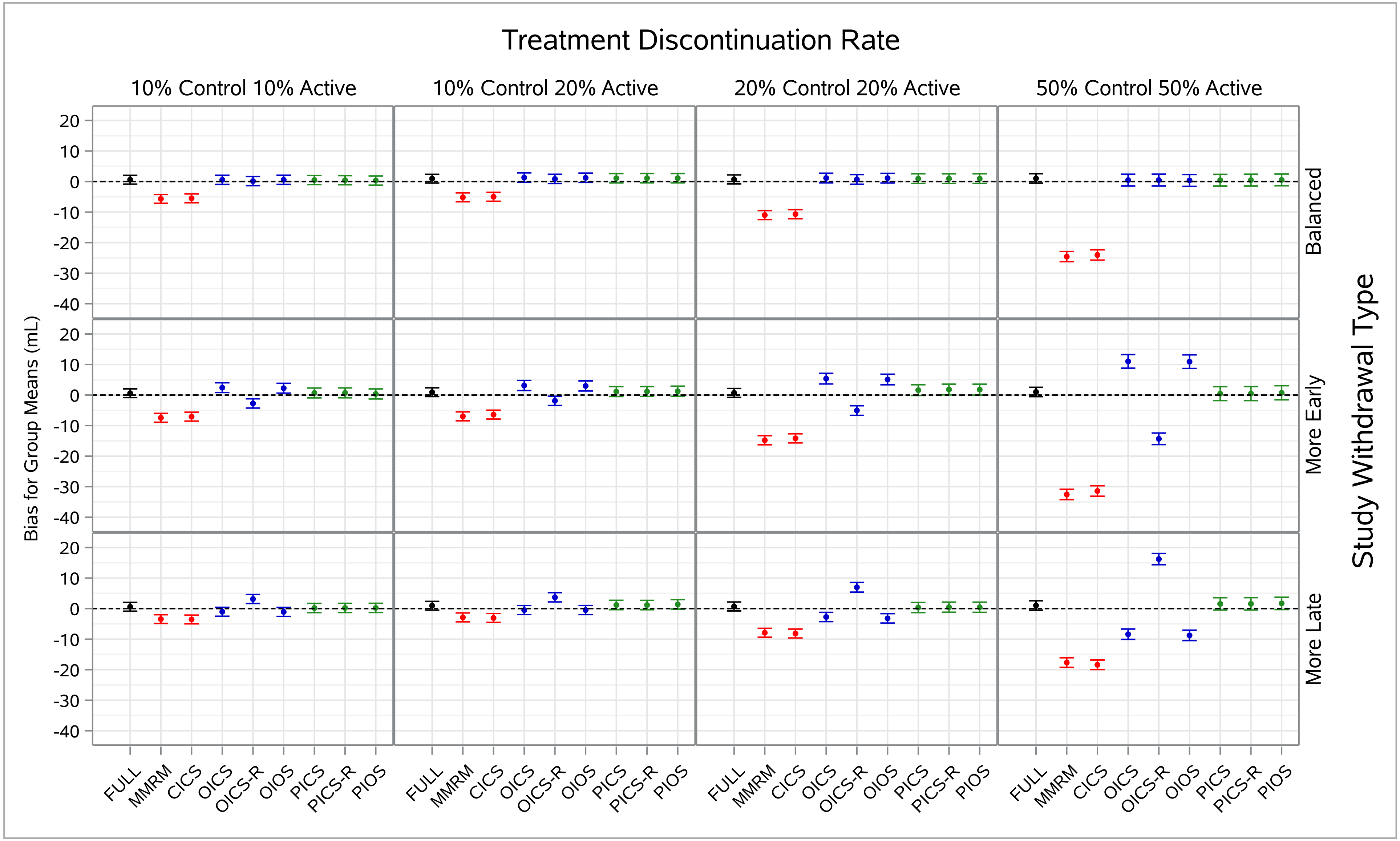

**SI Figure 4.10**
**Bias for Group Means - Active Group**

**Disc. Scenario: Same as Active**
**Disc. Mechanism: DAR**

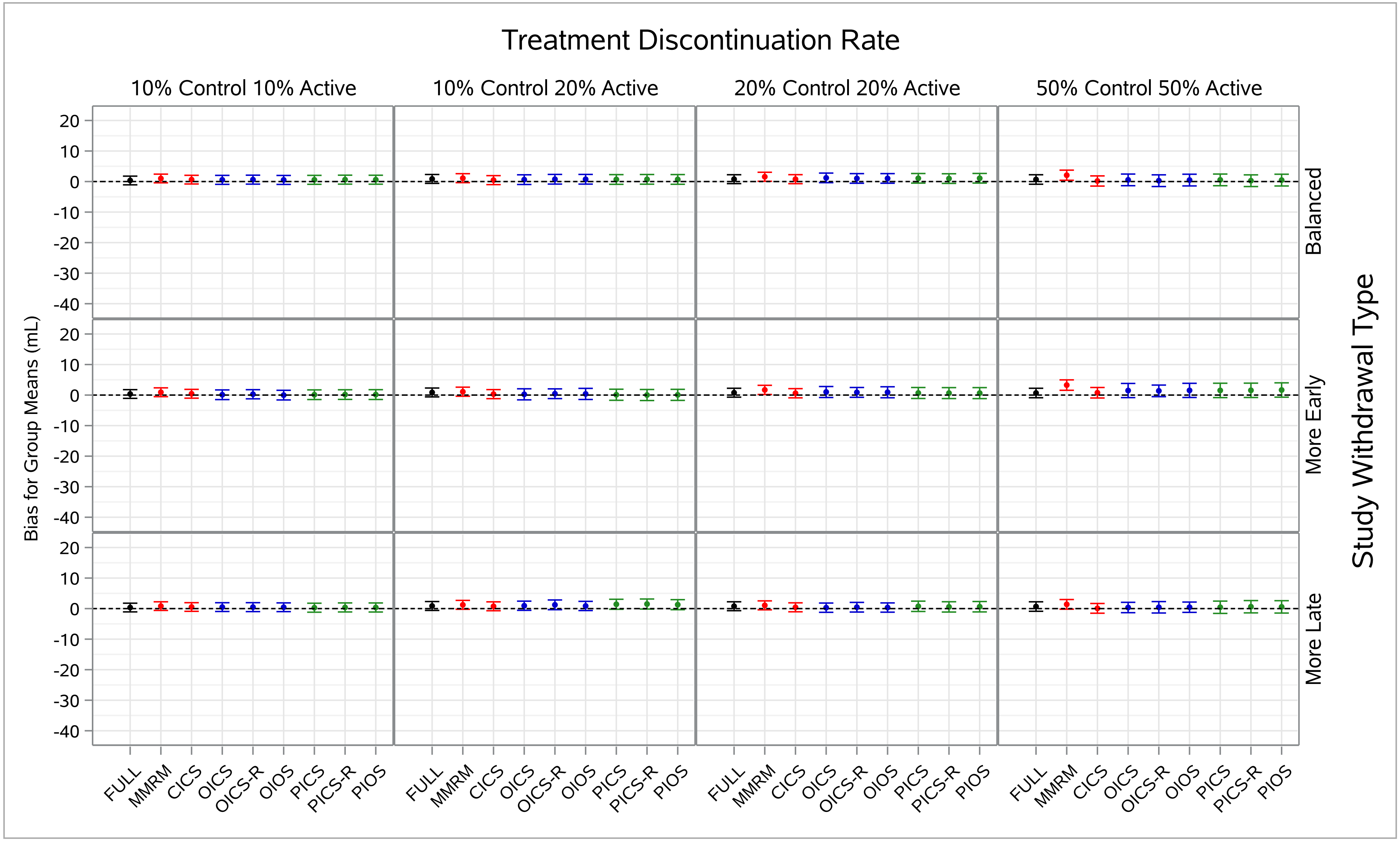

**SI Figure 4.11**
**Bias for Group Means - Active Group**

**Disc. Scenario: Same as Active**
**Disc. Mechanism: DNAR1**

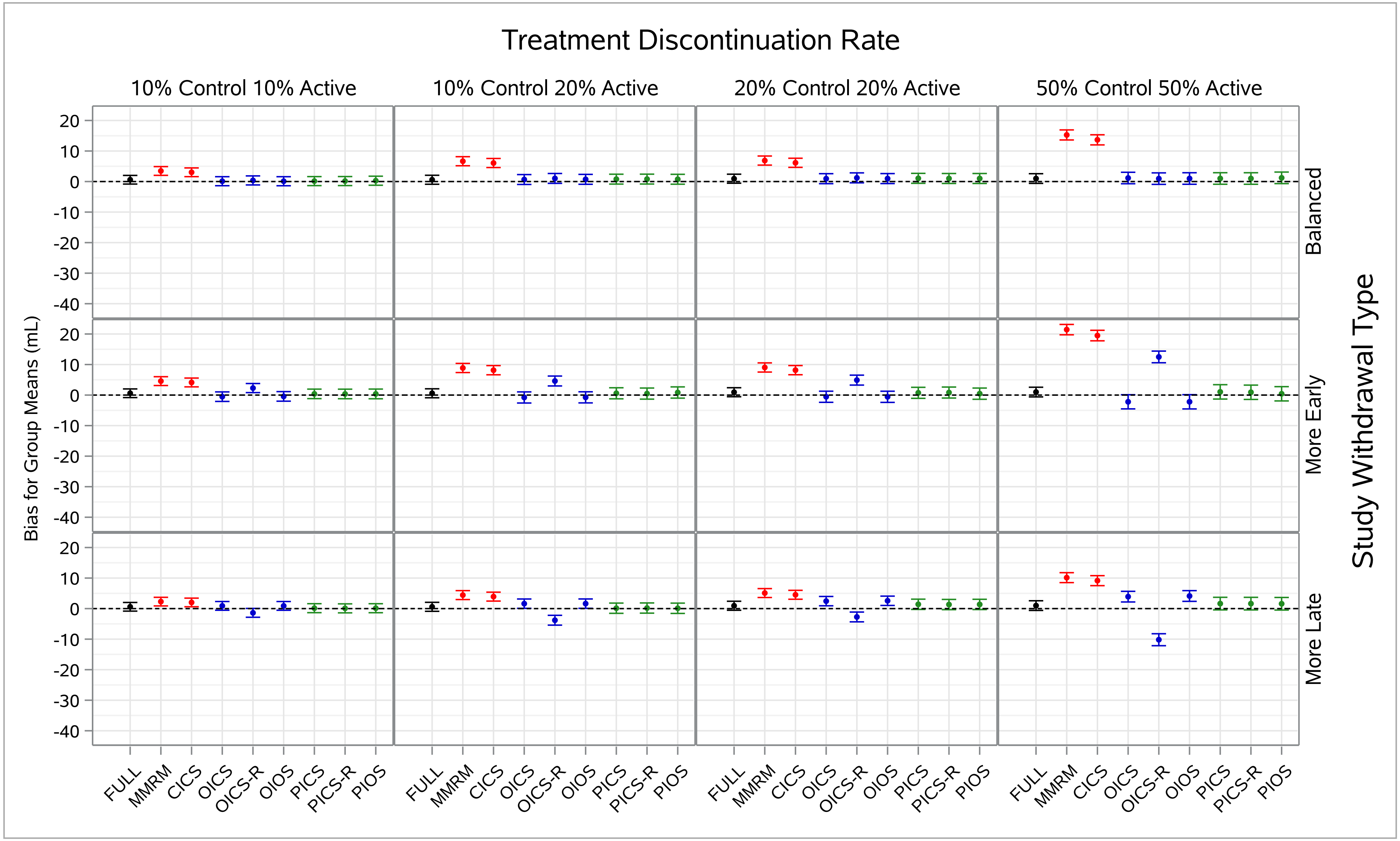

**SI Figure 4.12**
**Bias for Group Means - Active Group**

**Disc. Scenario: Same as Active**
**Disc. Mechanism: DNAR2**

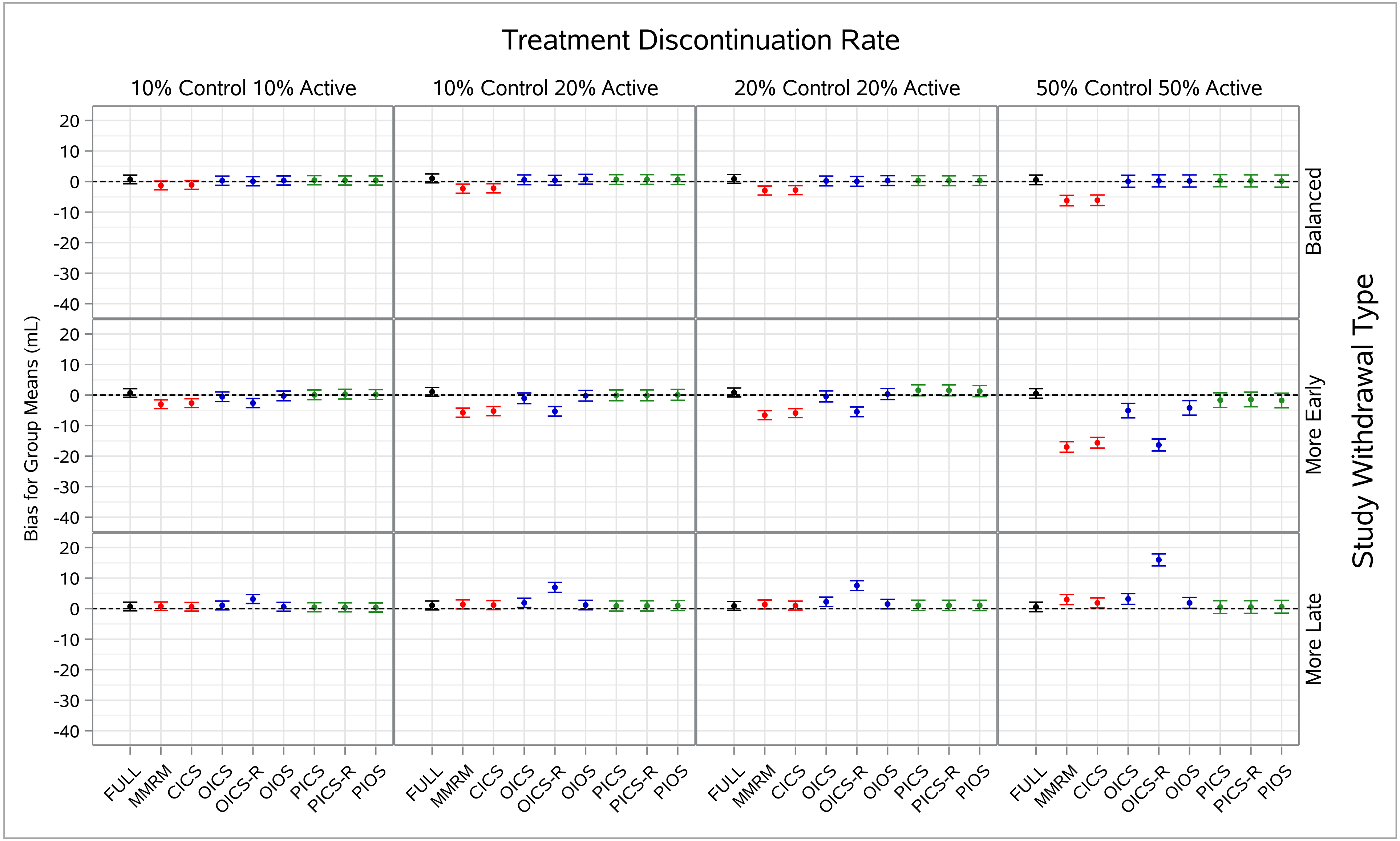

**SI Figure 5.1**
**Average 95% CI halfwidth for Group Means - Control Group**

**Disc. Scenario: Return To Baseline**
**Disc. Mechanism: DAR**

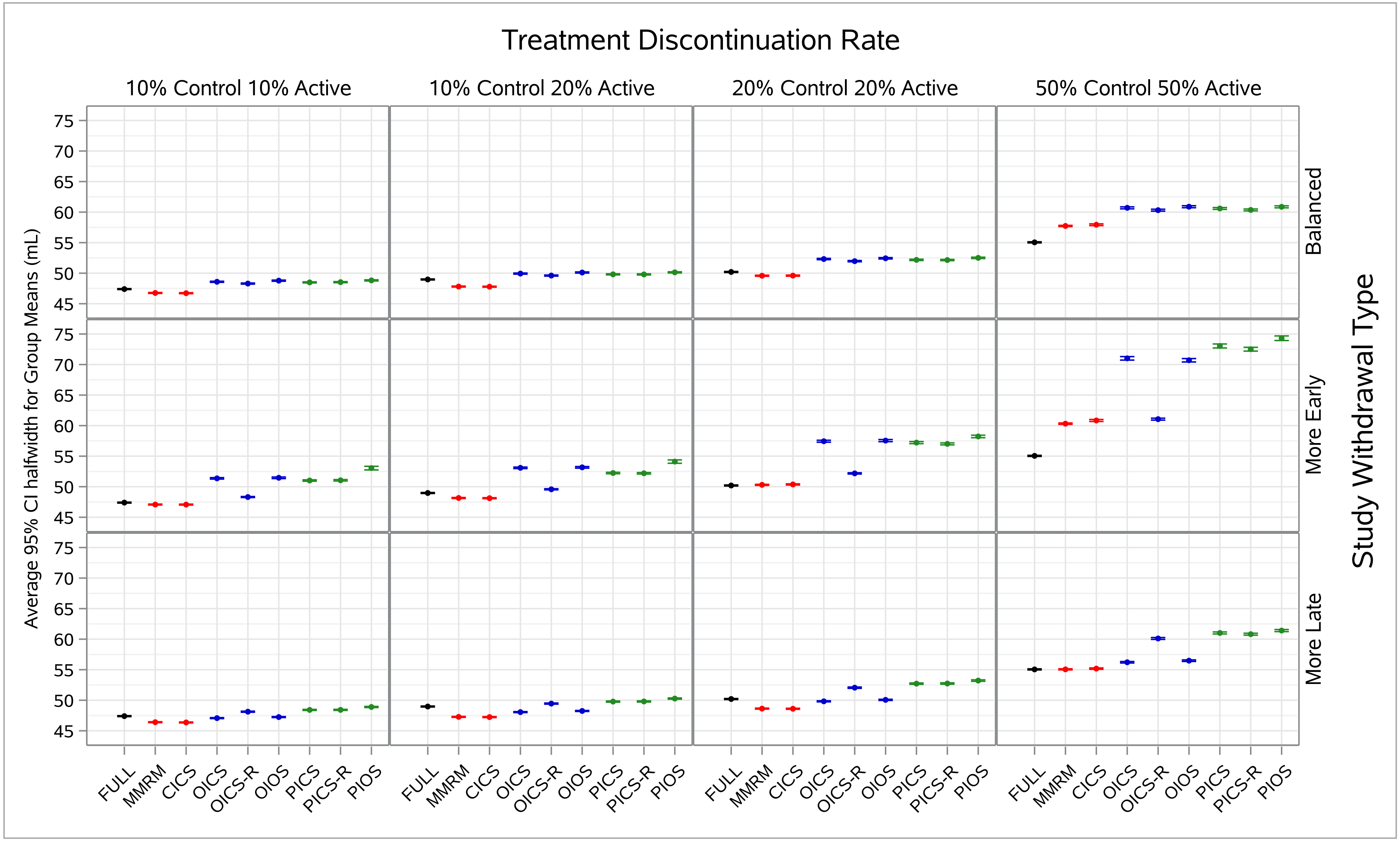

**SI Figure 5.2**
**Average 95% CI halfwidth for Group Means - Control Group**

**Disc. Scenario: Return To Baseline**
**Disc. Mechanism: DNAR1**

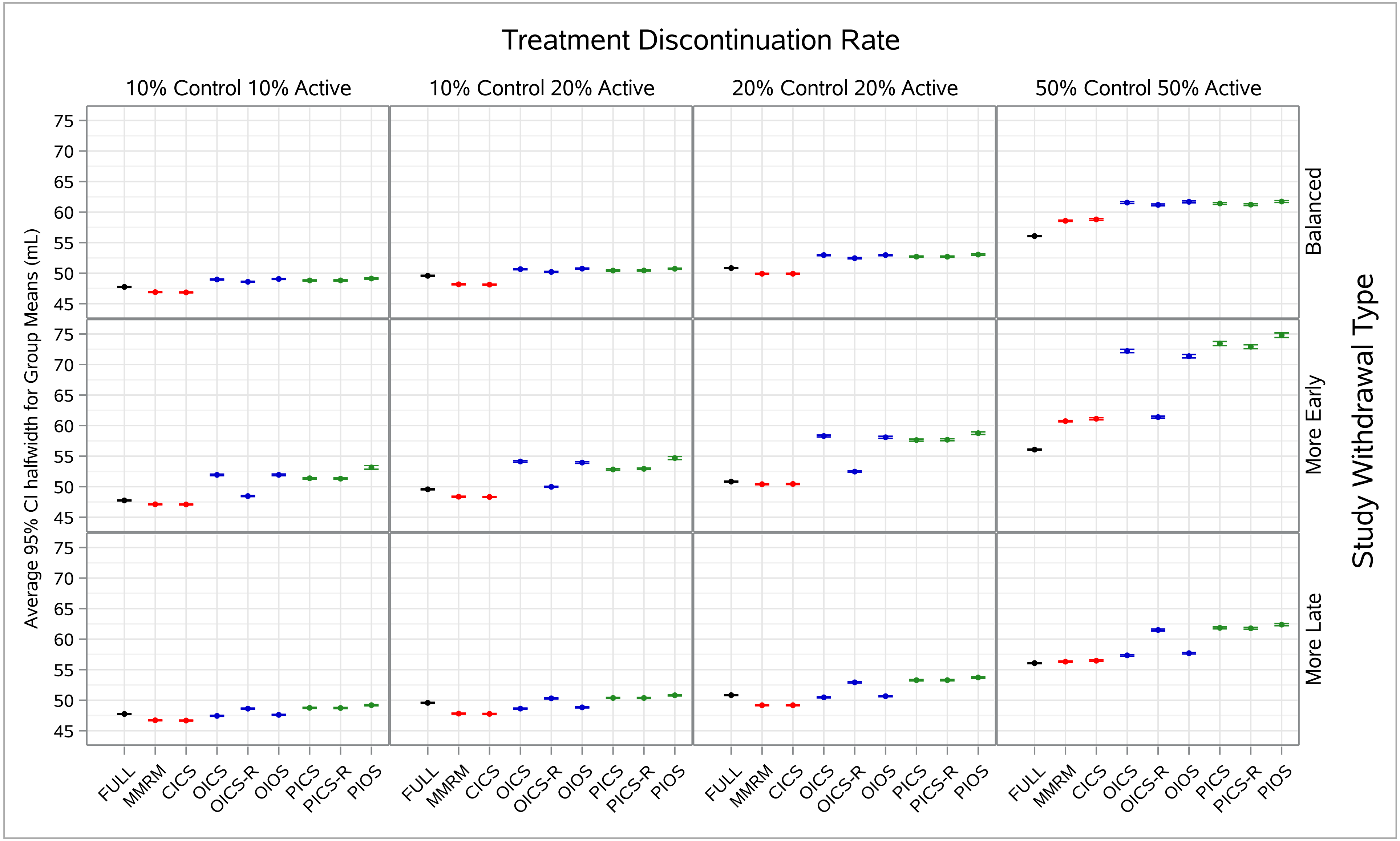

**SI Figure 5.3**
**Average 95% CI halfwidth for Group Means - Control Group**

**Disc. Scenario: Return To Baseline**
**Disc. Mechanism: DNAR2**

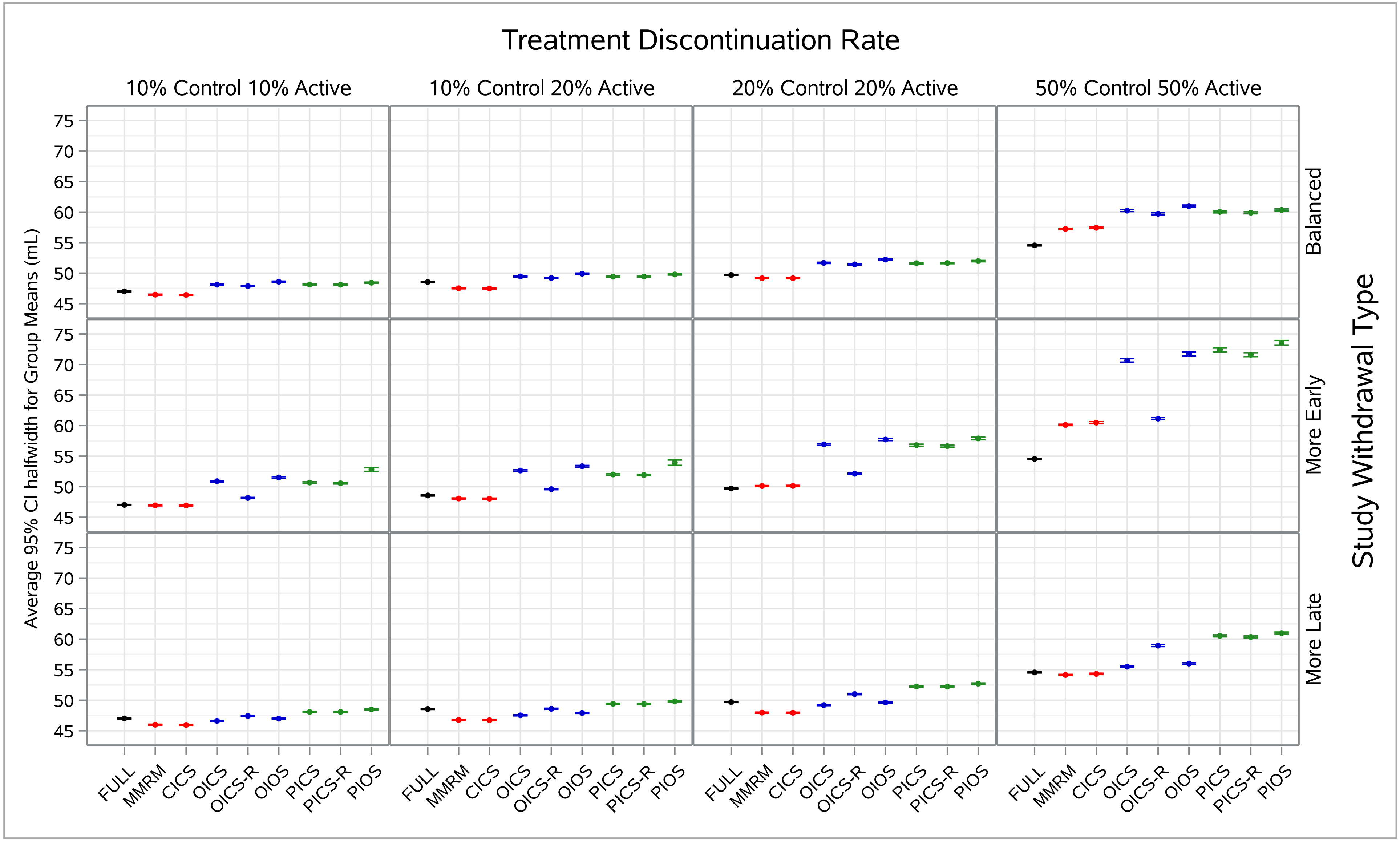

**SI Figure 5.4**
**Average 95% CI halfwidth for Group Means - Active Group**

**Disc. Scenario: Return To Baseline**
**Disc. Mechanism: DAR**

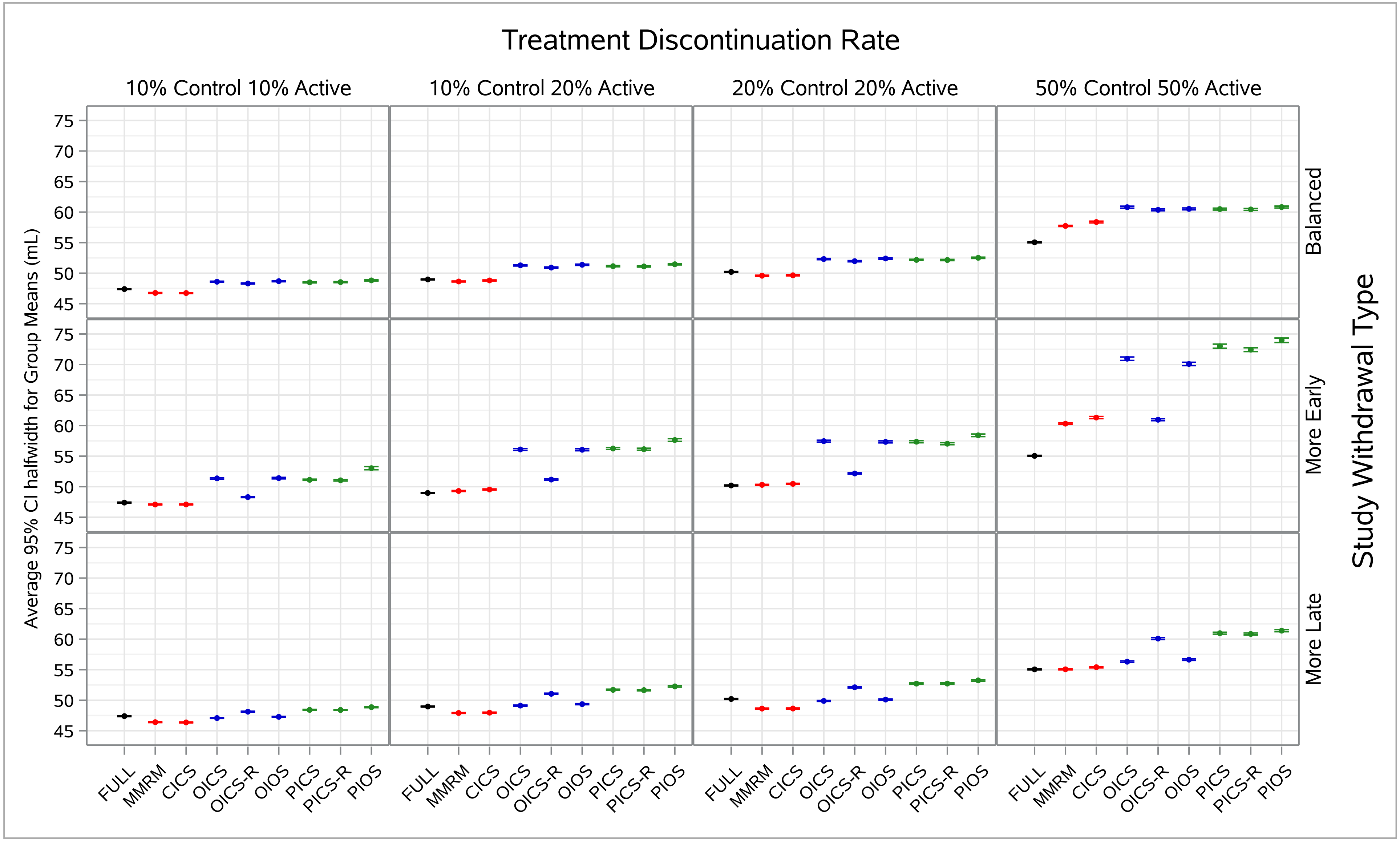

**SI Figure 5.5**
**Average 95% CI halfwidth for Group Means - Active Group**

**Disc. Scenario: Return To Baseline**
**Disc. Mechanism: DNAR1**

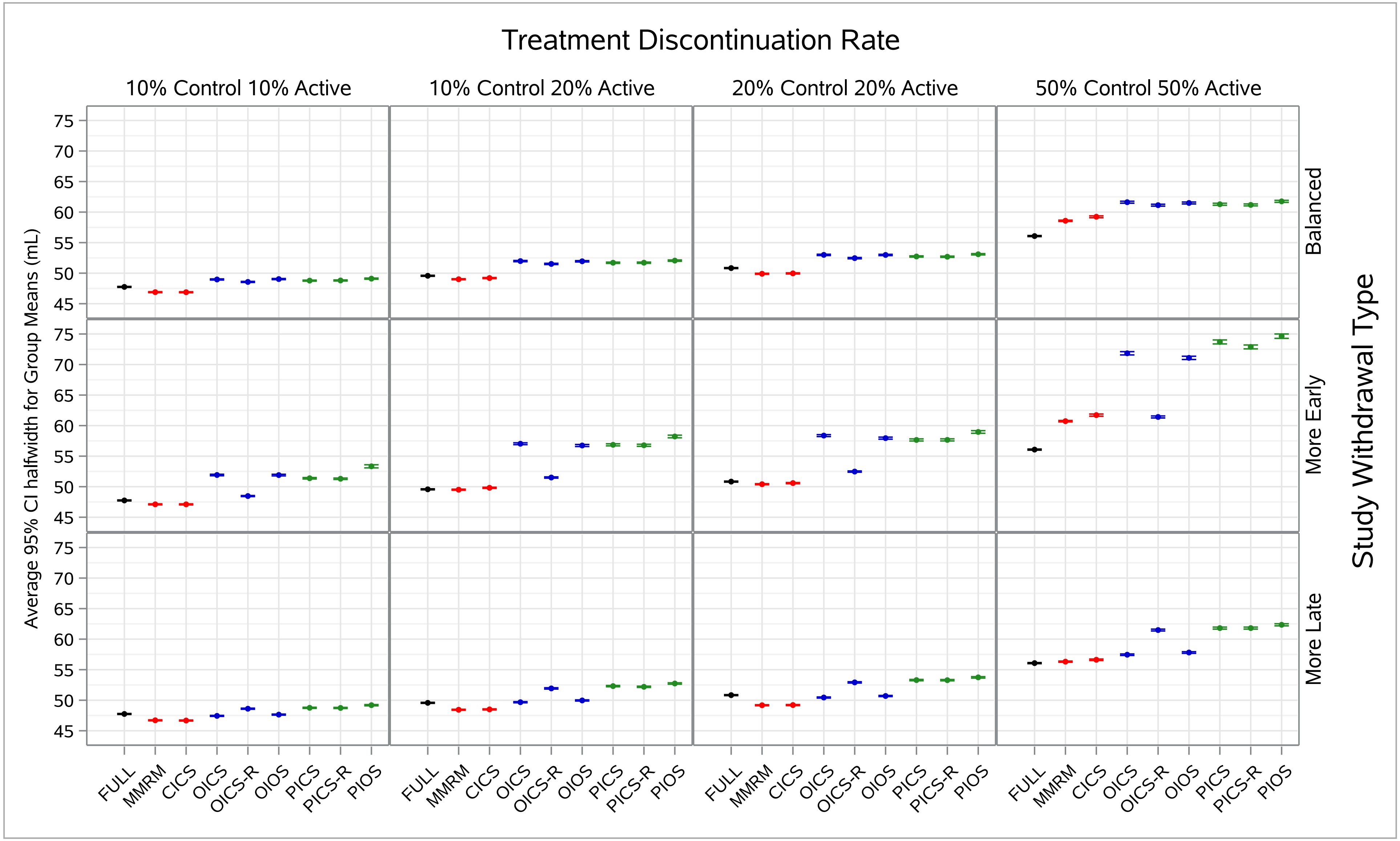

**SI Figure 5.6**
**Average 95% CI halfwidth for Group Means - Active Group**

**Disc. Scenario: Return To Baseline**
**Disc. Mechanism: DNAR2**

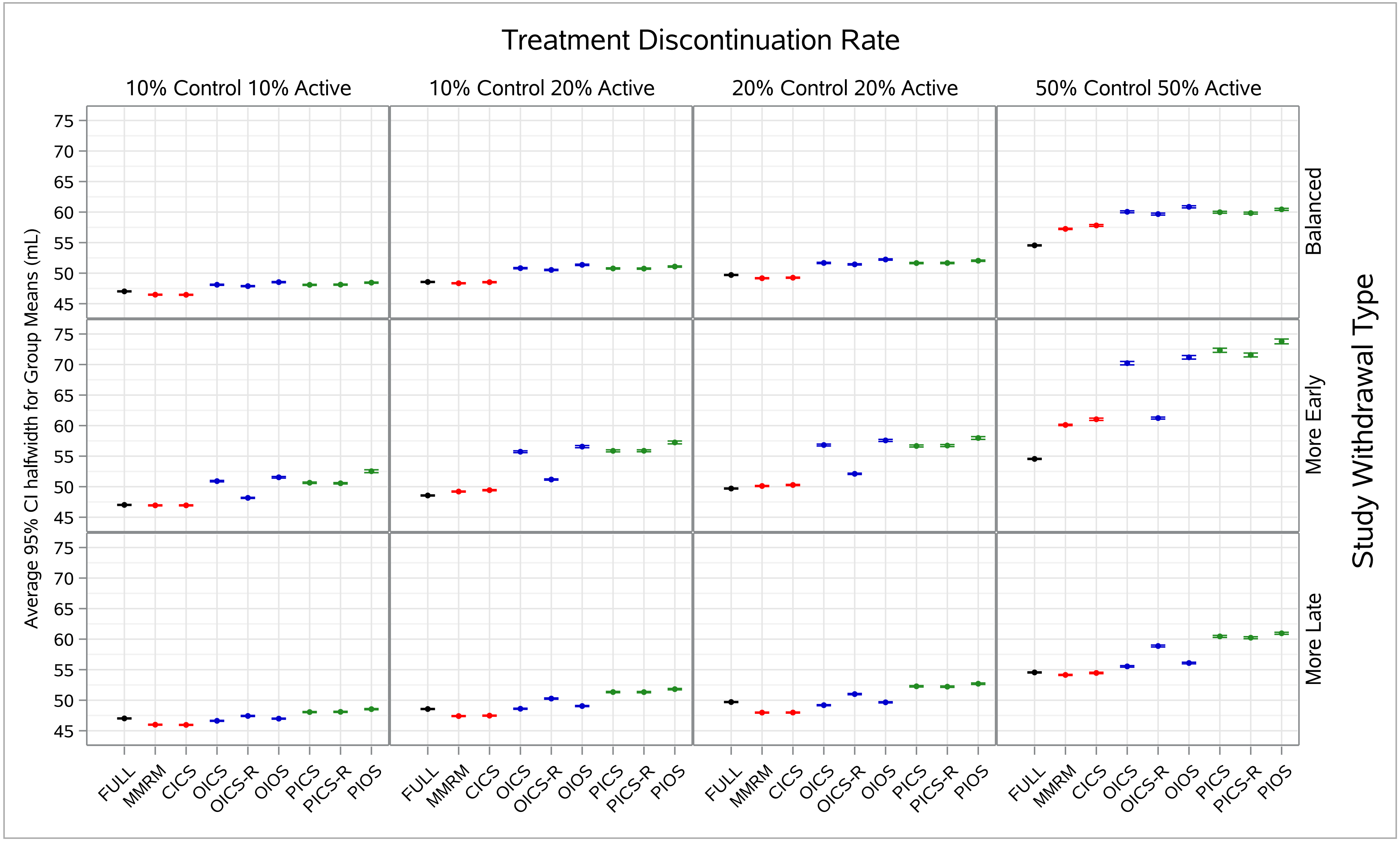

**SI Figure 5.7**
**Average 95% CI halfwidth for Group Means - Control Group**

**Disc. Scenario: Same as Active**
**Disc. Mechanism: DAR**

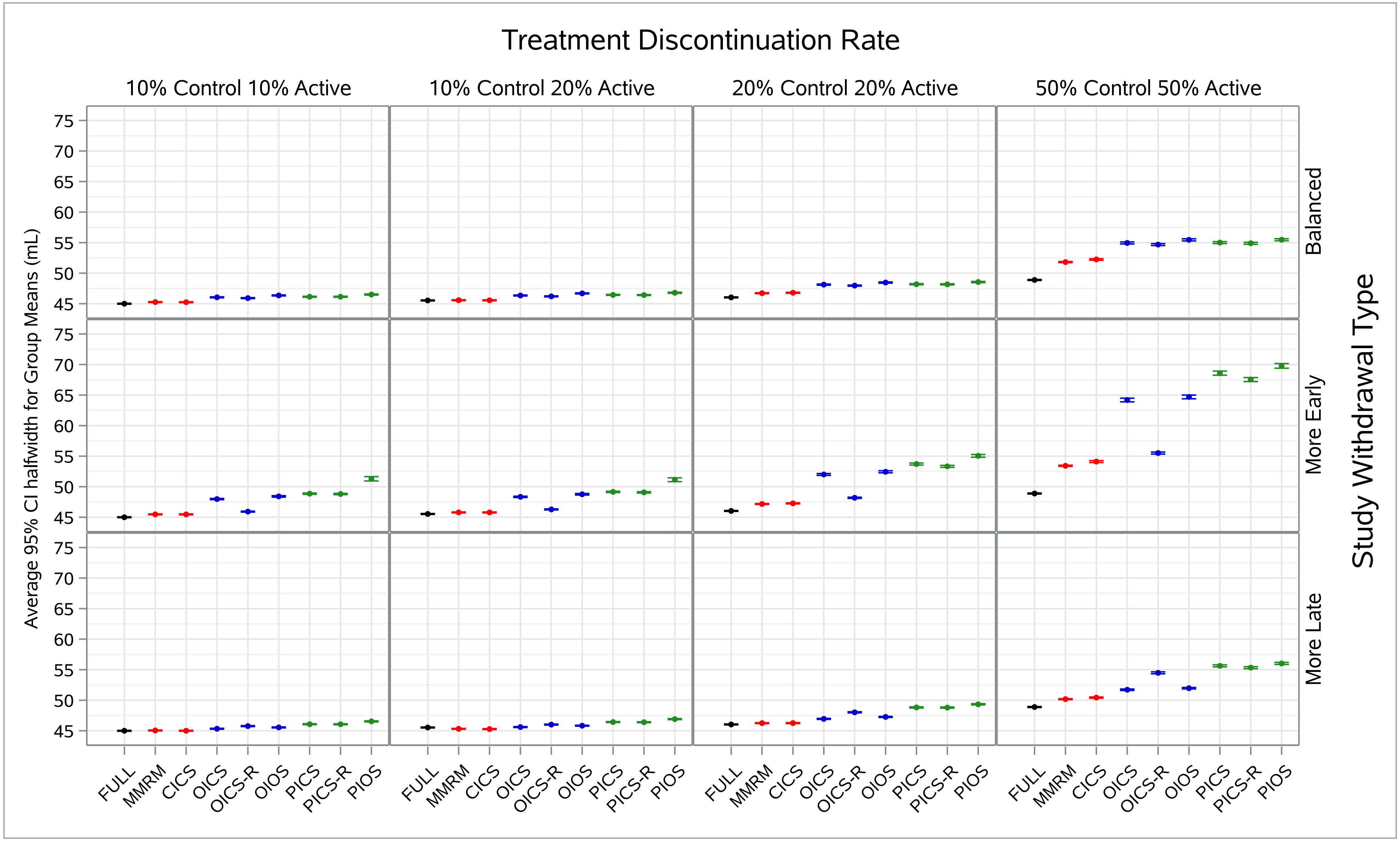

**SI Figure 5.8**
**Average 95% CI halfwidth for Group Means - Control Group**

**Disc. Scenario: Same as Active**
**Disc. Mechanism: DNAR1**

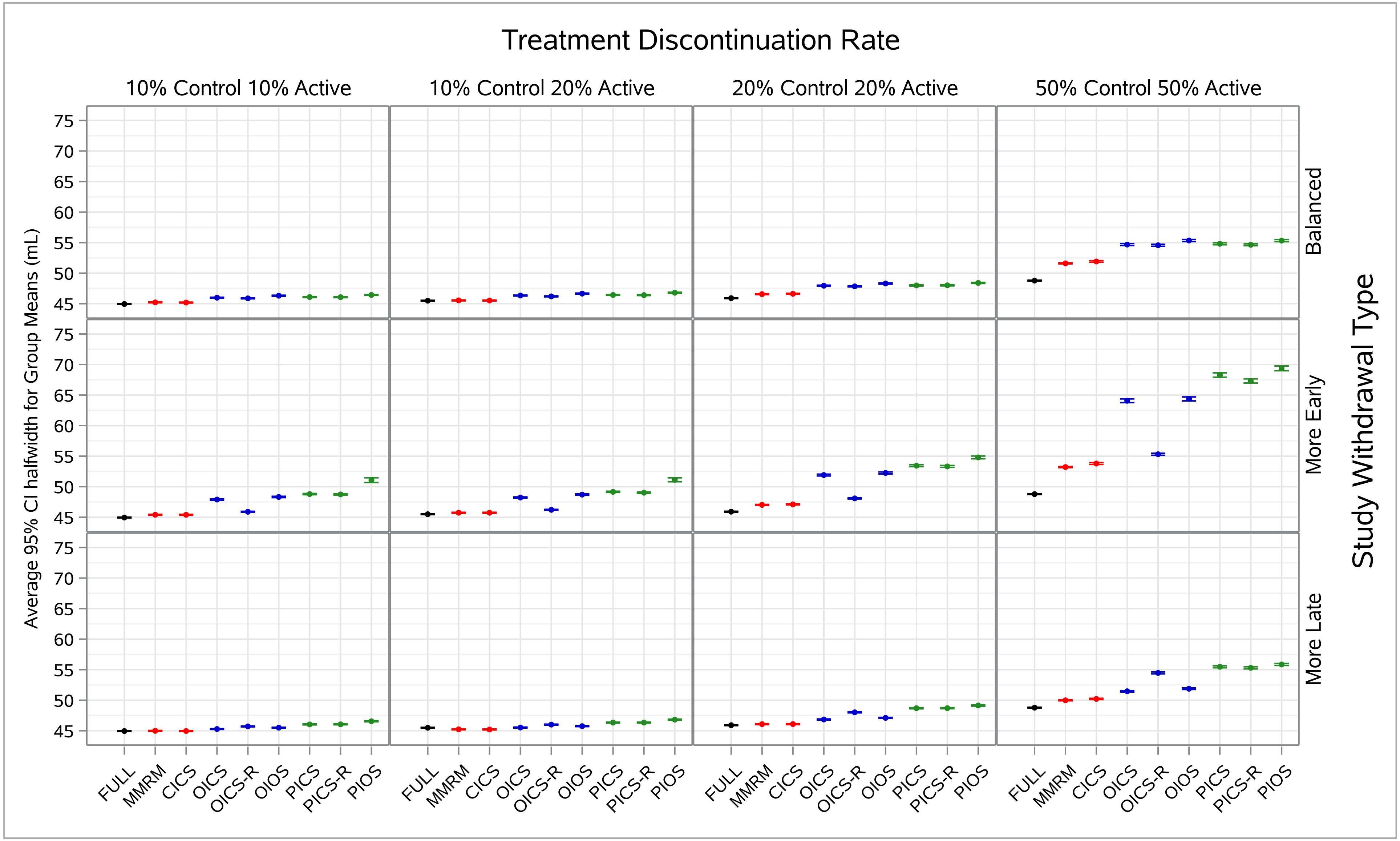

**SI Figure 5.9**
**Average 95% CI halfwidth for Group Means - Control Group**

**Disc. Scenario: Same as Active**
**Disc. Mechanism: DNAR2**

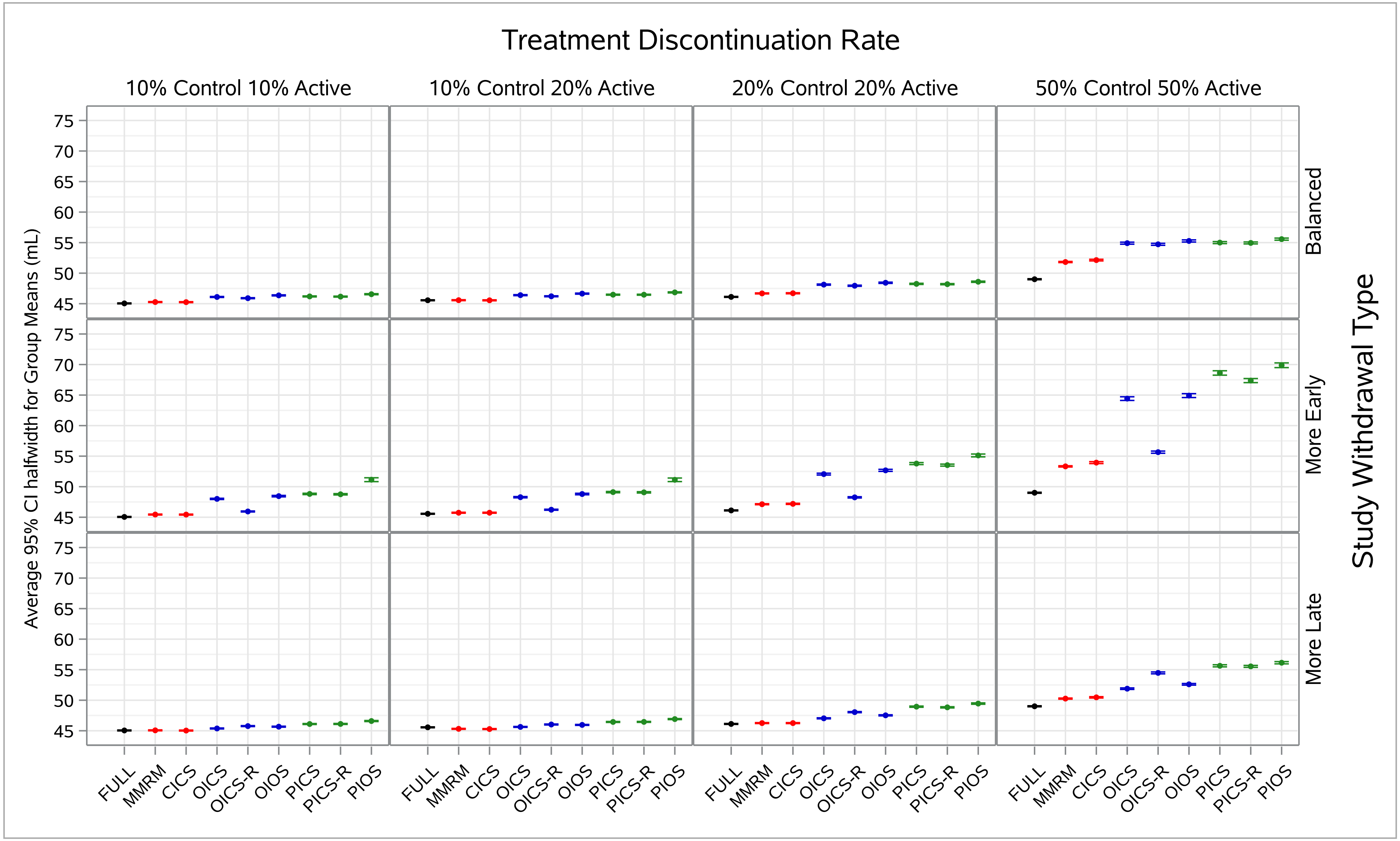

**SI Figure 5.10**
**Average 95% CI halfwidth for Group Means - Active Group**

**Disc. Scenario: Same as Active**
**Disc. Mechanism: DAR**

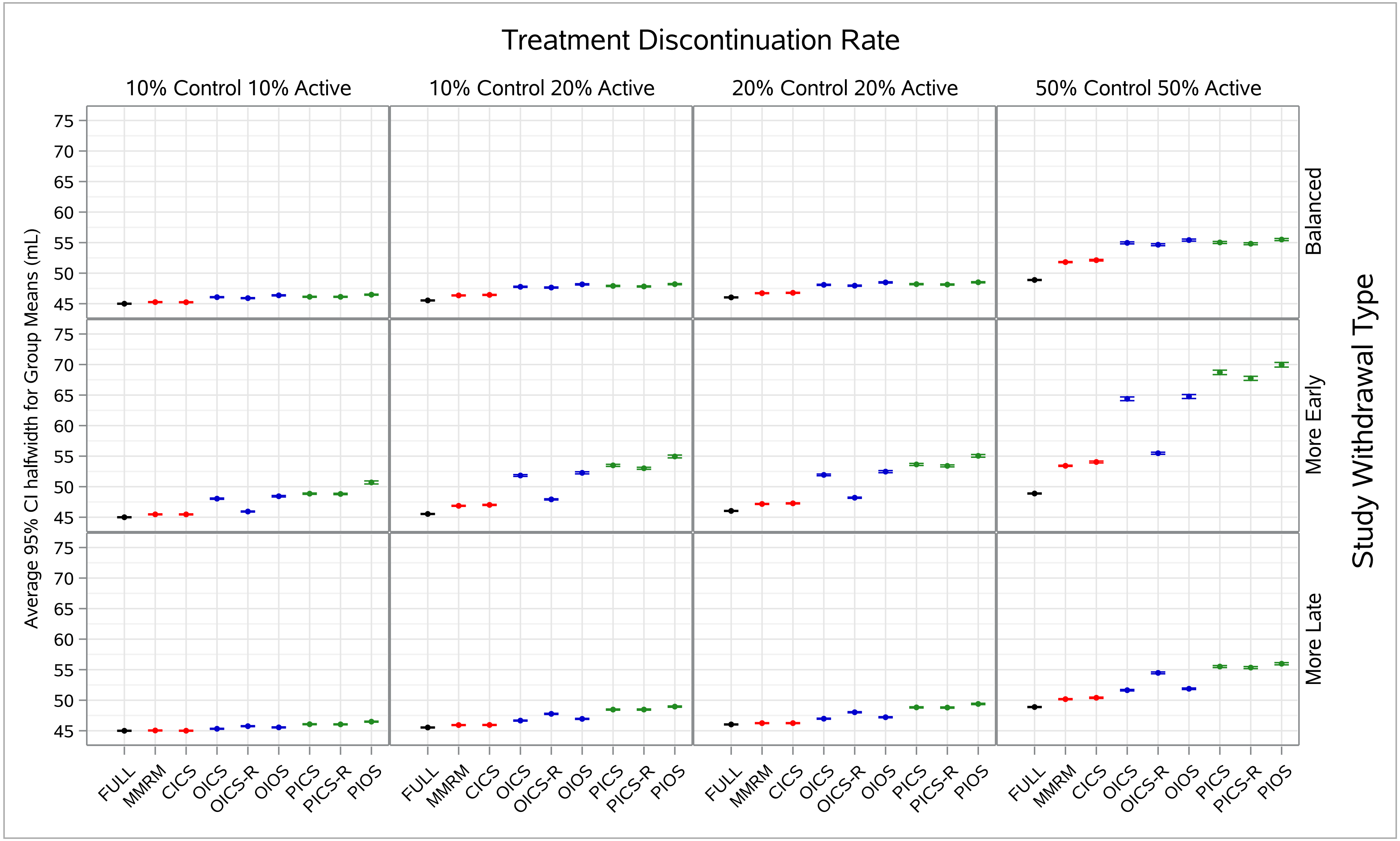

**SI Figure 5.11**
**Average 95% CI halfwidth for Group Means - Active Group**

**Disc. Scenario: Same as Active**
**Disc. Mechanism: DNAR1**

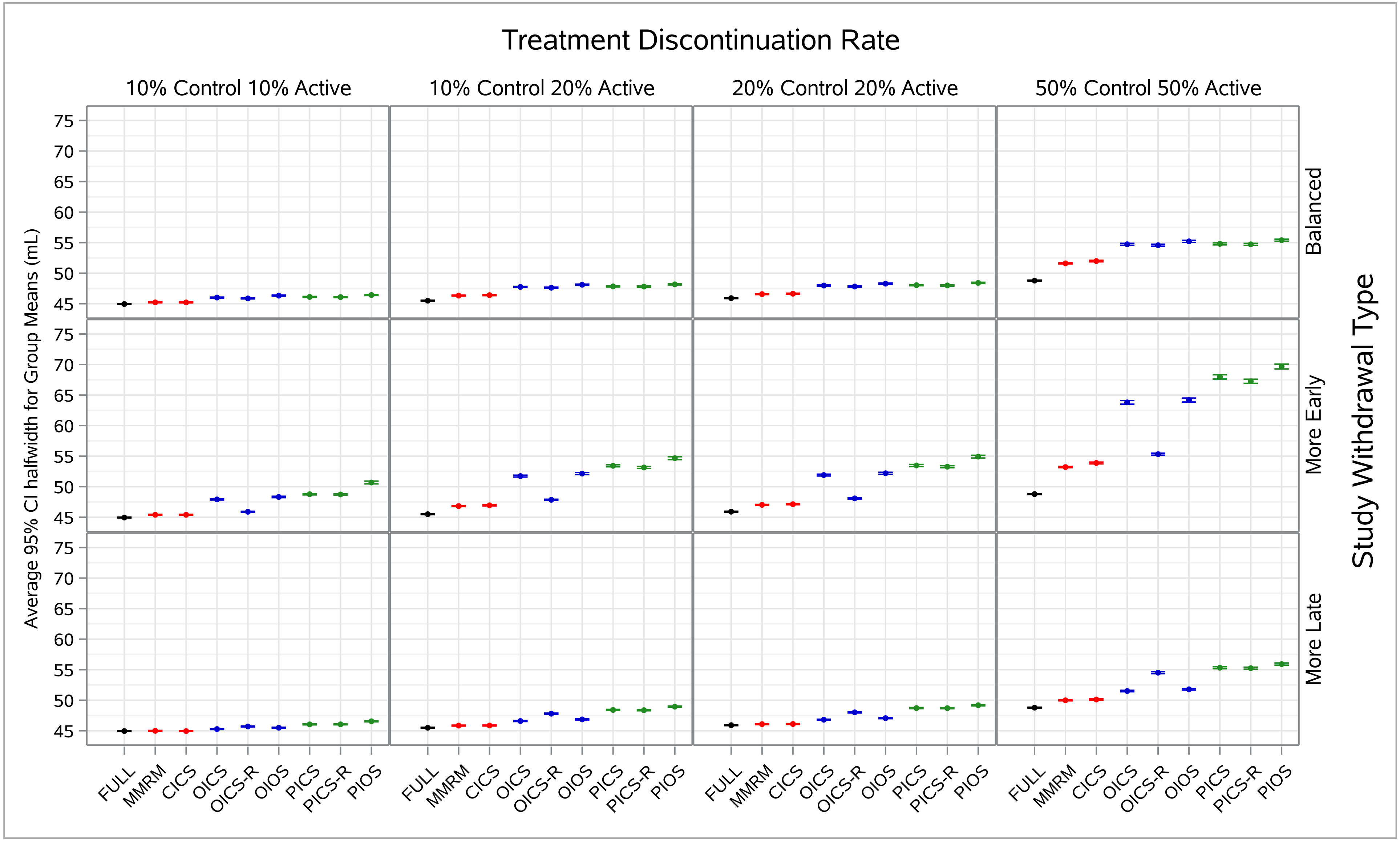

**SI Figure 5.12**
**Average 95% CI halfwidth for Group Means - Active Group**

**Disc. Scenario: Same as Active**
**Disc. Mechanism: DNAR2**

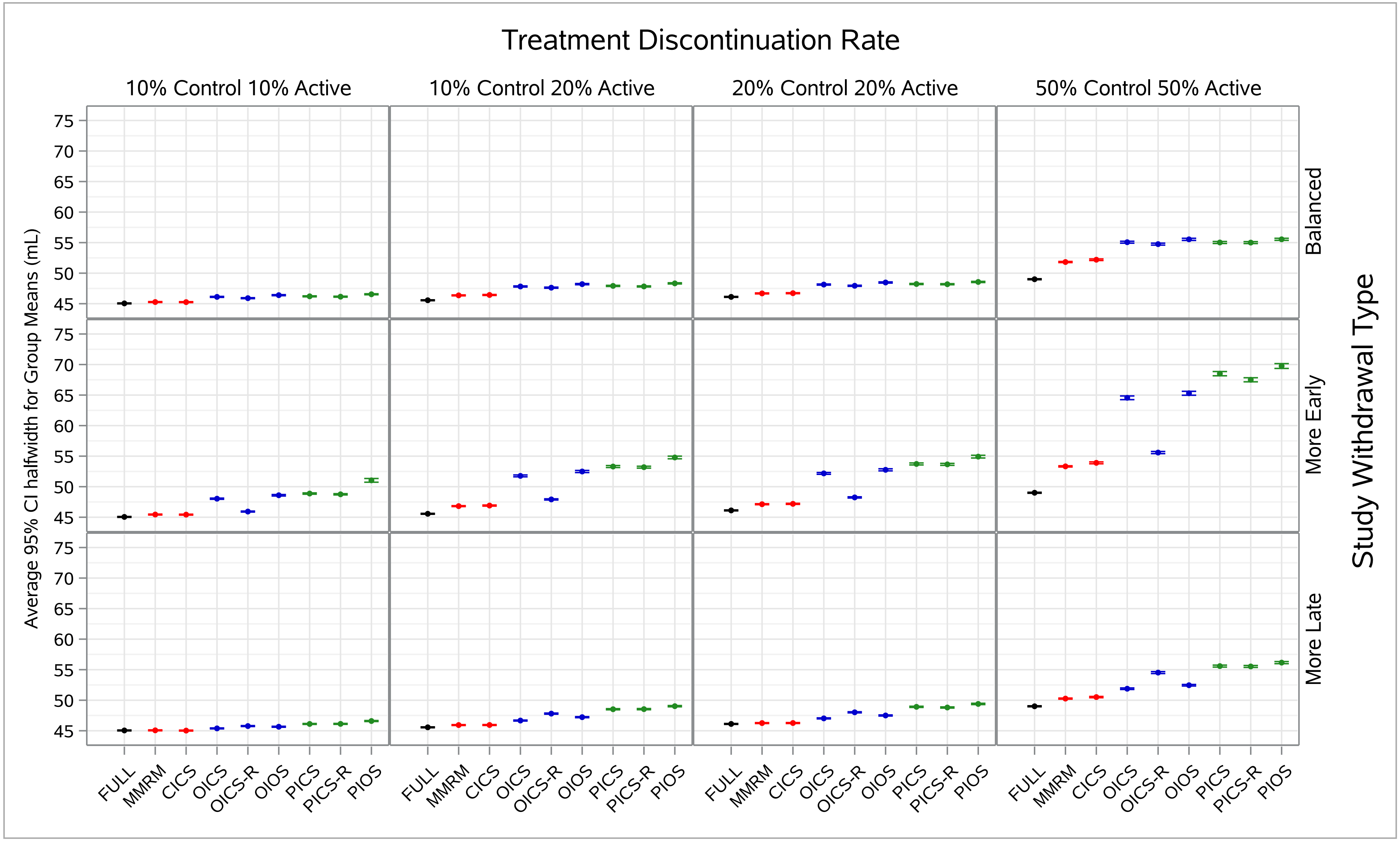

**SI Figure 6.1**
**95% CI Coverage for Group Means - Control Group**

**Disc. Scenario: Return To Baseline**
**Disc. Mechanism: DAR**

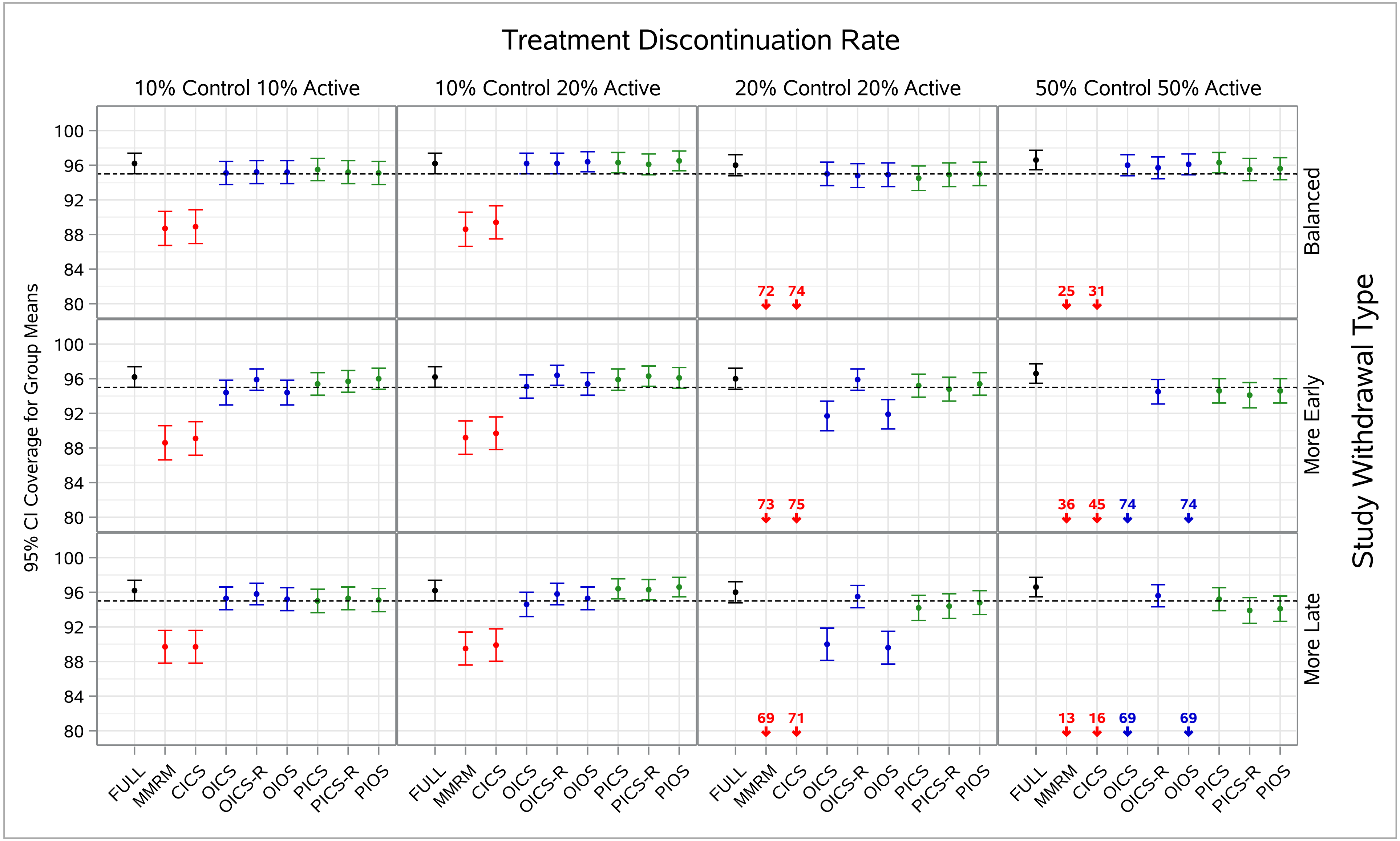

**SI Figure 6.2**
**95% CI Coverage for Group Means - Control Group**

**Disc. Scenario: Return To Baseline**
**Disc. Mechanism: DNAR1**

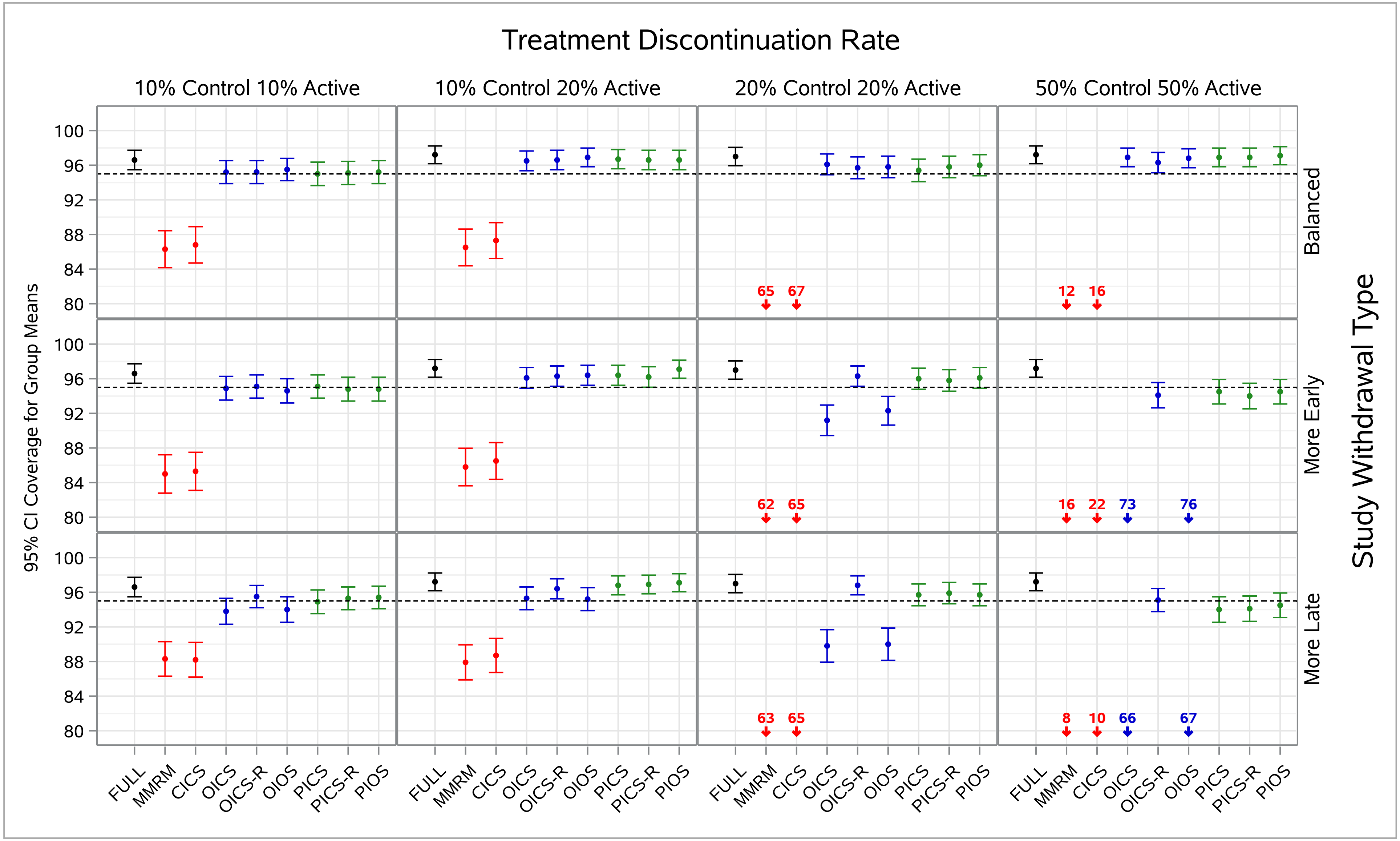

**SI Figure 6.3**
**95% CI Coverage for Group Means - Control Group**

**Disc. Scenario: Return To Baseline**
**Disc. Mechanism: DNAR2**

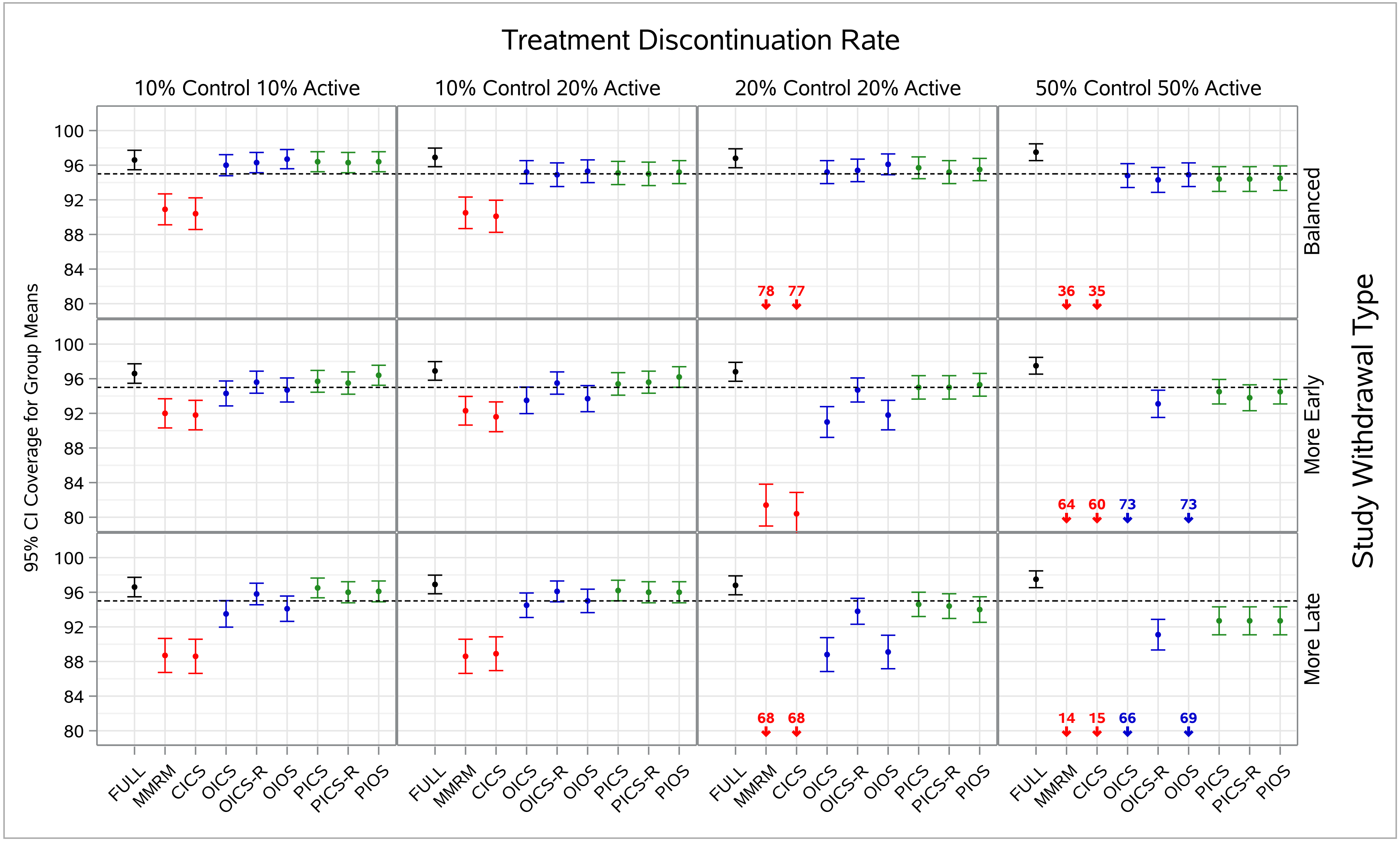

**SI Figure 6.4**
**95% CI Coverage for Group Means - Active Group**

**Disc. Scenario: Return To Baseline**
**Disc. Mechanism: DAR**

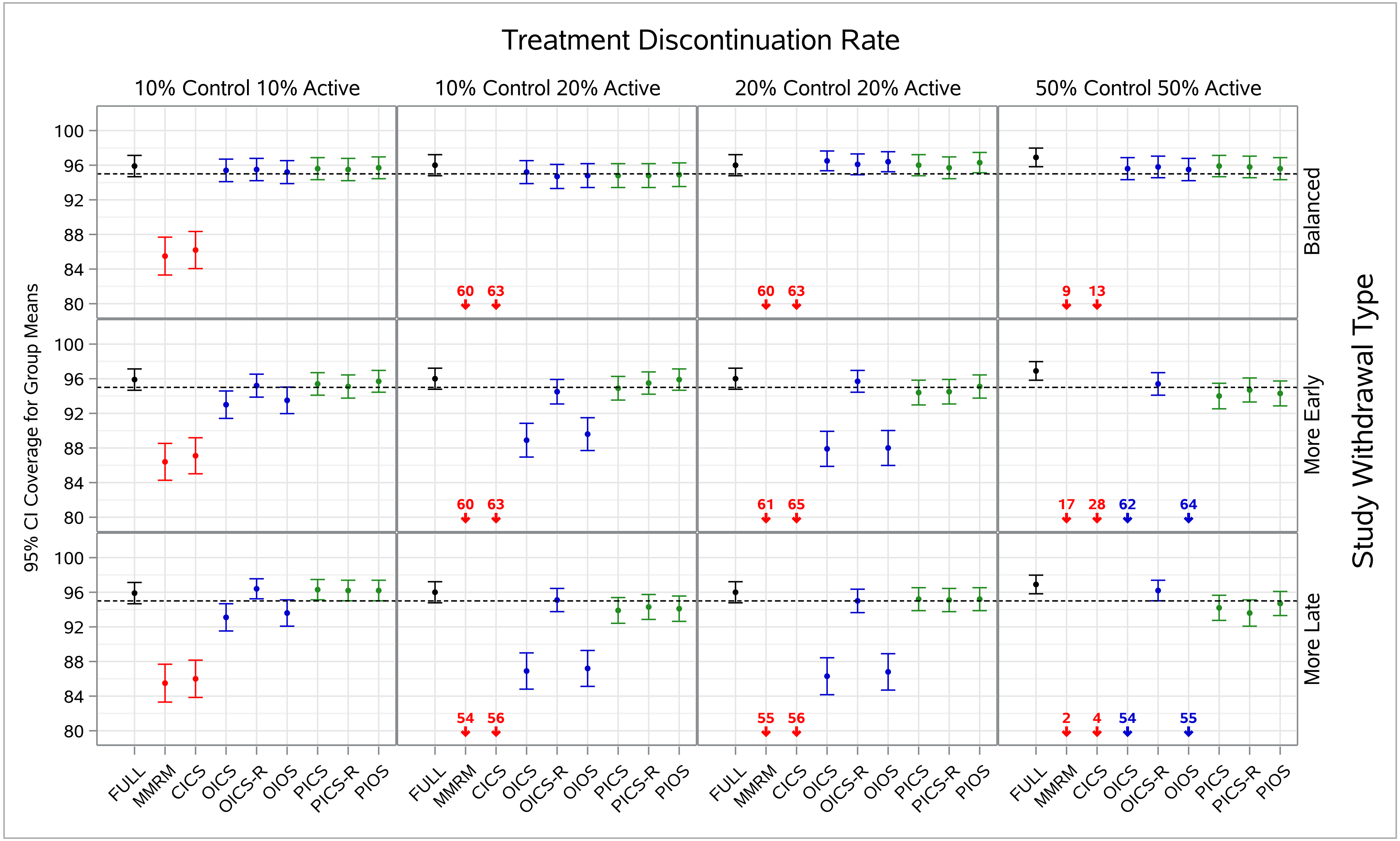

**SI Figure 6.5**
**95% CI Coverage for Group Means - Active Group**

**Disc. Scenario: Return To Baseline**
**Disc. Mechanism: DNAR1**

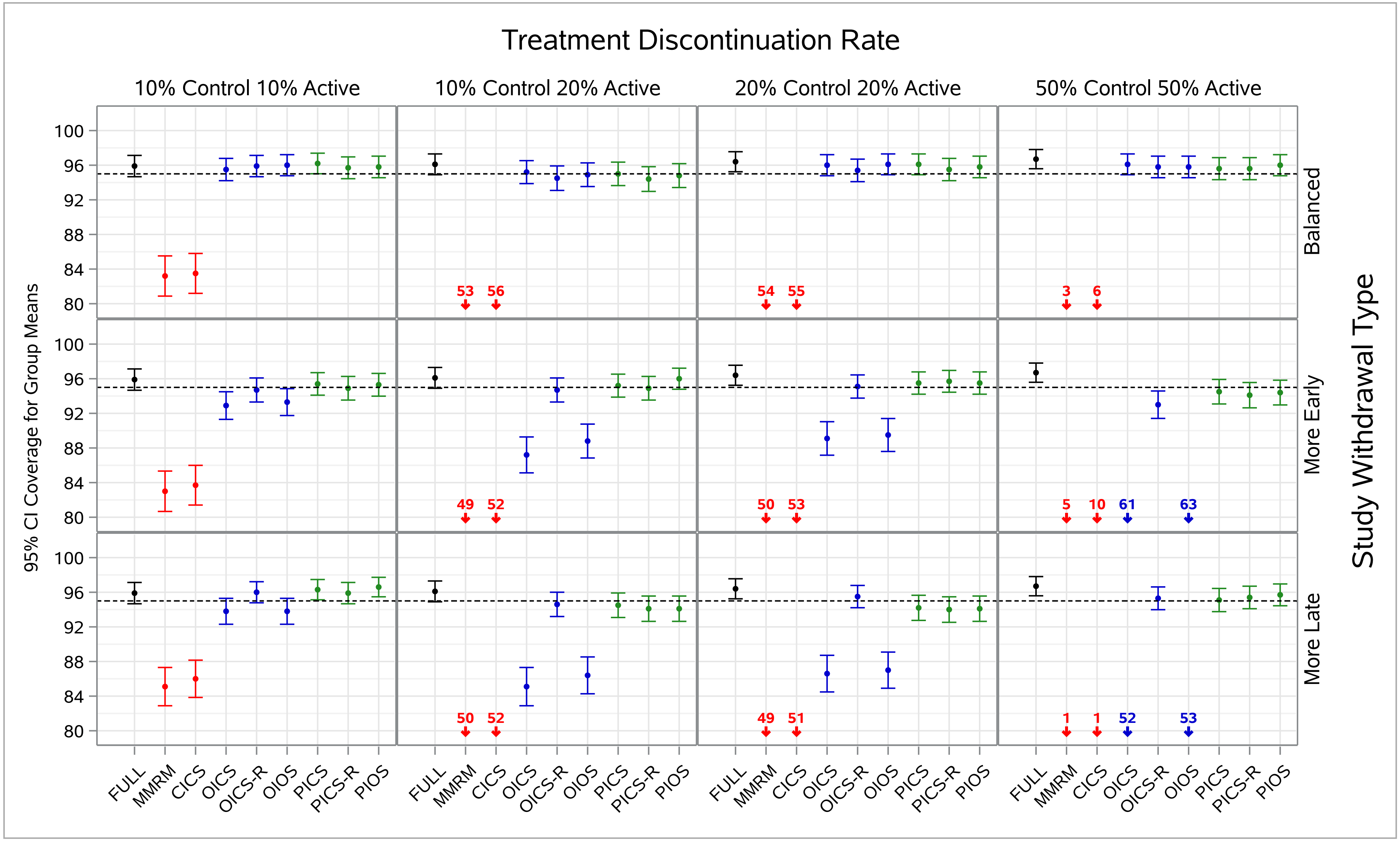

**SI Figure 6.6**
**95% CI Coverage for Group Means - Active Group**

**Disc. Scenario: Return To Baseline**
**Disc. Mechanism: DNAR2**

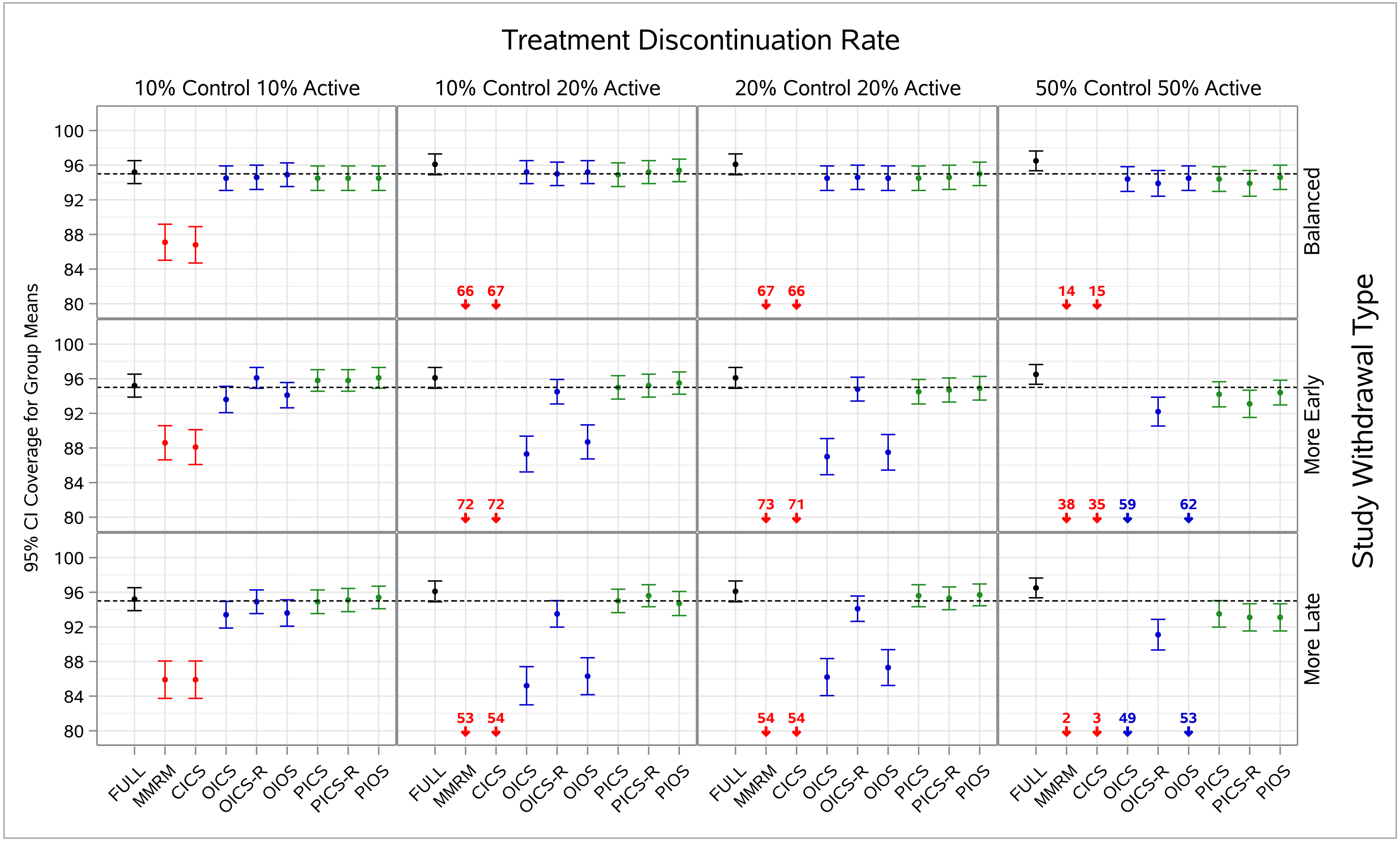

**SI Figure 6.7**
**95% CI Coverage for Group Means - Control Group**

**Disc. Scenario: Same as Active**
**Disc. Mechanism: DAR**

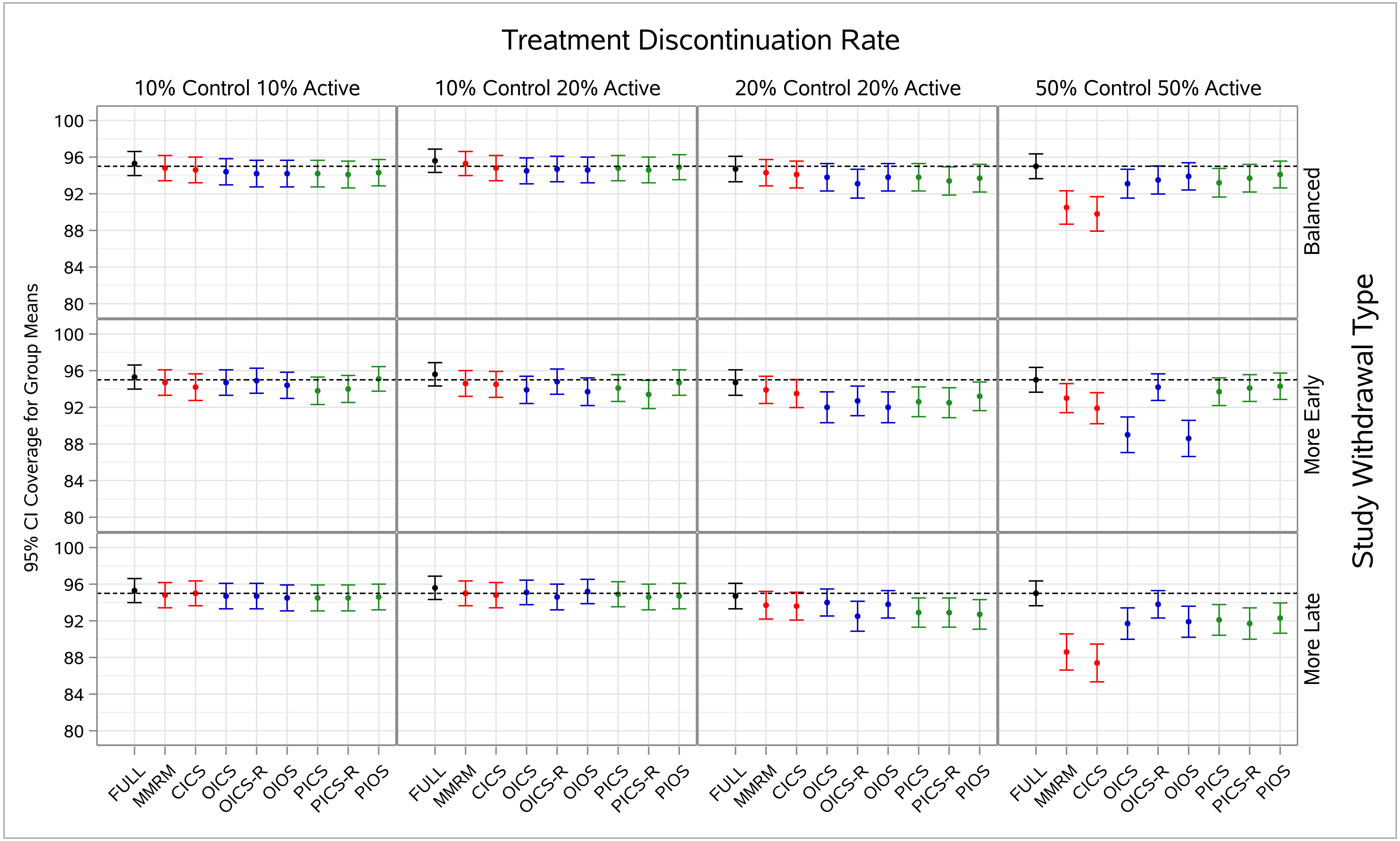

**SI Figure 6.8**
**95% CI Coverage for Group Means - Control Group**

**Disc. Scenario: Same as Active**
**Disc. Mechanism: DNAR1**

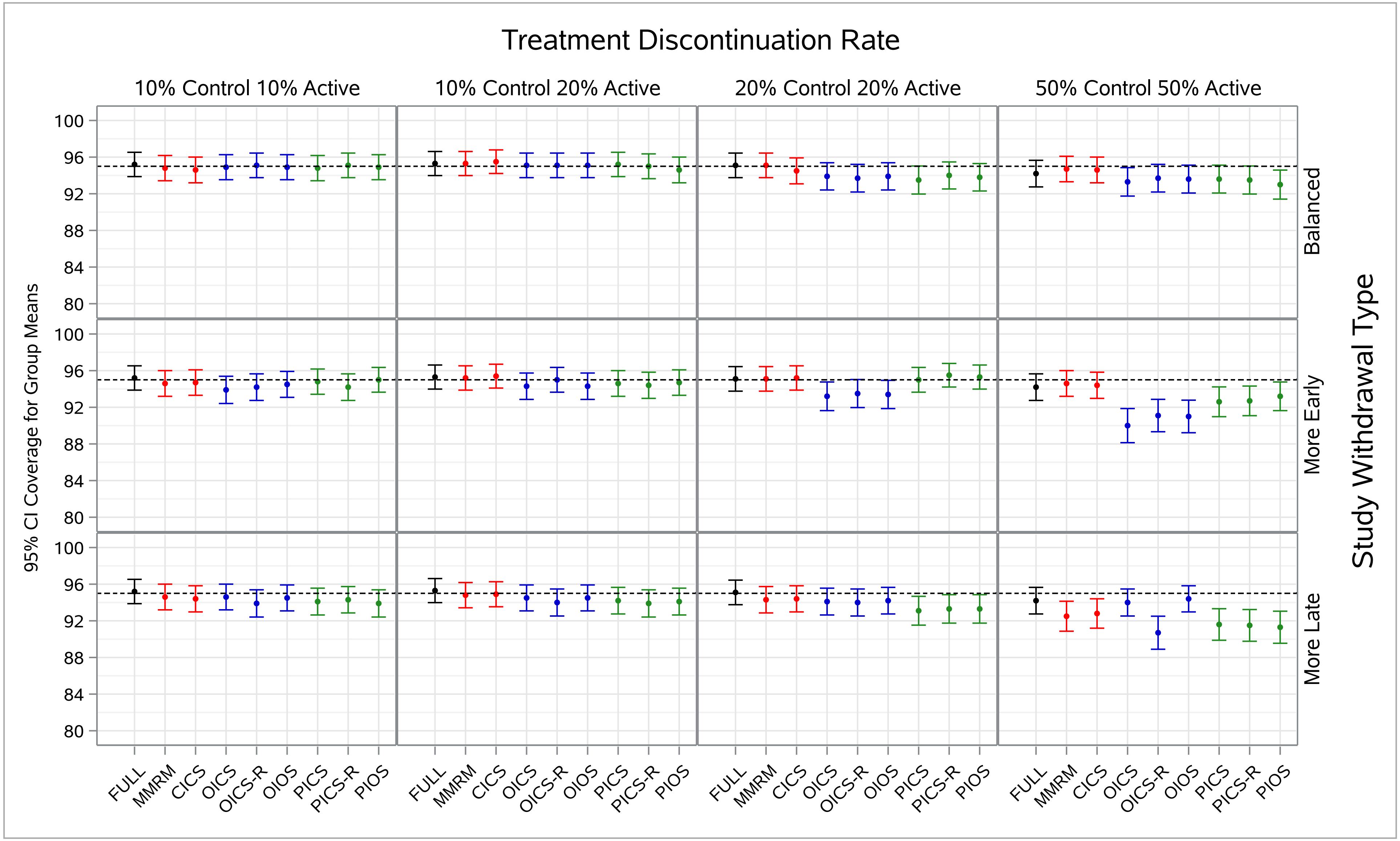

**SI Figure 6.9**
**95% CI Coverage for Group Means - Control Group**

**Disc. Scenario: Same as Active**
**Disc. Mechanism: DNAR2**

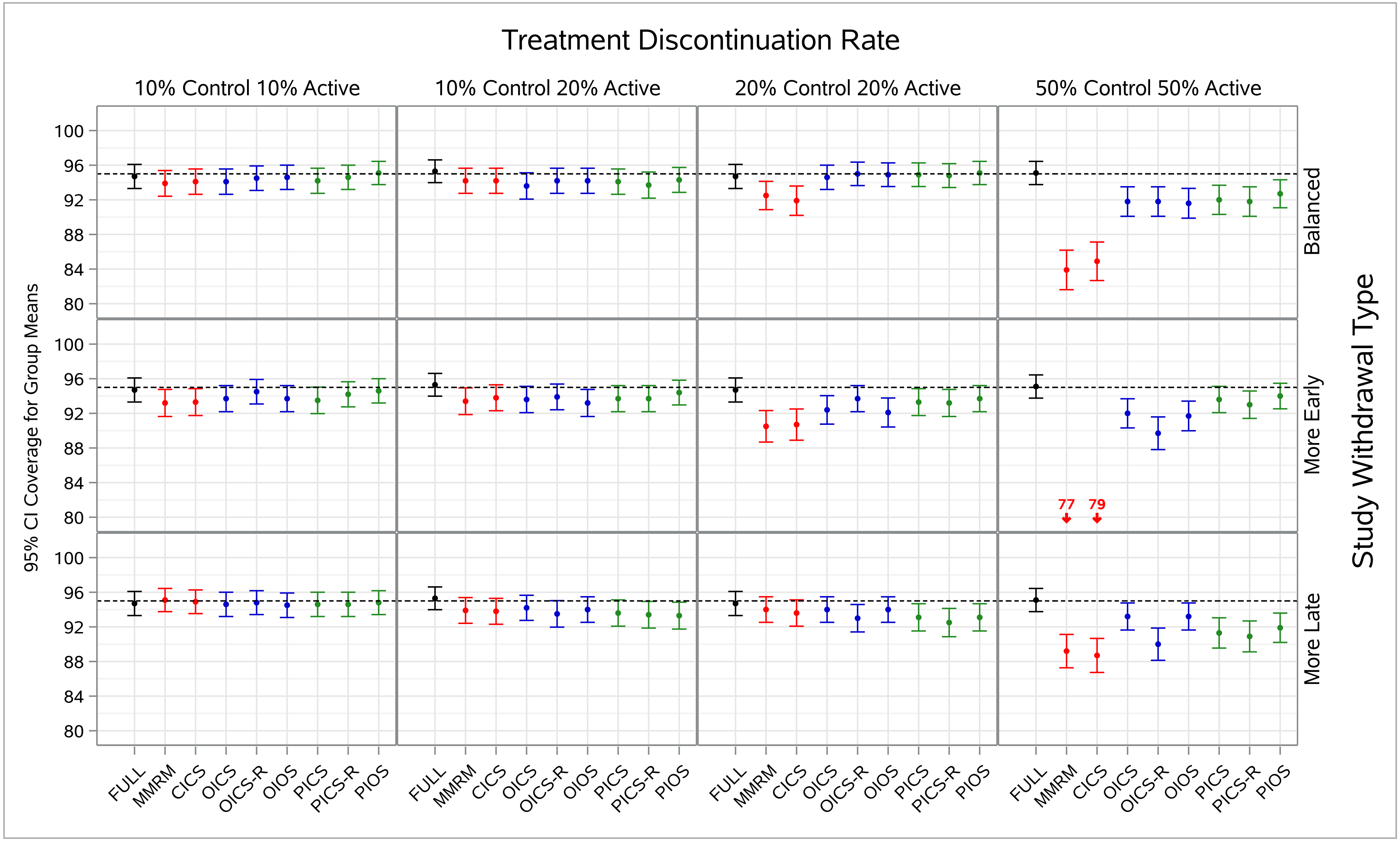

**SI Figure 6.10**
**95% CI Coverage for Group Means - Active Group**

**Disc. Scenario: Same as Active**
**Disc. Mechanism: DAR**

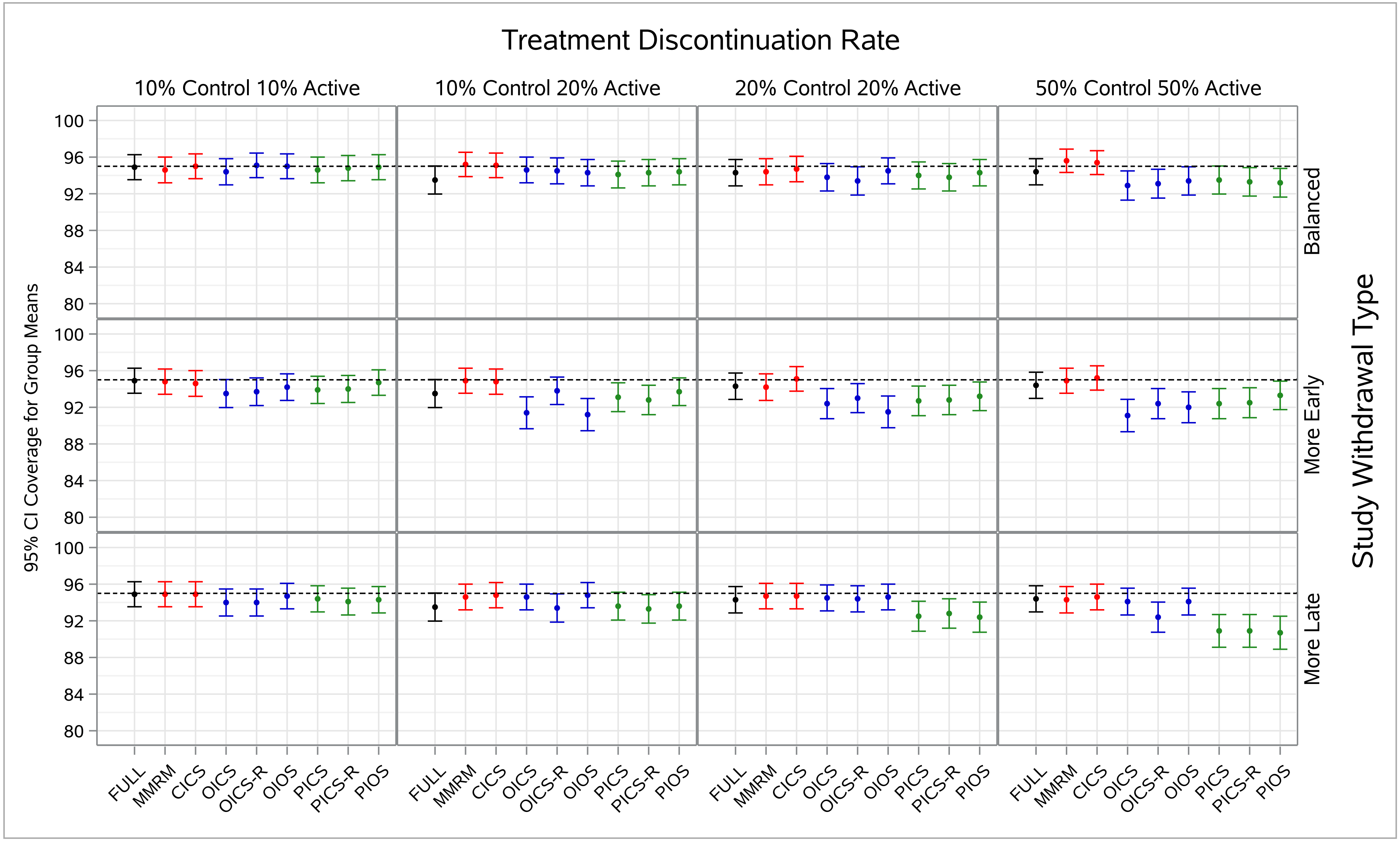

**SI Figure 6.11**
**95% CI Coverage for Group Means - Active Group**

**Disc. Scenario: Same as Active**
**Disc. Mechanism: DNAR1**

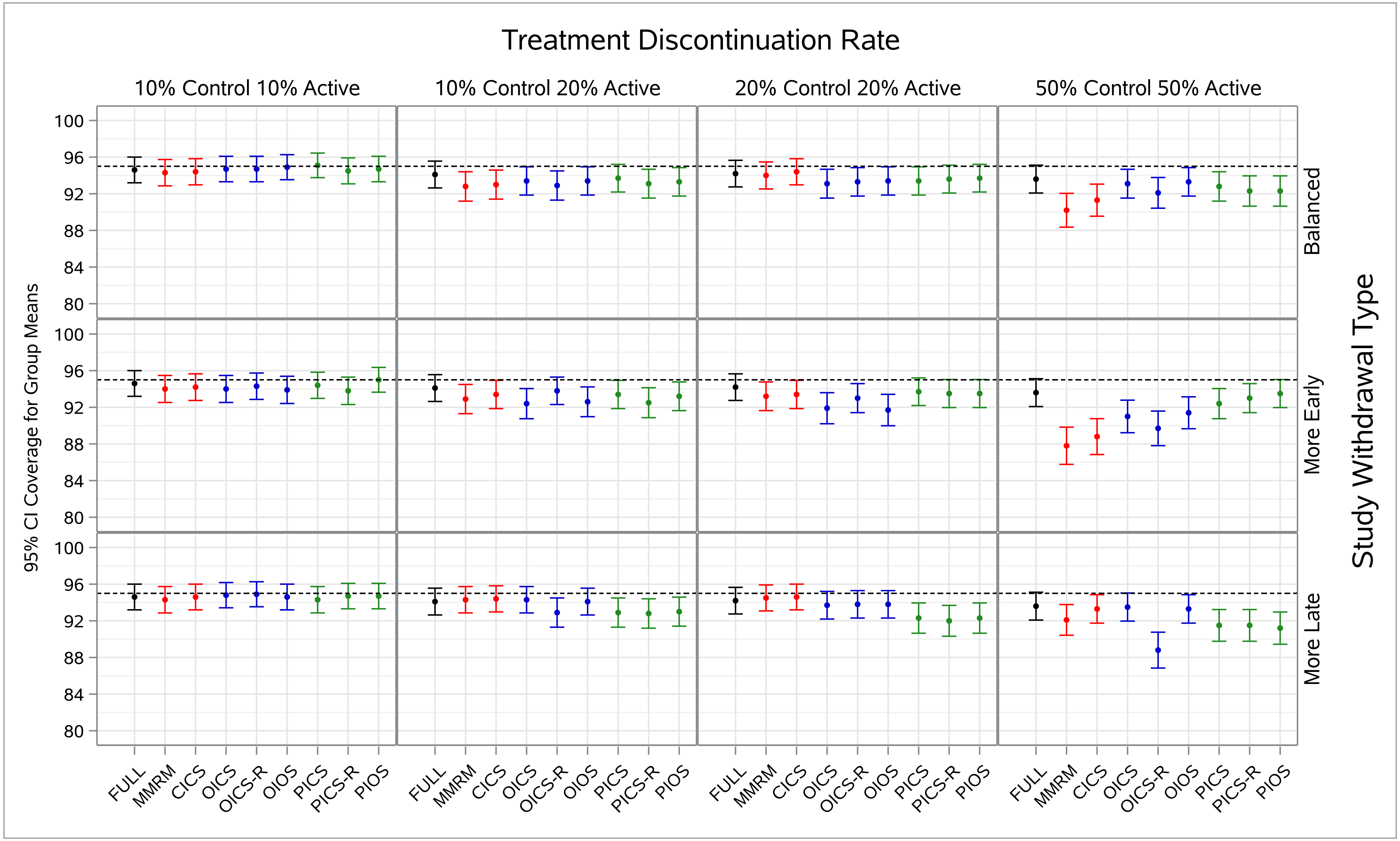

**SI Figure 6.12**
**95% CI Coverage for Group Means - Active Group**

**Disc. Scenario: Same as Active**
**Disc. Mechanism: DNAR2**

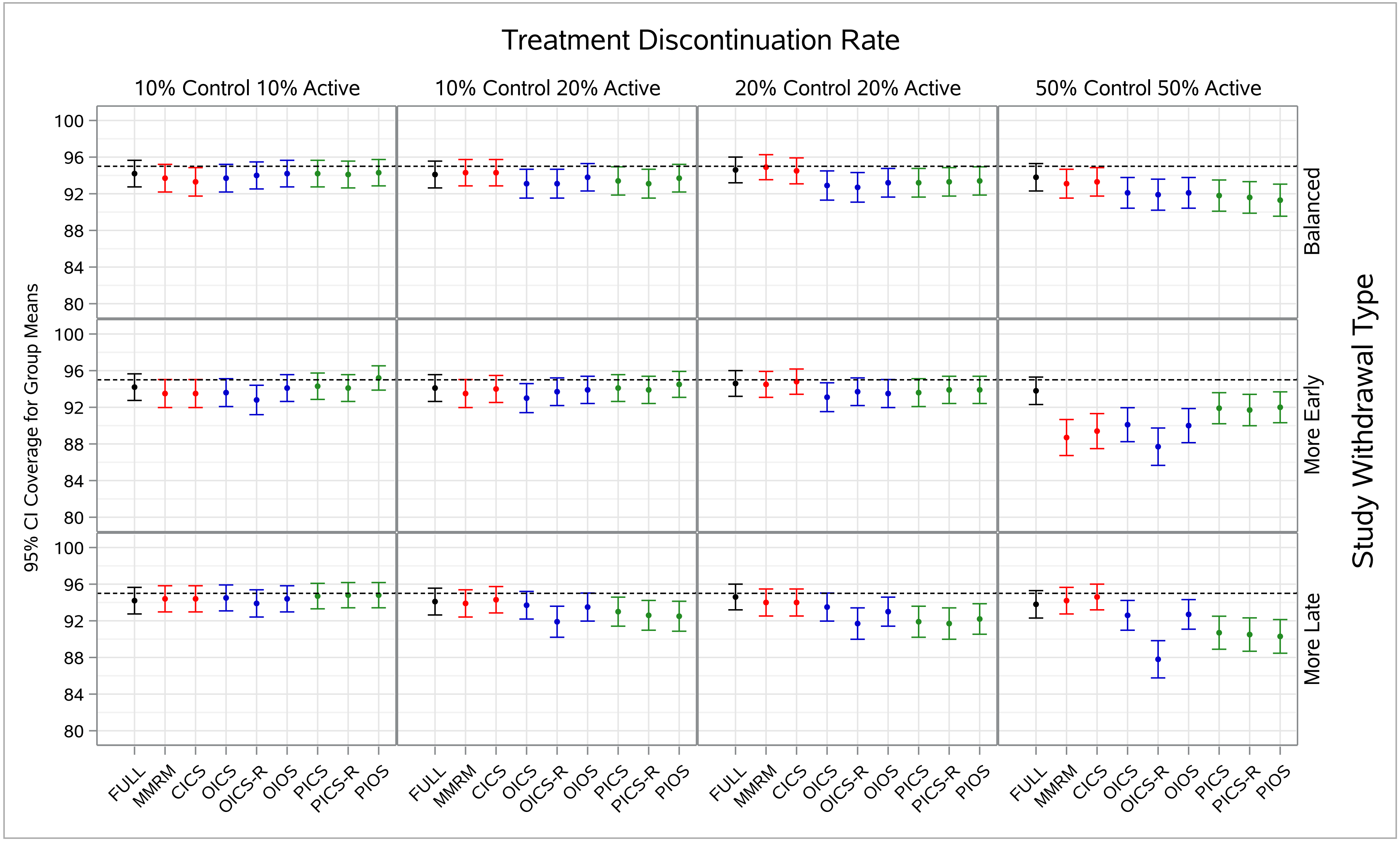